# Change Impact Analysis Based Regression Testing of Web Services

*Submitted in partial fulfillment of the requirements for the degree of*

**Master of Technology**

*By*

**Animesh Chaturvedi**

**(Roll no: 1010101)**

**Supervisor**

**Dr. Atul Gupta**

**Computer Science and Engineering**

**PDPM INDIAN INSTITUTE OF INFORMATION TECHNOLOGY, DESIGN AND MANUFACTURING JABALPUR**

**2014**

Dedicated to

***My Grand Father (Dada Ji) late Sri Ram Sevak Chaturvedi,***

***(Nana Ji) late Sri BL Pandey***

***And***

***My former Guide (in IET-DAVV) late Prof. Manohar Chandwani***

# APPROVAL SHEET

The thesis entitled "**Change Impact Analysis Based Regression Testing of Web Services"** by **Mr. Animesh Chaturvedi, Roll No. 1010101** is approved for the degree of Master of Technology in Computer Science and Engineering.

Examiners

____________________

____________________

____________________

Supervisor

____________________

Chairman

____________________

Date: ___________

Place: __________

# CERTIFICATE

This is to certify that the thesis entitled, "**Change Impact Analysis Based Regression Testing of Web Services"**, submitted by **Animesh Chaturvedi, Roll No. 1010101** in partial fulfillment of the requirements for the award of **Master of Technology** Degree in **Computer Science & Engineering** at PDPM Indian Institute of Information Technology, Design and Manufacturing Jabalpur is an authentic work carried out by him under my supervision and guidance.

To the best of my knowledge, the matter embodied in the thesis has not been submitted elsewhere to any other university/institute for the award of any other degree.

Date:
Place:

**Atul Gupta**
Associate Professor,
PDPM Indian Institute of Information Technology,
Design and Manufacturing Jabalpur,
E-mail: atul@iiitdmj.ac.in

# ACKNOWLEDGMENT

I would like to express my sincere gratitude toward my thesis supervisor Dr. Atul Gupta for his continues guidance and encouragement. Working under his supervision helped me to become an independent researcher. This research carried under his guidance will be a turning point of my research career.

I am grateful to Director/Prof. Aprajita Ojha and Academic Dean/Prof. Vijay Kumar Gupta, for motivating and helping me in academics during my graduation. I am also grateful to Dr. Pritee Khanna and Dr. Sraban Kumar Mohanty in Computer Science and Engineering for their valuable teaching at course work. Further, I would like to thanks to office-staff, especially to Mr. Avashesh Kumar Pal for their helps. I will always be indebted towards my family members and friends for their continuous support.

I wish to express my sincere thanks to all my friends and seniors, specifically Mr. Saurabh Tiwari, Mr. Santosh Singh Rathore, Mr. B. R. C. Reddy, Mr. Gyan Singh Yadav, Mr. Mitul Kumar Ahirwar, Mr. Amit Dhama, Mr. Allahbaksh Asadullah and Mr. Basava Raju for their constant moral support and academic help.

I want to thank MESOCA & ICSM 2013 committee specially Dr. Alexander Serebrenik, Dr. Anca Ionita, Mrs. Grace A. Lewis and Dr. Marin Litoiu for travel grant to present my paper in Eindhoven, The Netherlands.

I wish to thank Late Prof. Manohar Chandwani for sincere guidance, early feedbacks, and advices to do research. For valuable teachings and advices on my thesis work, I am thankful to Dr. Anjaneyulu Pasala, Dr. Srinivas Padmanabhuni and Dr. Hemant Mehta.

I am thankful to IIIT-DM-Jabalpur and IIT-Kanpur members for their support and appreciation. I am thankful to Dr Atul Gupta, Dr. Amey Karkare, Miss Shubhangi Chaturvedi and Miss Carolyn Walther for helping me in improving my English writing.

**Animesh Chaturvedi**

# ABSTRACT

Reducing the effort required to make changes in web services is one of the primary goals in web service projects maintenance and evolution. Normally, functional and non-functional testing of a web service is performed by testing the operations specified in its WSDL. The regression testing is performed by identifying the changes made thereafter to the web service code and the WSDL. In this thesis, we present a tool-supported approach to perform efficient regression testing of web services. By representing a web service as a directed graph of WSDL elements, we identify and gathers the changed portions of the graph and use this information to reduce regression testing efforts. Specifically, we identify, categorize, and capture the web service testing needs in two different ways, namely, Operationalized Regression Testing of Web Service (ORTWS) and Parameterized Regression Testing of Web Service (PRTWS). Both of the approach can be combined to reduce the regression testing efforts in the web service project.

The proposed approach is prototyped as a tool, named as Automatic Web Service Change Management (AWSCM), which helps in selecting the relevant test cases to construct reduced test suite from the old test suite. We present few case studies on different web service projects to demonstrate the applicability of the proposed tool. The reduction in the effort for regression testing of web service is also estimated.

**Tool link** for more detail

https://sites.google.com/site/animeshchaturvedi07/research/awscm

# TABLE of CONTENTS

# LIST of FIGURES

# LIST of TABLES

# CHAPTER 1

# INTRODUCTION

## 1.1 Web Service

A web service is a composable unit of web application, which can interact with other applications using Simple Object Access Protocol (SOAP) messages. Various operations are performed by the web service are specified using an XML based interface called Web Service Description Language (WSDL). SOAP is a protocol specification for exchanging structured information in XML for a web service, over HTTP or SMTP. A web service operation is a unit of functionality to be provided as a part of web service, which is interface by a WSDL. A web service can be a standalone or a part of a web application that can be independently developed, deployed, tested, and put to operation. Web services can be implemented in variety of ways like using object-oriented programming in Java or C++, or using procedural languages like C. Many upcoming computational domains like Cloud Computing and Service Oriented Computing typically rely upon web service framework.

Figure 1.1 shows the communication involved in a web application that is making use of web services. The client part of the web application generates soap request messages to invoke an operation of web service residing on web server. The operation code, written in programming language in C, Java etc, responds to the client request with the soap response messages.

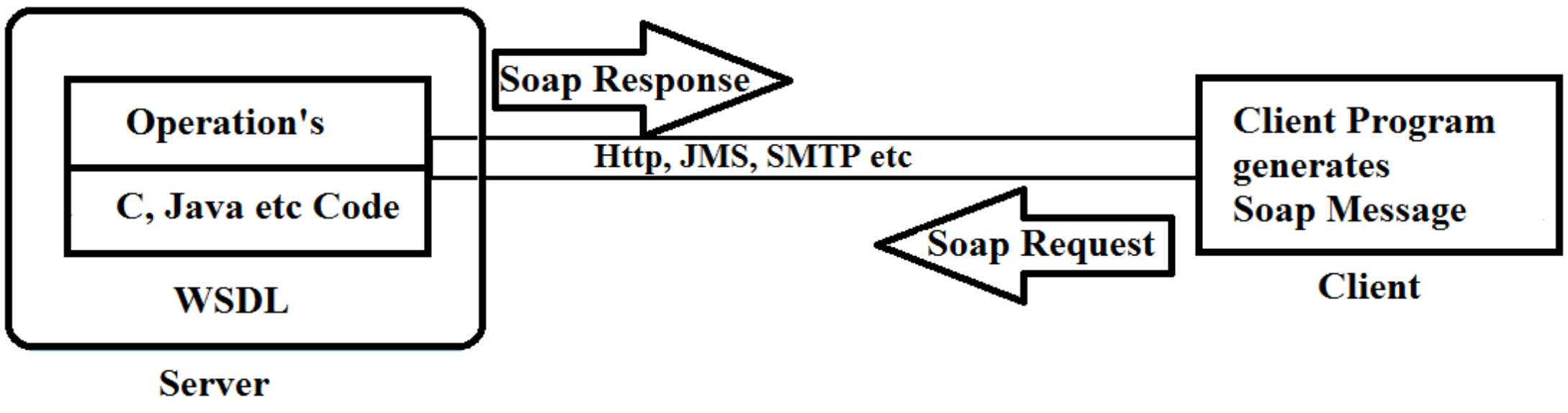

**Figure 1-1 Web Service communication**

With the success of service-oriented architecture, web services have become important components of web applications. The advancement in the field of web services has tremendously helped to deal with issues like interoperability and portability that helped the domain like Grid Computing, Cloud Computing, and virtualization to realize. For example, for Grid services deployment on Grid environment as described in [61]. Moreover, Cloud systems offer services at two different layers: Infrastructure Layer, Virtual Machine Layer. Usually the Infrastructure Layer designed according to the Service Oriented Architecture (SOA) model, exploiting the web services technology [62]. In order to manage the platform with a service interface they adopted WSAG, a framework that implements the WSAI (Web Services Agent Integration) project to supports the communication between Web Services and agents [62]. An excellent example of WSDL usage is Eucalyptus; the popular deployed cloud computing software platform [59] with Eucalyptus uses WSDL on Github [60].

## 1.2 Web Service Testing

Functional and non functional testing of a web service is normally conducted by generating automated test cases for the operations by parsing the WSDL of the web service [7][8]. During testing, the web service communicates with the client using Soap request and response messages. Here we define some terms as given below.

**Test case** is the set of input data execution conditions and expected outcome to test specific aspect of a software artifact under test.

**Test sequence/step** is a sequential or step-by-step invocations of web service operations for a given test data.

**Test suite** is a collection of test cases.

**Test template** is a test sequence/step, which can be used to generate multiple test cases by augmenting it with suitable test data.

**Test Case Assertion**: It allows to create simple and complex assertion on any property from project to individual test sequence (or test steps) in a test case for request/response, JMS, JDBC or Security-related activities and can be grouped and Boolean logic. An example shown in fig 1.2 test cases of operation 'readingFile' it shows green if test assertion success and red if test assertion fail. It failed when our asserted string 'xyz123' is not present in soap response and failed when Service Level Agreement (SLA) violated when we asserted for response time 2 milliseconds and actual response time is 41 whereas rest all test assertions for other test case are passed.

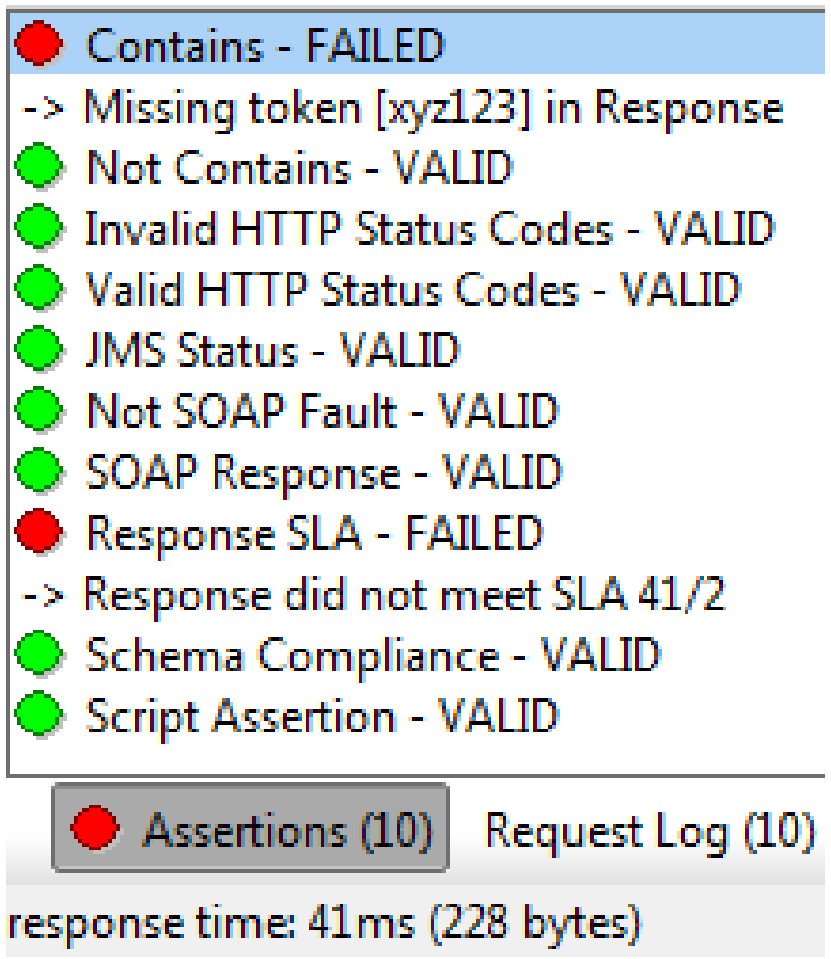

**Figure 1-2 SOAPUI Test assertion for test sequences/steps.**

The web service testing also follows the standard software testing approaches. Following definitions are according to the IEEE Standard Glossary of Software Engineering Terminology [63].

**1. White Box Testing:** "A system or component whose internal contents or implementations are known. Pertaining to an approach that treats a system or component" [63]. Similarly, for a web service when we have code and want to test the internal structure of a code, tests may be conduct to insure the operations done according to the requirement. The regular static testing techniques examples are Statement coverage, Conditional testing, Path coverage, Loop testing. In this thesis we presented web service example supported approach/algorithm of unit based testing for white box testing.

**2. Black Box Testing:** "A system or component whose inputs, outputs, and general function are known but whose contents or implementation is unknown or irrelevant. Pertaining to an approach that treats a system or component whose inputs, outputs, and general function known but whose contents or implementation are unknown or irrelevant" [63]. We perform black-box testing of a web service using the interface definition as specified in WSDL.

**Tool Support available:** There are many web service-testing tools available for web service testing. In this thesis, we have used SOAPUI and JMeter. We can perform the regression testing of web service code with the traditional methods of program regression testing [2][3][12][15][16][17][18]. On the other hand, we can also perform the black box testing of web service using WSDL [7][8] with the help of testing tools like SoapUI and JMeter. In this thesis, we have developed the tool support that considers both traditional regression testing and web service techniques.

## 1.3 Regression Testing

Regression testing is a kind of software testing that aims at discovering bugs in software after changes made to it. This insures that changes did not introduce any new bugs in the code. Here the main purpose is to determine whether a change in one part of the software has affected the other parts of the software.

Change impact analysis is a process of gathering changed portions in the code while the project is going through its evolution phase. Traditionally, regression testing is performed based on change impact analysis by mapping the results changed portions of the code to their respective test cases. While increasing the probability of detecting new errors because of changes made in the code. The regression testing aims to complete the mentioned task within the time constraint.

According to the terminologies described in IEEE standards on software engineering [63], “selective retesting of a system or component to verify that modifications have not caused unintended effects and the system or component still complies with its specified requirements. Testing required to determine that changes to a system component has not adversely affected functionality, reliability or performance and has not introduced additional defects”. Some of the issues and their strategies that are typically followed during regression testing include.

1. Tester might have avoided the failure but cannot get to the underlying cause of it.
2. Side effects of fixed bug, this means bug might itself fix automatically but this fix might create other bugs.
3. If two or more tests cases are similar, remove the least effective test case.
4. Collect all the test cases that are continuously passing.
5. Changes are small or large to code, all the affected portions required regression testing.
6. Affected portions due to changes required to be trace for cost effective regression testing.

Note: Regression testing is different from regression analysis.

## 1.4 Motivation

Regression testing and retesting of a modified web service can be very costly if it is conduct by running all the test cases, as it tends to generates many test cases. The cost of regression testing can be reduced significantly by identifying and testing only the modified portions of the web service. This avoids the costly construction of new test cases and the unproductive rerunning of existing test cases when it can guarantee that the unmodified code of web service will produce the same results as it is produced previously. Web service maintenance can be optimize by doing selective testing of recently modified and inserted portions of the code only with the assurance that new and old code would confirm to the changes in requirements. The change impact analysis is the process of gathering the affected portions of the code, and is considered as one of the challenging problems of program comprehension [56][57][58]. We use change impact analysis to gather the old and new parts of the modified code.

This thesis works on approach that if testing scenario is properly designed then it is more cost effective in regression testing. We used two types of scenarios for traceability information similar that can find affected components with their associated test cases for regression testing. If this process can be automated in a tool, then it can save manual efforts, by detecting the changes in the web service automatically and selects the relevant test cases to perform efficient regression testing of web service.

State of art is add-on by categorizing various types of necessary web service analysis. Analysis can be optimized by doing selective maintenance of inserted and modified features, with the guarantee that new and old code works according to requirements. This observation is re-explored in the context of WS - the thesis proposes two change impact analysis techniques based on operations and parameters of a WS. These techniques are useful to bridge the gap between code level and WSDL parsing based WS testing. Even in some cases without having any detailed knowledge of code or program statements.

For a changed WS from version N-to-N+one, require to undergo from regression testing. An approach is required to gather the change impact analysis on the WSDL and WS code. Further, a mapping technique is required between changes and testing environment. For example, suppose a simple WS that have undergone changes from version WS_1 to WS_2. During ORTWS, we capture the changed operations of WS_1 that still exist in WS_2. During PRTWS, we capture method/operational flow changes in WS_1 to create WS_2. We will select the relevant test case from two different types of existing test suites for ORTWS and PRTWS, to construct two different types of Reduce Regression Test Suite (RRTS). For WS_2 validation, these two RRTS can be executed.

Motivation about approaches is that all these steps are automatic. State of the art for regression testing of WS is add-on by the following contribution

• We proposed various ‘Subset WSDLs’. We performed web service analysis with subset WSDL in a tool to perform automated regression testing of WS. All affected portions are discriminated from unaffected portions that results to retest with reduce number of test cases. Affected portions information helps to slicing the WS code in subset to reduce analysis efforts.

• Two proposed techniques ORTWS and PRTWS, to identified changes in consecutive versions of WS and mapping those changes to their respective scenarios.

• We capture the changes both at WSDL and at code. Thereafter these changes are used in operational and parameter based regression testing. Our objective is to automatically manage the WS changes and mapping them to their respective test cases. Mapping between change impact analysis to the test suite will help in selecting the set of regression test cases.

• This thesis will provide WSDL based WS regression testing technique with or without having detailed knowledge of code. This will result in efficient WS regression testing by facilitating the standard WSDL based WS testing.

• We demonstrated few case studies that will help in the better understanding.

## 1.5 Problem Definition

Regression Testing of Web Service (RTWS) can be optimized by doing the Operationalized and Parameterized analysis for the modified and inserted portion only with the guaranteed that new and old code will meet the changes in requirements. With the proper analysis we can avoids the costly construction and re-run of the useless test cases, which leads us to reduced regression testing efforts.

In our approach, we capture the web service testing needs in two different settings, namely, Operationalized Regression Testing of Web Service (ORTWS) and Parameterized Regression Testing of Web Service (PRTWS) as shown in fig 1.3. We also show that the two types of testing can also combined to reduce the regression testing efforts further. ORTWS is performed by generating the test cases for each operation defined in the WSDL. PRTWS is performed by identifying one of the input parameters as the 'primary' parameter generates test cases by varying the values of the other input parameters while keeping the 'primary' parameter at a fixed value.

The proposed approach is prototyped as a tool, named as Automatic Web Service Change Management (AWSCM), which helps in selecting the relevant test cases along with test sequences/steps from the old test suite.

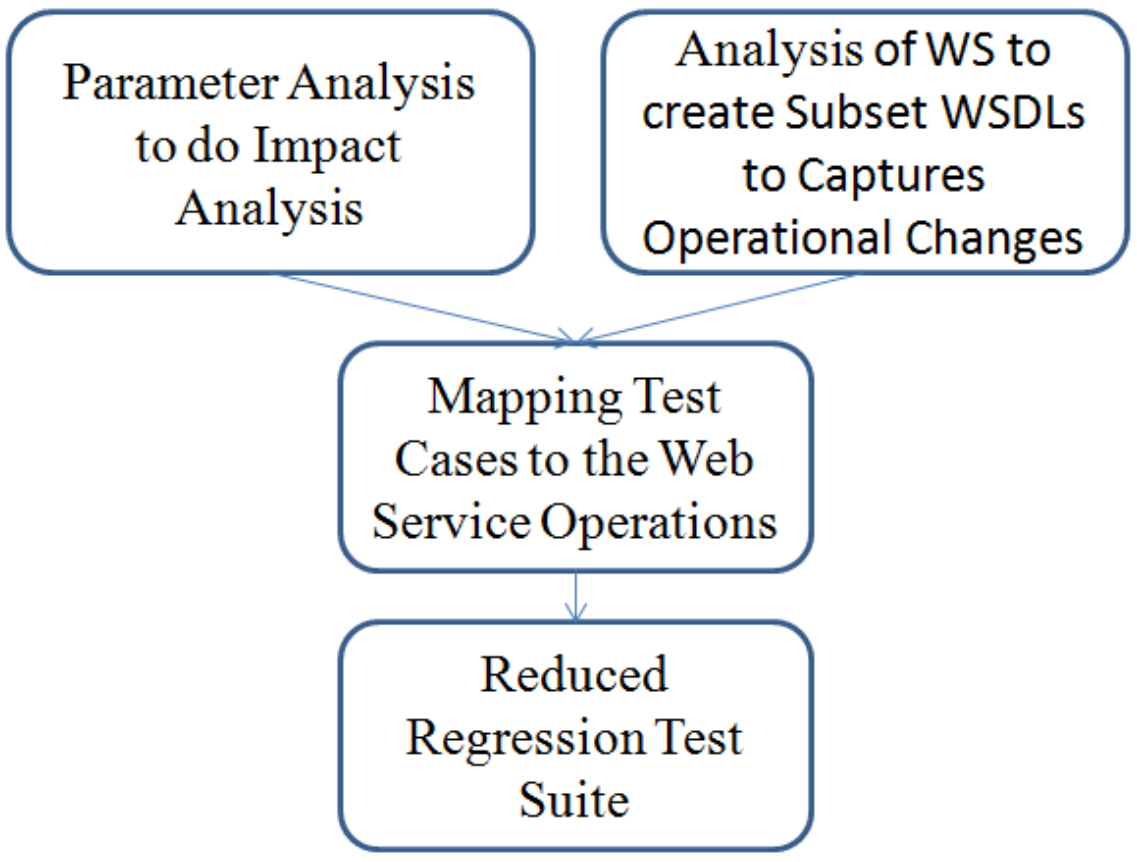

**Figure 1-3 Introduction to approach of ORTWS & PRTWS**

## 1.6 Thesis Organization

The rest of the thesis organized is as follows:

**Chapter 2:** In this chapter, we review the basic concepts and the work that has been done in order to deal with the problem of change impact analysis based regression testing of web service. Starting from the most relevant historical research, thereafter specifically we review some results in the field of regression tests, cost-effectiveness regression testing approaches, regression testing of web services, types of web service testing and tools support. Towards the end, we compare and contrast our approach with the existing solutions.

**Chapter 3:** In this chapter, for the Operationalized analysis of web service we present our approach to perform The Operationalized Regression Testing of Web Service (ORTWS). In the process of ORTWS, we created the Subset WSDL with the operations and mapped them their respective test cases. In the end, we demonstrated some case studies for this approach.

**Chapter 4:** In this chapter, for the Parameterized analysis web service we proposed The Parameterized Regression Testing of Web Service (PRTWS). In the process of PRTWS, we created a ‘primary’ parameter testing scenarios and mapped them to their respective test cases. In the end, we demonstrated some case studies for this approach. Afterward we perform both ORTWS and PRTWS for ‘BookService’ project.

**Chapter 5:** In the chapter, we presented our tool Automated Web Service Change Management (AWSCM) that we have developed to prototype the two approaches presented namely ORTWS and PRTWS, presented in Chapter 3 and Chapter 4 respectively. We highlight various applications of proposed tool AWSCM.

**Chapter 6:** In this chapter, we draw our conclusion and indicated the future directions.

# CHAPTER 2

# RELATED WORK

Regression testing aims to test the changes made in the code to ensure that the changes are not causing any side effects. This can be ensured by re-running all the valid old test cases in the old test suite, which however, can be prohibitively costly. A better approach is to detect the changed portion of the code and selects the test cases that can map the changed and unchanged portions of the code.

Bennett and Rajlich in [57] had given a roadmap to conduct software maintenance and evolution, such that program comprehension is the prerequisite of impact analysis thereafter software is validated with regression testing. For manitenance and evolution process we are following the same roadmap as described in [57]. Gold and Bennett in [56] had given WS comprehension issue and key research topics. We inferred from both [56][57] that there is huge scope of research in web service changes maintenance, specifically for technical and business applications.

David Binkley in [1] described about reducing regression testing cost by an algorithm that uses semantic differences and similarities between old and new programs. David

Binkley in [2], described program that generates reduced test cases by identifying components that need tests after getting differences and similarities between the old and new programs. We are reiterating the same approach in context of web service, to get differences and similarities to map them  to the test cases.

W. T. Tsai et al. in [3] proposed a scenario-based functional regression testing for ripple effect analysis based on end-to-end integration test scenarios. A scenario based OO test framework in [4], for re-testing according to tester's need to specify the modified scenarios so that new test scripts can be generated to test the modified feature. We are also describing the similar type of scenario in context of web services, to make mapping between the test suite and web service code.

## 2.1 Web Service Testing

W. T. Tsai et al. in [7] described four ways to extending WSDL to facilitate web service testing. Research paper [7] had adviced tester to perform web service testing using WSDLs. X. Bai et. al. [8] desribed WSDL based test case generation for web service testing. Now, almost every web service testing tool had infered [8] to perform web service testing. Similarly in this thesis we are considering WSDL based testing in our approaches of regression testing of web service.

Sana Azzam et al. in [28],  presented an  analysis on the basis of various testing approaches of different web service to compare automated testing tools (SoapUI, PushToTest, Jmeter and WebInject). This paper is evidence that SoapUI and JMeter were important tools and thus we are using them in our case studies. In [28] there is no comparitve study between tools for the regression testing of web service. This gives us evidence that change impact anlysis based regression testing is not incoporated with SoapUI and JMeter till now.

## 2.2 Regression Testing

Regression testing can be categorized into four categories, namely, retest-all, test case selection, test suite reduction, and test case prioritization. Leung and White in [11] found that selective strategy is more economical than the retest-all strategy. In our thesis, we are also trying to do selective regression testing of web service instead of retest all. In the test suite reduction of regression testing, Harrold et al. in [12] identified changes and then eliminated redundant as well as, obsolete test cases in the test suite. According to the paper [12], we are also constructing the reduced test suite for efficient regression testing. Chen et al. in [13] identified, exact subset of the test suite that must rerun to test a new version of a system for selective regression testing. Chen and Lau in [14] found redundancy in test suites, when some of its proper subsets can still satisfy the same testing objective.

## 2.3 Cost-effectiveness Regression Testing Approaches

Rothermel and Harrold in [16] used Control flow graphs of program and its modified version to select test cases that execute changed code from the original test suite. We are inferring from [16], to use web service graph visualization and automatically gather the changes to create the subset selection of the old test suite.

Vokolos and Frankl in [17] describe textual differencing technique for regression testing by comparing code of old and the new version of the program. Similarly, to gather the operation code changes, we also used textual difference. A. Orso et. al in [18] proved that cost-effectiveness improves with the size of the program under test. We also find the same result, as the size of our case study project increases the cost effectiveness of our approach is also increasing.

## 2.4 Regression Testing of Web Service

M Ruth et al. in [20] described a mechanism that transforms the service code into a local program for applying a safe regression test suite techniques, with an end-to-end manner using a code transformation-based approach. In another technique, M. Ruth et al. in [23] described a safe regression test selection techniques for Web services. They do not automated their proposed approach and even more they had not used standard way of WSDL based automated web service testing like in SoapUI and JMeter.

Abbas et al. in [21] Regression testing of web service based applications with the safe regression test suite using finite state automaton technique, which is white box. They had proposed an automaton based approach to find the safe regression test suite but not giving any implementation support.

M. D. Penta et. al in [22] described a tool-supported approach for user to run and re-run test suites repeatedly for functional and non-functional expectation. T. A. Khan in [24] described model-based regression testing of web service. Both [22] and [24] are not using standard practice of WSDL based web service testing like in SoapUI and JMeter.

Askaruinisa and Abirami in [25] developed a prototype-based approach for generating web service test cases, using WSDL-S and Object Constraint Language (OCL), technique of Orthogonal Array Testing (OAT). They had developed their prototype tool, but without giving any standardized approach that can be implemented and integrated with SoapUI and JMeter. In this thesis, we are considering different approach of regression testing i.e. change impact analysis.

P. Bhuyan et. al. presented a survey in [29] on different type of issues in Regression Testing of SOA by comparing different approaches. The paper [29] giving some evidence that there is a requirement of automated change impact analysis based approach for regression testing of web service.

## 2.5 Types of Web Service Testing

Software Metrics is a measure of some property of a piece of software or its specifications. Regression testing of Web service performs by manipulating information provided by WSDL and web service code. We are modifying previous metric [8] in more precise, formal, and well-defined way for Web service testing. Following are the Web service testing metric based on the WSDL information.

**T1 Web Service Policy Testing (WSPT):** It is a type of non-functional testing, which test behavior of a policy assertion that represents a requirement, a capability, or other properties like binding security policy assertion. MIME media type name of application,

MIME subtype name, Required parameters, Optional parameters, Encoding considerations, Security considerations, Interoperability considerations, Published specifications, Applications which uses this media type, Additional information (File extension, Fragment identifiers, Base URI, Macintosh File Type code, Intended usage, Author/Change controller).

**T2 Message Part Testing (MPT):** This is to test the WSDL capability to post Soap message correctly and get the Soap response correctly for that we required testing message part communication.

**T3 Operation Code Testing (OCT):** It is a type of functional testing that is used to test the operation logic. Logic of operation must be tested thus proper output will be generated by Web service according to the requirement of business. A proposed technique of doing this is further described and explains in details at chapter 3.

**T4 XSD-Input/Output Testing (XSDIOT):** It is a type of functional testing, performed to provide correct input and get output results from the service. This can do by manipulating XSD of web service. It is of three type Input dependency type, Output Dependency type and Input-Output Dependency type. A proposed technique of doing this is further describe and explains in details at chapter 4.

**T5 Web Service Composition Testing (WSCT):** It might be possible that Web service combine to another component of web world i.e. one Web service is calling other Web service, Website or some other web component. Therefore testing of component calling by web service is also required.

In this thesis, main emphasis is given to, T3 Operational Code Testing (OCT) in chapter 3 and T4 XSD Input - Output Testing (XSDIOT) in chapter 4 respectively. Operation and input-output parameter is an important part of web service as business logic is coded in operation and communication takes place with input-output.

## 2.6 Tools Support

Works done in this thesis relate to the field of change impact analysis tools are as follows. Luuk P. described various output formats for content returned by many Diff tools and Diff XML [30]. X. Ren et al. in [31] proposed Chianti tool to analyze two versions of a Java program, and reported in terms of affected tests. Anjaneyulu Pasala et al. in [37] proposed a prototype tool called InARTS for the regression test suite selections based on analyzing the dynamic behavior of the application. Jiang Zheng and others presented I-BACCI process for change identification and then performing regression test selection. As I-BACCI is conducted in the scenario of black-box dynamic link library components [32], COTS-based Applications [33] and source code is not available [34], [35]. Hence, I-BACCI is relevant to the regression testing of web service, specifically when source code is not available to web service tester. Neha Rungta and P. Suzette et al. proposed DiSE [41] that performs an inter-procedural analysis and iDiSE [42] combines static and dynamic calling information for change impacted. Xiang Fu et al. presented a WSAT tool [64][63] for the formal analysis of WS, tool works for the intermediate representation, synchronizability with reliability analyzes and handling of XML data manipulation.

We learned from above tools that they had gathered the changes done to the code and used those changes with some proposed mapping technique between test cases to the code. Using change mapping we can construct the reduce regression test suite from the old test suite.

As a standard tool feature of SoapUI, Jmeter, WebInject and Diff tools are as follows

**1. SoapUI:** Its easy-to-use graphical interface and enterprise-class features, soapUI allow you to easily and rapidly create and execute automated functional, and load tests. In a single test suite of soapUI provides complete test coverage - from SOAP and REST-based Web services, to JMS enterprise messaging layers, databases, Rich Internet Applications [47].

**2. JMeter:** JMeter has features to load and performance testing. Jmeter test different types of server/protocol like: Web-HTTP, HTTPS, SOAP, FTP, Database via JDBC,

LDAP, Message-oriented middleware (MOM) via JMS, Mail - SMTP(S), POP3(S) and IMAP(S), Native commands or shell scripts, TCP, Portability Java purity, fully multithreading by separate thread groups, good GUI design allows faster test plan building and debugging, caching and offline analysis/replaying of test results, highly extensible core [48].

**3. WebInject:** WebInject consists of the WebInject Engine (test runner) and an optional User Interface (GUI). The WebInject Engine executes from command line or GUI. This tool can be used for automated testing of web services where test cases are written in XML files and result reports are generated in HTML (for viewing) and XML with pass/fail status, errors, response times are the result after testing [49].

**4. Diff Tool:** This tool contains a features of xml diff through which we can get change between two different xml file (old and changed). Example of this tool is Liquid XML Studio [50], Netbeans Diff [51].

**Usability of above tools for different types of testing:** We made some analysis on the four tools through which we can be helpful in web service testing. We gathered tool that can perform particular type of testing i.e. for a particular type of testing by a particular tool could be beneficial and demonstrated. Following are the different types of testing techniques that can be perform with the help of NetBeans, SoapUI, JMeter WebInject, and Diff tool.

**1. Performance Testing:** For performance testing, we can use WebInject, SoapUI, and JMeter. The tools can be used to compute and give us user-friendly output result of response time of web service.

**2. Load Testing:** For load testing we can use SoapUi and JMeter with this we create, even the most advanced load tests quickly and easily we can apply various measures parameter of load. SoapUI can also be integrates seamlessly with LoadUI, which provides a vastly superior load testing experience. With a visual, drag-and-drop interface, loadUI allows you to create, configure and redistribute your load tests interactively and in real-time.

**3. Functional Testing:** Development platforms (like Eclipse and NetBeans) and testing tools (like SoapUi, JMeter and WebInject) are used for functional testing because both are easy to implement. However, netbeans/eclipse must require project source code to perform functional testing.

**4. Assertion Testing:** A test sequence/step can have assertions that can perform better functional testing, and execute many assertions sequentially together. SoapUi and JMeter tool can be used for the assertion testing as it is very user friendly and implements many type of assertion on the request and response on web service and perform the testing of all assertions together one by one sequentially.

**5. Regression Testing of Web service with the help of Diff tools:** A Diff tool (like Liquid XML Studio or Netbeans Diff) contains a features of diff through which we can get changes between two file (old and changed). We can perform best manual change gathering with the help of a Liquid Xml (Xml editing tool) of web service. These changes can be used manually to do the selective regression test cases and apply them to the testing tool. To automating this feature, is the problem domain were our thesis is projected.

## 2.7 Summary

As described in the last paragraph of the previous section, regressions testing can be done using diff tool after gathering the changed portions of web service and map them to their respective test cases. This thesis will demonstrate automated tool "Automated Web Service Change Management" (AWSCM) for above mention process of change impact analysis based regression testing of web service. We gathered in our literature survey that lots of research work is required in regression testing of web service, specifically in context of, change impact analysis, test suite reduction, and test case prioritization. There is a need of an automated approach to perform change impact analysis based regression testing of web service.

In this thesis, we are considering both white box and black box testing together. Thesis gives special consideration for change impact analysis mapping on WSDL. Moreover, we are using standard tool (like SoapUI, JMeter, NetBeans and Eclipse) to convey our approach of regression testing of web service. We are doing selective regression testing of web service instead of retest all [11] to construct the reduced test suite with the proper subset selection [14] for efficient regression testing [12]. We are inferring from [16], to use web service graph to visualize and automatic gathering of changes to create the subset selection of the old test suite. Textual difference [17] is the used by us to gather the operation code changes. We are finding the same result as mentioned in [18] that as the size of our case study project increases the cost effectiveness is also increasing. Even more, we are considering two different approach of regression testing process based on change impact analysis similar to the tools JRipple [58] and InARTS [37]. JRipple and InARTS provide the program organizations, impact analysis and change propagation for two activities of incremental software changes. Similarly, our tool AWSCM also comprehends the change impact analysis for the web service regression testing.

In the real world project like in industry, where thousands of test sequences/steps are required to execute repeatedly creates confusion. A sincere approach of change impact analysis based regression testing of web service is required.

# CHAPTER 3

# OPERATIONIZED REGRESSION TESTING OF WEB SERVICES

Testing is one of the most important to validate the software according to its requirment. Normally, web service testing is performed by generating test cases for operations specified in the WSDL [7][8]. Web service testing involves communication (with soap requests and responses) between the web service and its clients. Web service testing can be performed using both the traditional white box testing as well as black box testing. White box testing of web service makes use of operation code; whereas the black box testing can be performed, using the operation definition specified in WSDL with software tools like SoapUI [47] and JMeter [48].

The Operationalized analysis based regression testing is performed by getting change impact analysis on the operations. Typically, changes made in web services can be of two types, either changes in WSDL or changes in operation code. In this chapter, we proposed an approach of identifying the changed portions of a web service. Web service is modeled as a directed graph of its constituent components. In web service graph we gather change information in between the subsequent version. Change information of a web service is used to perform the Operationalized analysis with our proposed approach Operationalized Regression Testing of Web Service (ORTWS)

In this approach, the changes in the web service identified in three ways, insertions, deletions, and modifications of its one or more operations. We gathered the changes in between the old and updated version of WSDL. The proposed algorithm compares the two versions of WSDL to identify the inserted, deleted, and/or modified operations. Subsequently this information is used to select test cases for efficient regression testing.

## 3.1 The Structure of a Web Service

A web service can have an abstract view highlighting the operations with input and output, or a detailed view highlighting each operation with further details of message, port, binding, and service information. Abstract and detailed views are also referring as web service graphs, figure 3.1 presents the two views of the web service. The abstract view is describing the policy definition as root, operations as child of the root, and input-output as leaf nodes (which are specified in XML Schema Definition (XSD)). The detailed view elaborates each parts of WSDL to give complete description of the WSDL with message, port, binding, and service information. The figure 3.2 shows a small web service named as SaaS_1 with one operation ‘Index’ for indexing files in a folder.

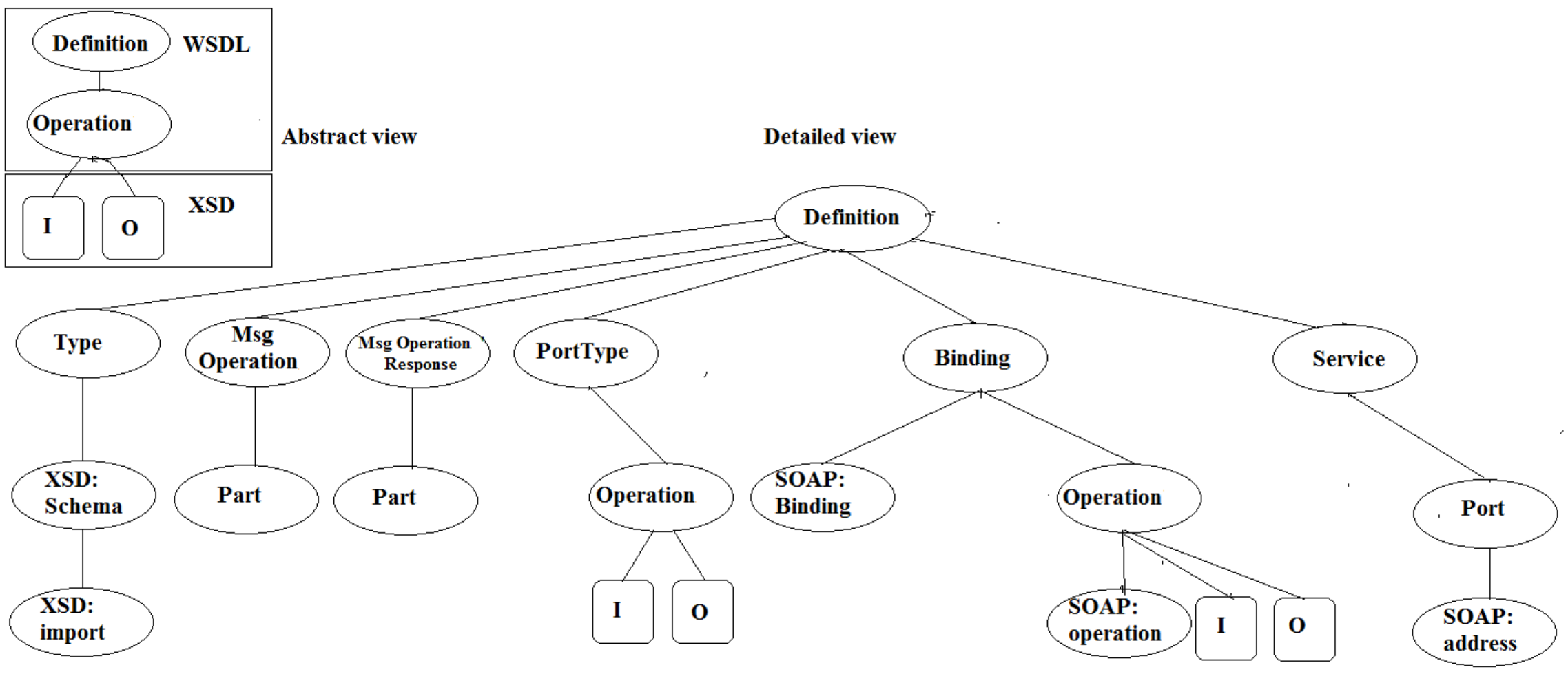

**Figure 3-1 The abstract and the detailed form of web service**

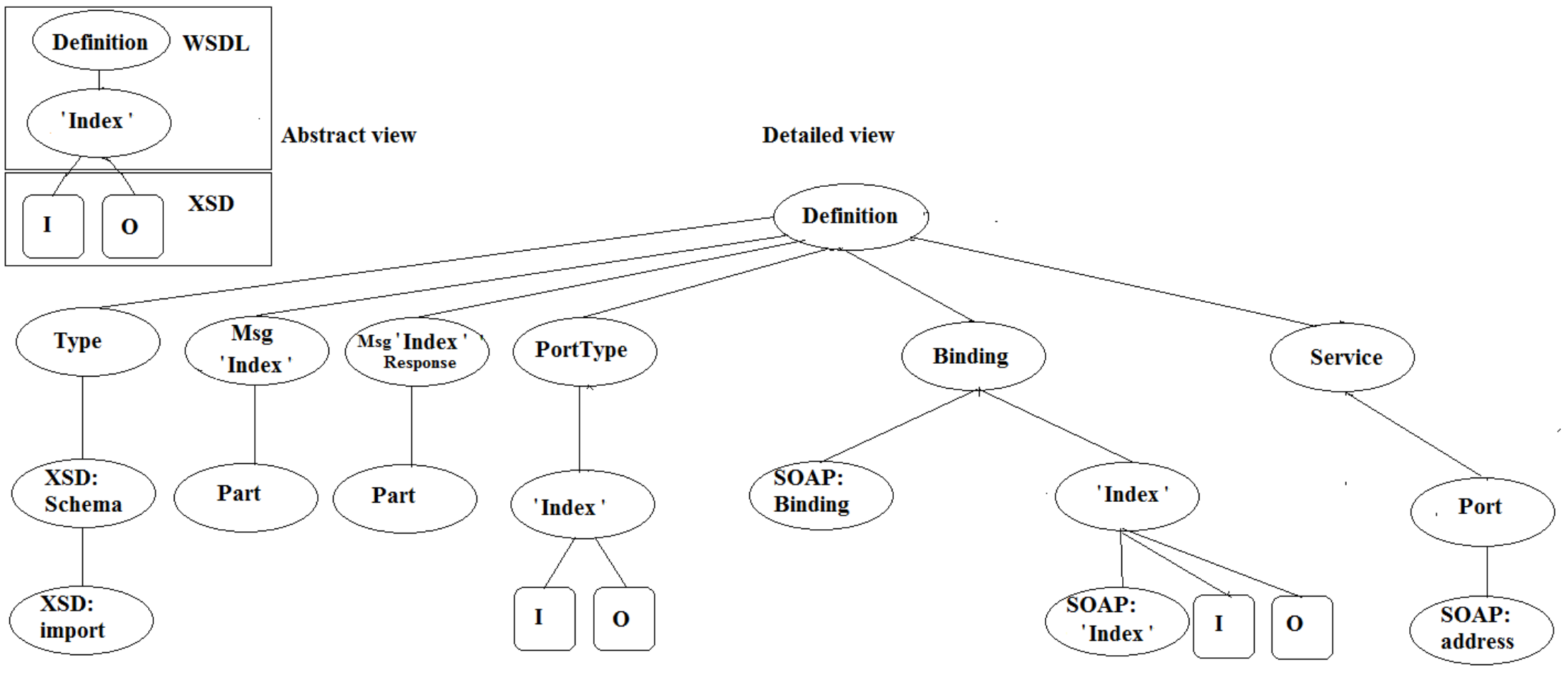

**Figure 3-2 The abstract and the detailed views of web service SaaS_1**

## 3.2 Change Impact Analysis of Web Service

Change impact analysis identifies the regions of changes in the web service using automated diff gathering tools. ‘SaaS’ is simple web service developed by us, which is written in java, has five operations, namely, ‘index’, ‘searching’, ‘readingFile’, ‘editFile’ and ‘edit’. SaaS that have undergone many evolutions, which are demonstrated in different sections in this thesis.

Consider the web service SaaS_1 modified to create SaaS_2 with two additional operations, searching of file and reading of text files, namely, ‘Searching’ and ‘readingFile’. The difference (using diff tools) between SaaS_1 and SaaS_2 is shown in Figure 3.3, where deletions are in red color, insertions are in green color, and modifications are in blue color.

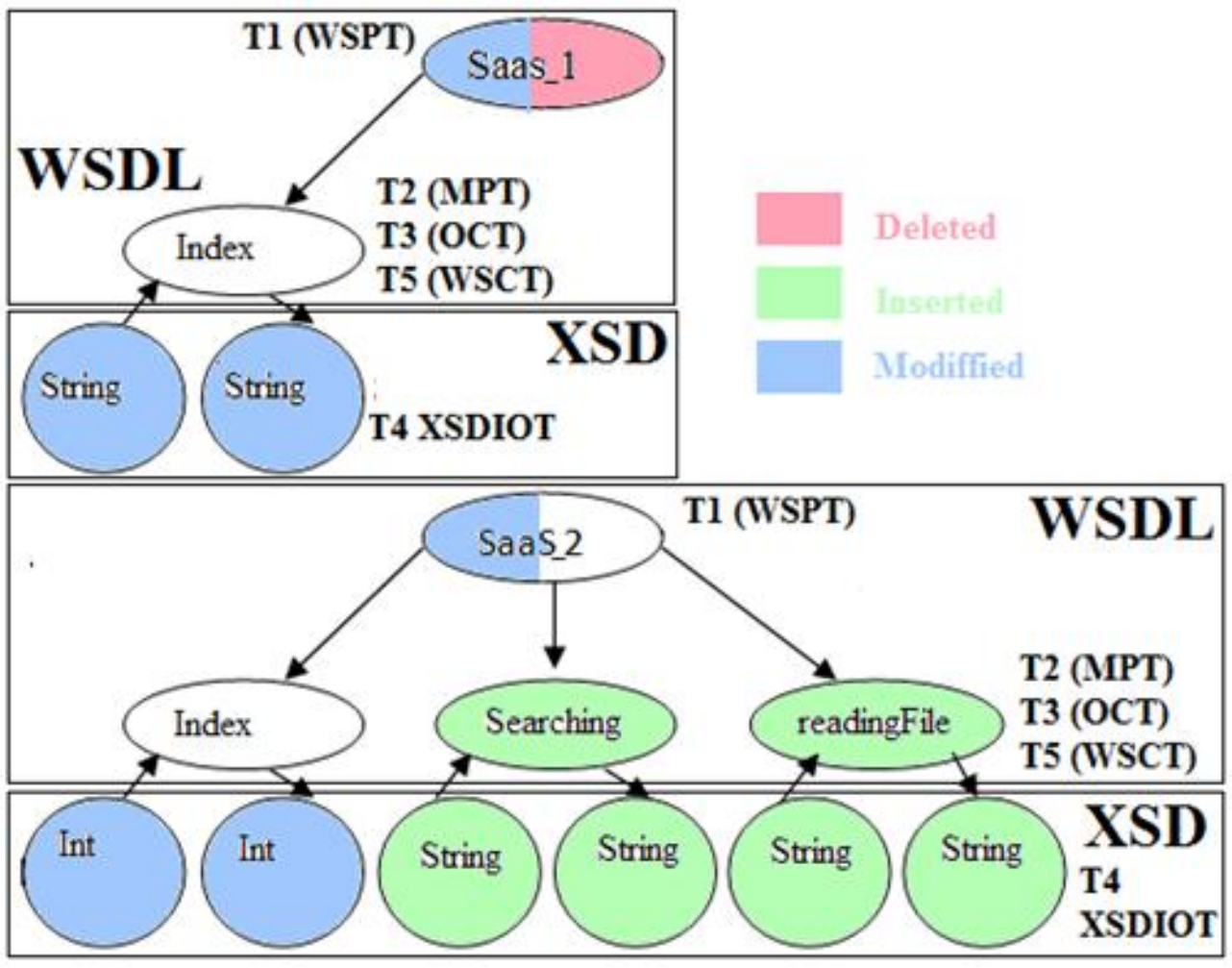

**Figure 3-3 Changes highlighted in WSDL of SaaS_2 compared to WSDL of SaaS_1**

## 3.3 The Regression Testing Approach

The approach gathers the information of changed operation by identifying the changes in WSDL and the operation code using a ‘diff’ utility. This information is then used to construct the Subset WSDL. Subset WSDL contains the subset of operations that have undergone any changes namely, inserted, deleted and/or modified operations. These operations are map to their respective test cases for proper subset selection. Specifically, the approach carry out in three steps: in first step, we compute Subset WSDL to capture subset operational changes. In the second step the test cases of old test suite are map to the original operations of the Subset WSDL. In third step, reduced regression test suite is constructed by selecting the test cases for subset operations of the web service. In the following sections, we elaborate on each step of our approach including the illustrative example.

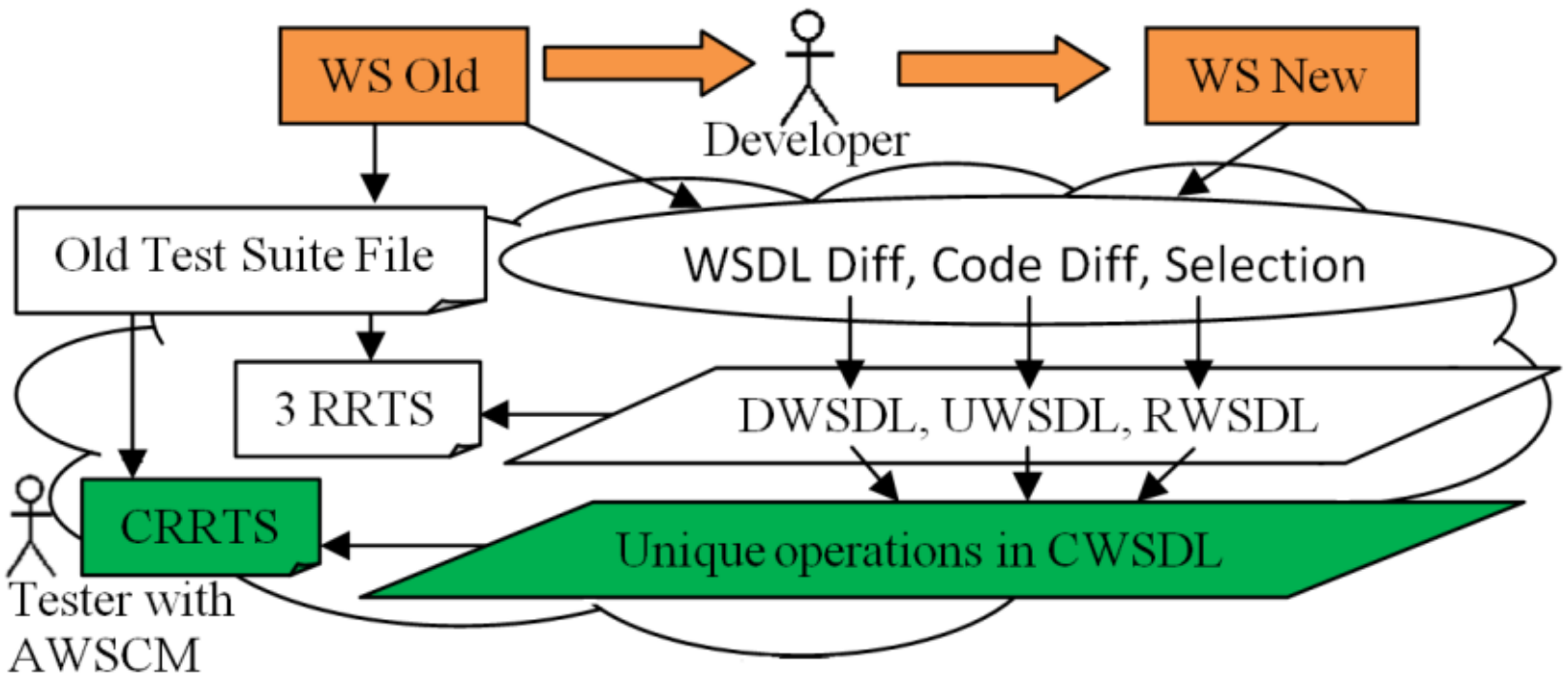

**Figure 3-4 Overview of proposed approach for regression testing of web services.**

### 3.3.1 Subset WSDLs to Capture Subset Operational Changes

We proposed a Subset WSDL is a WSDL containing subset operations of the operations in subject WSDL (as input). The Operationalized static analysis comprehends WS by parsing the WSDL and its operation code, to construct different Subset WSDLs are commonly named as SWSDL. We used SWSDL to perform selective regression testing.

**Algorithm for Computing Subset WSDL**

An approach to visualize, validate and construct a Subset WSDL (SWSDL) from subset operations of subject WSDL (as input). Automatically gathering operations in WSDL will help in change management. Semantically slice all the required data and all the part (message, xsd schema, port, binding, service etc) from the subject WSDL.

Algorithm gathers various parts of WSDL to construct a new subset WSDL i.e. SWSDL. First, extraction of xml starting element with <?xml….?> thereafter Definition with tag <definitions…> OR <wsdl:definitions….> (depending on WSDL). Second, extraction of all required Types for the operations. Schema (XSD: describes the data structure I/O) inside <types> element tag of WSDL. Third, extract all the Message part data for required operation in between <message…>…. </message>. Every operation has its corresponding unique message, which describes the information needed to perform the operation. Message contains one or more logical parts (a part corresponds to a unique message type attribute inside Type). Fourth, extraction of all the Ports required for the operation in between <portType …</portType> for communication of operations with the messages i.e. PortType contains address or connection point. Fifth, extraction of all Bindings required for the operation to name its input output used in Type i.e. binding defines the operations and its SOAP binding style (RPC/Document) and transport (SOAP Protocol). Sixth, extraction of Service parameter location required for system functions that expose to the Web based protocols (WS client to communicate it to web service via Port Binding and URL). At the end, extract end of Definition </definitions>. Note: Some terminologies are different between WSDL1.1 and WSDL2.0 i.e. PortType change with Interface. Moreover, in WSDL 2.0 message part is not used.

To arrange all the parts semantically according to the WSDL standard and input WSDL so that client can properly communicate with the web service code via WSDL, which is in algorithm of '**ConstructSWSDL**'.

File **ConstructSWSDL**(File *wsdl*, String[] *requiredOperations*) {

String *cStartDef, cXSD, cMsg, cPort, cBinding, cService, cEndDef*

Array of String[] o*peration*

[*requiredOperation* ∈ operationOFwsdl | *requiredOperation* = operation for new SWSDL required to be constructed ]

*cStartDef* = **constructStartDefinition**(*wsdl*)

*cXSD* = **constructXSD**(*wsdl, requiredOperation*)

*cMsg* = **constructMessage**(*wsdl*, *requiredOperation*)

*cPort* = **constructPort**(*wsdl*, *requiredOperation*)

*cBinding* = **constructBinding**(*wsdl*, *requiredOperation*)

*cService* = **construcService**(*wsdl*)

*cEndDef* = **constructEndDef**(*wsdl*)

*SWSDL* = *cStartDef* + *cXSD* + *cMsg* + *cPort* + *cBinding* + *cService* + *cEndDef*

**return** *SWSDL* }

During operation analysis, to compute an intermediate form of WSDL (SWSDL), the changes made in the web service are identified and captured in two different categories, namely, changes in WSDL and changes in an operation code. These changes are capture in two Subset WSDL, namely, DWSDL and UWSDL. Additionally, we can also capture the need of selective re-testing using another Subset WSDL, named as Reduced WSDL (RWSDL). These Subset WSDLs are then combined to form a Combined WSDL (CWSDL), which then used for regression testing of the web service.

The Difference WSDL is constructed from the diff between two versions of WSDL. DWSDL contains operations that have undergone some changes at WSDL. The Unit WSDL is constructed with the operations that have undergone change in code. UWSDL contains operations that have undergone some changes at operation code. The Reduce WSDL is constructed with user selective operations of WSDL. RWSDL contains additional selected operations. Combined WSDL is constructed by taking the union of the operations in the D/U/R WSDL such that it contains only unique and non-redundant operations.

**{**DifferenceWSDL ∈ NewWSDL | (DifferenceWSDLOperation = NewWSDLOperation – OldWSDLOperation) ∧ (operation have same semantics i.e. port, binding, schema, message part for that operation)}

{UnitWSDL ∈ NewWSDL | (UnitWSDLOperation = changed operation at code) ∧ (same semantics)}

**{**ReduceWSDL ∈ Input WSDL | (ReduceWSDLOperation = user selected operations) ∧ (same semantics)

**{**CombinedWSDL ∈ Input WSDL | (CombinedWSDLOperation = unique operations merge) ∧ (same semantics)}

## 3.3.2 Mapping Test Cases to the Web Service Operations

We compute the mapping of old test suite with the older version of web service operations. This mapping helps us to selects only those test cases, which required to be exercised because of modified. For added operation additional test cases require to be generated where as for the deleted operations are not to consider for the modified web service.

Figure 3.5 shows the correspondence between web service elements and testing information. An operation in the WSDL represents a piece of functionality that can be exercised by a corresponding set of test cases. Test sequences/steps exercise the different portions of the code for its functional correctness.

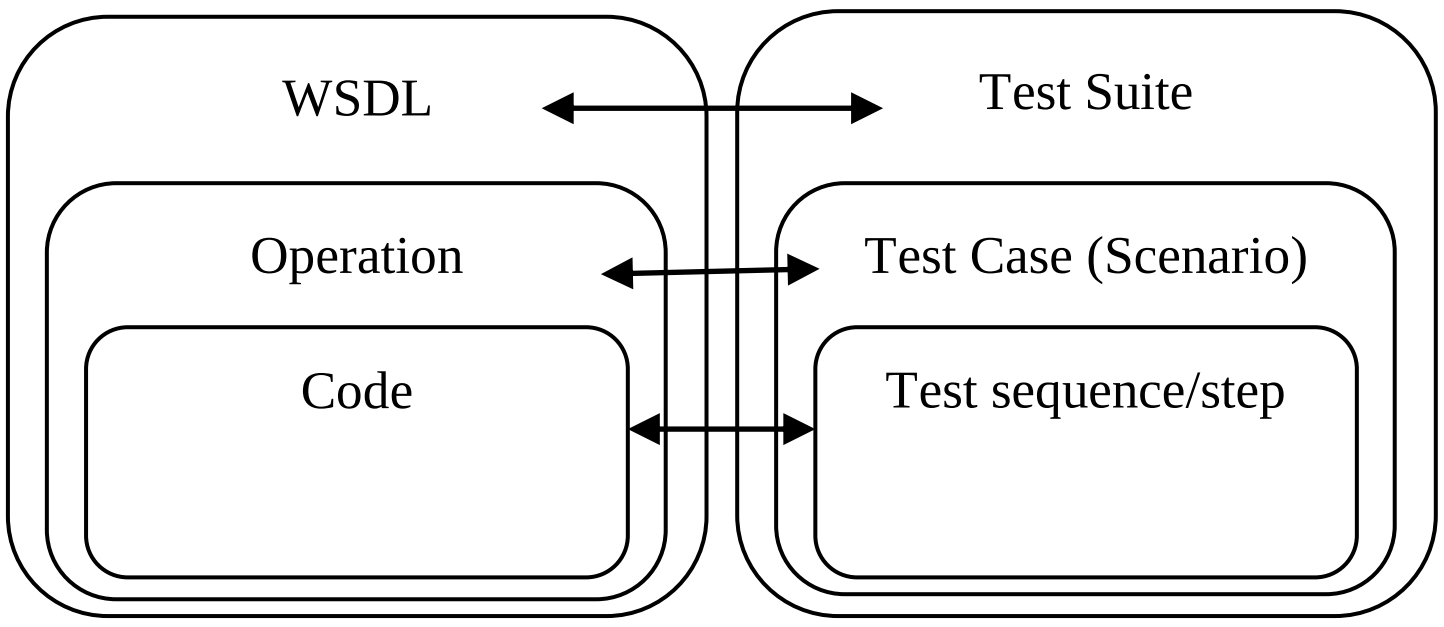

**Figure 3-5 ORTWS mapping test case to the web service operation**

Testing of web service operations can be performed with the help of tools like SoapUI, WebInject and JMeter. Description of the mapping between a web service operations with the set of test cases via test scenario is in figure 3.6, where every operation of the web service is tested against a number multiple usage scenarios. Each scenario corresponds to the test cases for operations under investigation.

We can use (D/U/R/C) WSDL in at-least two ways to perform efficient regression testing. First way is to apply (D/U/R/C) WSDL one by one to the web service testing tools (like JMeter or SoapUI) which generates testing templates for inserted operation. Empty testing template helps to write new test data for the inserted operations. Second, we can retrieve reduced test cases from the old test suite on the basis mapping of test cases with the (D/U/R/C) WSDL.

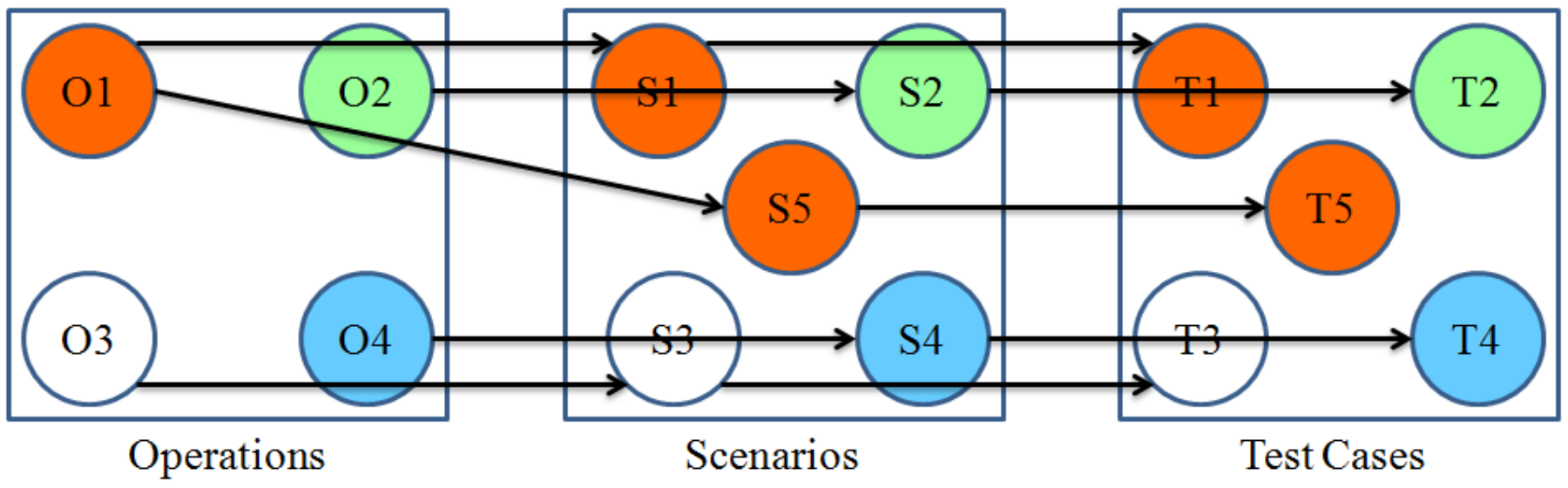

**Figure 3-6 Mapping of operations with testing scenarios and test cases**

Changes in the web service could occur in three ways i.e. deletion, insertion and modification.

1. Deleted operations will not present in the next version of web service, such operations are not required for regression testing.
2. Insertion of an operation can be identified from WSDL change between two versions and should be considered for regression testing of web service. These operations are captured in DWSDL and RWSDL is used in figure 3.7.
3. Modification could occur in two places at the WSDL and at the code. First, modification at WSDL means change in XSD i.e. when for a particular operation only input, output, or both are change. These operations are captured in DWSDL and RWSDL shown in Figure. 3.7. Second, if modification at operation code, then these operation are captured in UWSDL in Figure. 3.7. UWSDL had change impact analysis information and we can map them to the test suite in terms of test sequences/steps inside test case.

### 3.3.3 Computing Reduced Regression Test Suite

We automatically gathered the information about the changes in the particular operations of web service. Test those particular operations to validate requirements by generating the Reduced Regression Test Suite (RRTS) from old test suite and subset WSDL. These RRTS contained test cases only for those operations, which were in the subset WSDL. Further test sequences/steps are selected with its respective test case of the operation for the construction of RRTS. Test data flow analysis is done by intra-procedural dependencies for the changes in the code. Rerun test sequences/steps according to intra-procedural analysis. After identifying changes at the operation code, we can construct UWSDL and RRTS according to the changes.

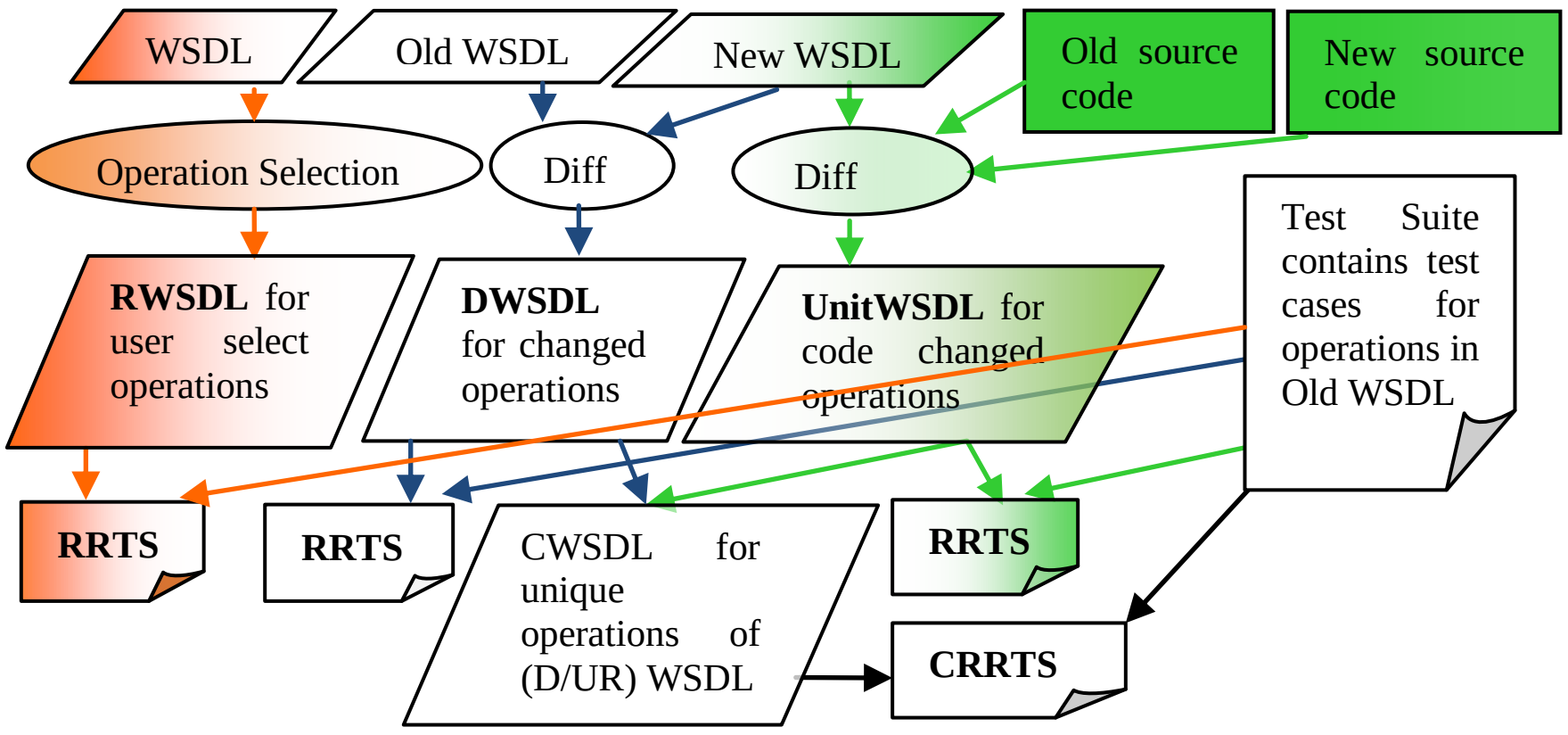

**Figure 3-7 Three RRTS are constructed by automated selection of the test cases of old test suite for the operations in the RWSDL, DWSDL and UWSDL.**

We construct Combined RRTS (CRRTS) for the operations in CWSDL. CRRTS contains all the unique test cases and eliminates redundant test case. Hence, we get reduced number of useful test cases to perform cost effective regression testing of web service.

The proposed approach of change impact analysis based regression testing of web service is shown in figure 3.8. With above testing technique a tool can be create for regression testing of web service or this new feature can be added to the existing tools of web service testing. Here web service diff is required, which can be easily gathered using available open source. This approach is prototyped with our proposed tool AWSCM. Utilities and libraries used by AWSCM to gather changes with semantic textual difference are JDiff [41], Predic8 [48] and Regex [49]. AWSCM works according to the algorithm described in chapter 5.

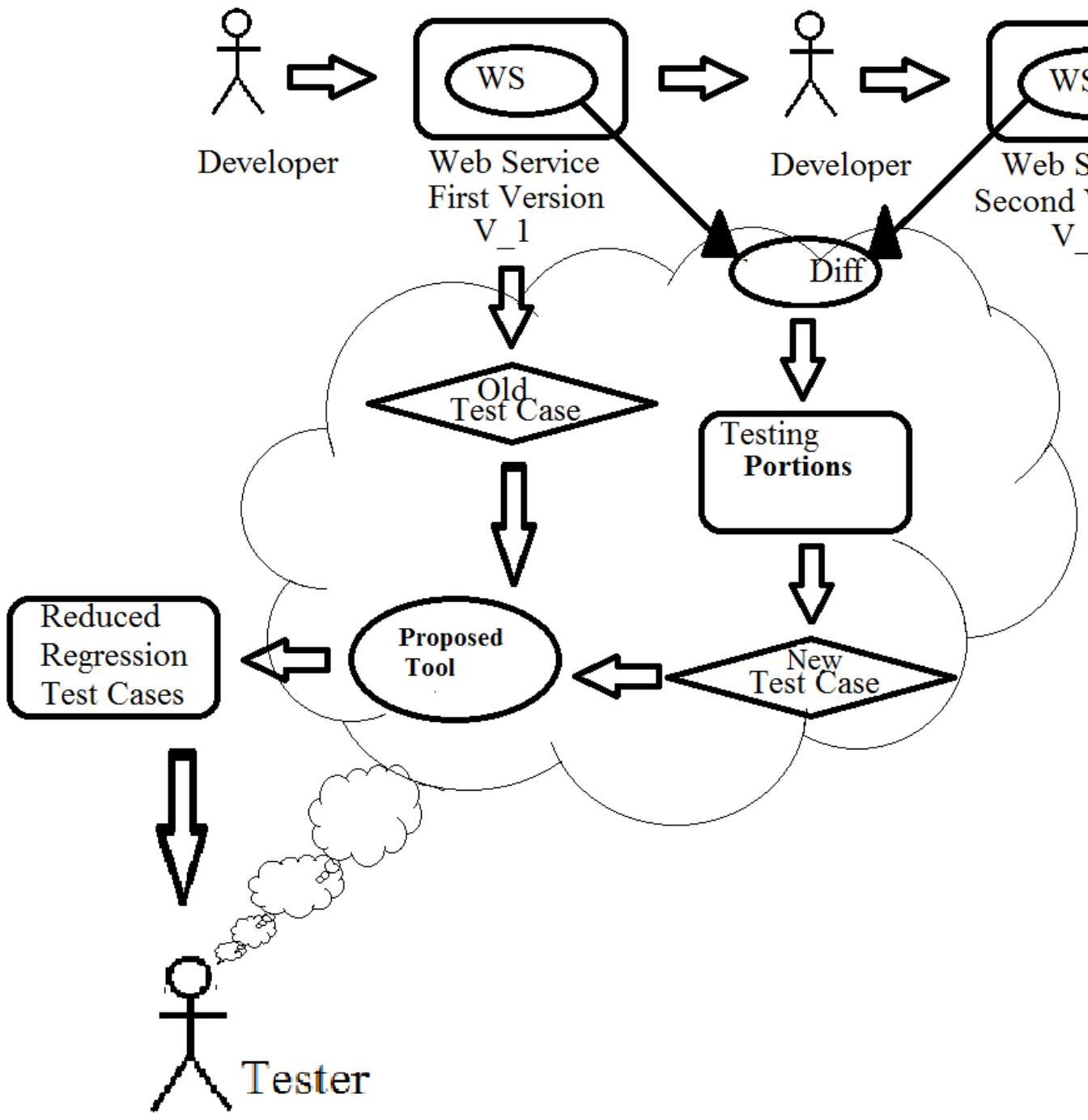

**Figure 3-8 Identify testing portion in web service to generate the reduced regression test suite.**

### 3.3.4 SaaS Example Revisited

Consider that the SaaS_2 is again change to upgrade it as Saas_3 by following

1) Insert a new operation 'editFile'.
2) Change both input and output type of 'Index' operation from 'String' to 'Integer'. Figure 3.9 shows above changes in Saas_2 to make Saas_3. This information can discover with the diff tool.

Step 1: Code portion diff gives us the differences in subsequent web service versions, with which we can easily see that deletion, insertion and modification in operations/policies to a web service. We find 'Searching', 'readingFile' as the unmodified operation, not required to test them. Whereas we find that 'editFile' and 'Index' are the inserted and modified operations respectively. Operation 'Index' and 'editFile' were in DWSDL whereas UWSDL contains only 'Index' operation.

Step 2: Using mapping we are selecting the proper test cases for the operation 'Index' from the old test suite.

Step 3: Construct a reduce regression test suite with the above selected test cases of 'Index', and with a test template for the 'editFile'.

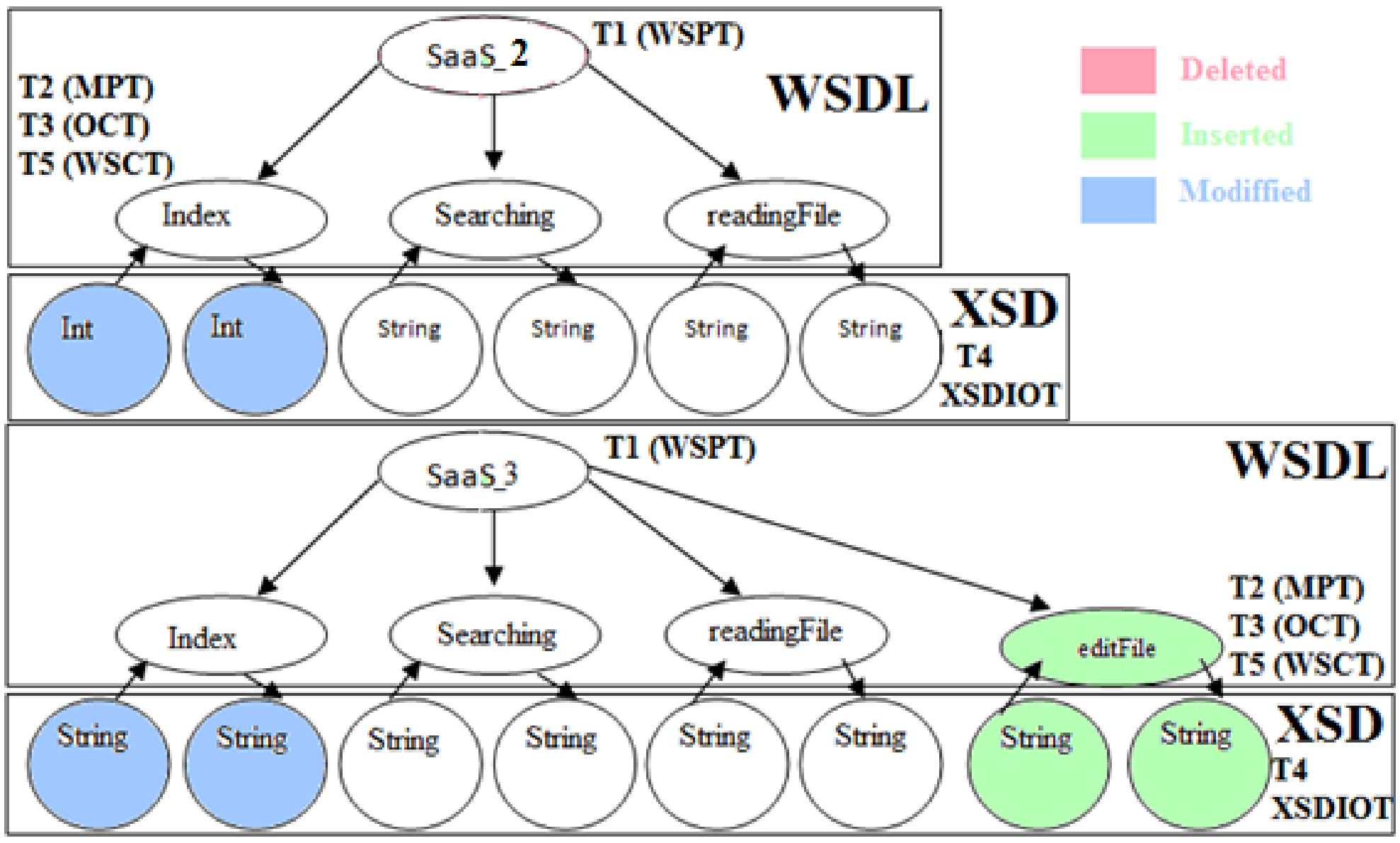

**Figure 3-9 WSG describing deletion of policy in red color, insertion of 'editFile' in green color and modification of input-output type of 'Index' in blue color.**

## 3.4 Illustrative Case Study

The result of the case studies, AWSCM supported case studies on WSDL based web service projects, namely, Eucalyptus_CC (cloud controller CC_wsdl is available on Github), Amazon Web service, SaaS web service (in different versions) and some other web service projects. The case study conducted using the proposed tool AWSCM for constructing the reduced test cases from old test suite shown in Figure 3.11. 3.12.

Eucalyptus and Amazon web service available on Internet and they are the part of evolving project. Eucalyptus_CC used in all (D/U/R/C) WSDLs constructions and Amazon web service WSDL, only RWSDL was constructed. We have done a case study on the three parts i.e. for DWSDL, UWSDL, and RWSDL are as follows.

The case study for Eucalyptus_CC Web Service is conducted over two versions at the Github branch of (/maint/master 3.1, master) of cc_wsdl and codes (cluster/handlers.c) selected from the repository [24]. In figure 3.11 the application of the DWSDL. SoapUI generated reduced test case templates for inserted operations, namely, 'DescribeSensors', 'BundleRestartInstance' as shown in Figure 3.10 (a). We can put additional test data in those templates to test newly inserted operation for the eucalyptus according to DWSDL and UWSDL of Eucalyptus_CC. Additional features of selective re-testing using RWSDL is in chapter 5 in section 5.5.

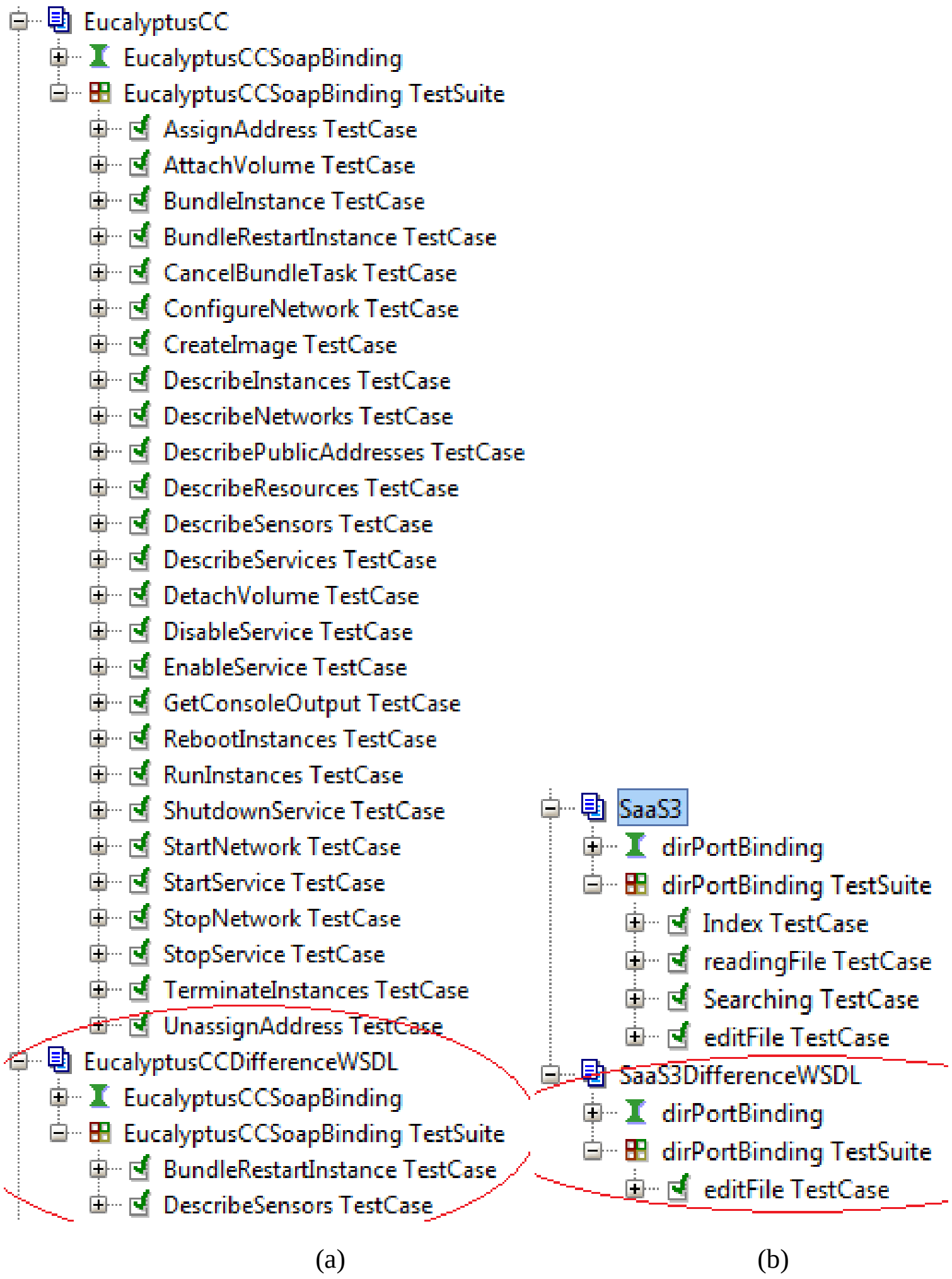

**Figure 3-10 Snapshot of SoapUI showing old test suite and RRTS (a) Using DWSDL Eucalyptus_CC, (b) for changed operation 'editFile' in SaaS**

The case study for SaaS in figure 3.12, SaaS_3 was containing operations, namely, 'Index', 'readingFile, 'Searching' and 'editFile'. The operation 'editFile' changed at the code to make next version of SaaS i.e SaaS_4. Construction of UWSDL from WSDL and the source code of the two versions (SaaS_3 and SaaS_4) as input to the AWSCM. The AWSCM constructed the RRTS (which contains test cases of only 'editFile' operation as shown in figure. 3.10 (b)) from UWSDL and the old test suite given as input. Similarly, we constructed the DWSDL with consecutive version of WSDLs for SaaS. Similarly we constructed the RWSDL for user selected operation of new WSDL to generate the RRTS from old test suite.

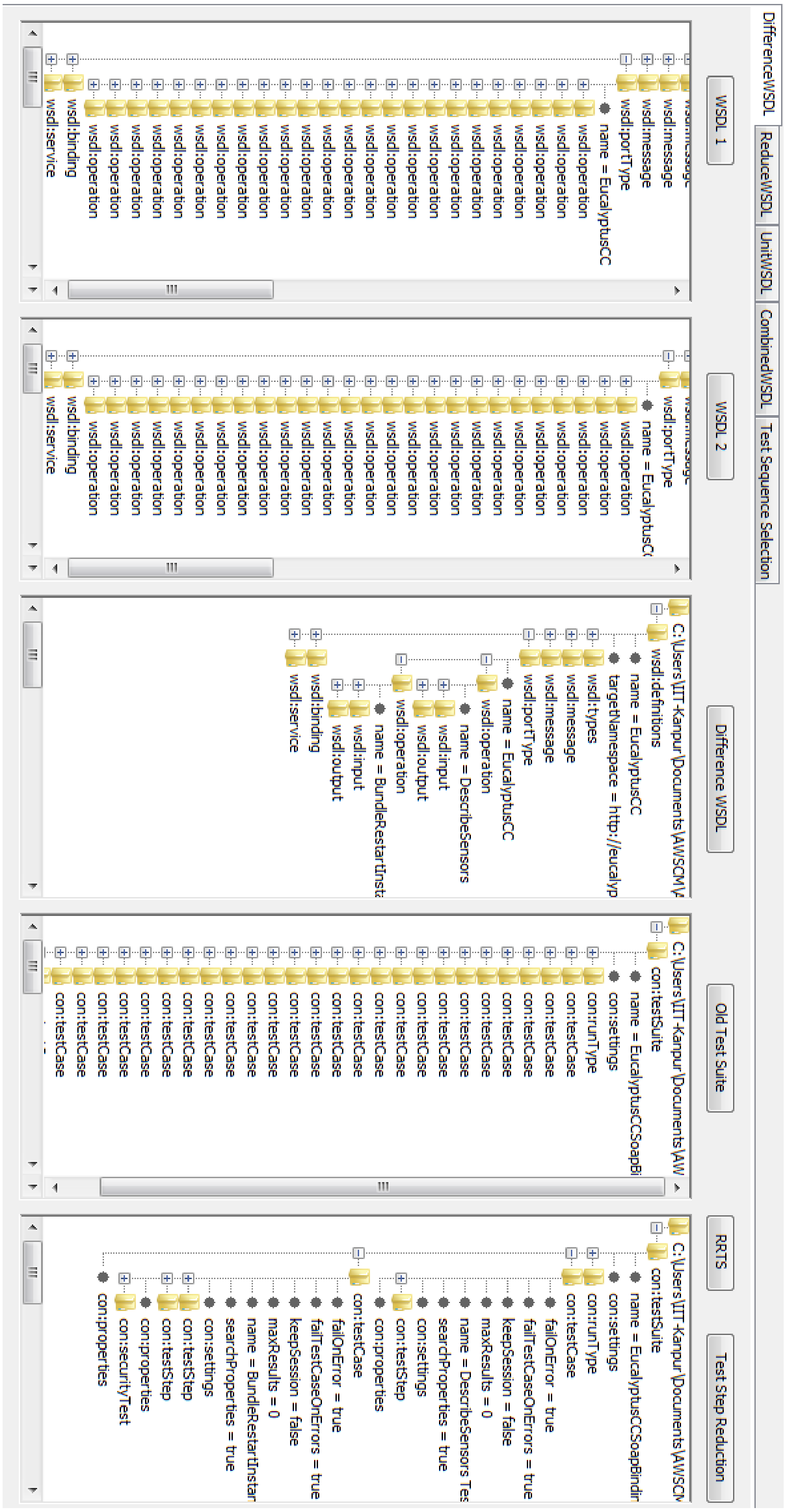

**Figure 3-11 DifferenceWSDL of AWSCM for EucalyptusCC.**

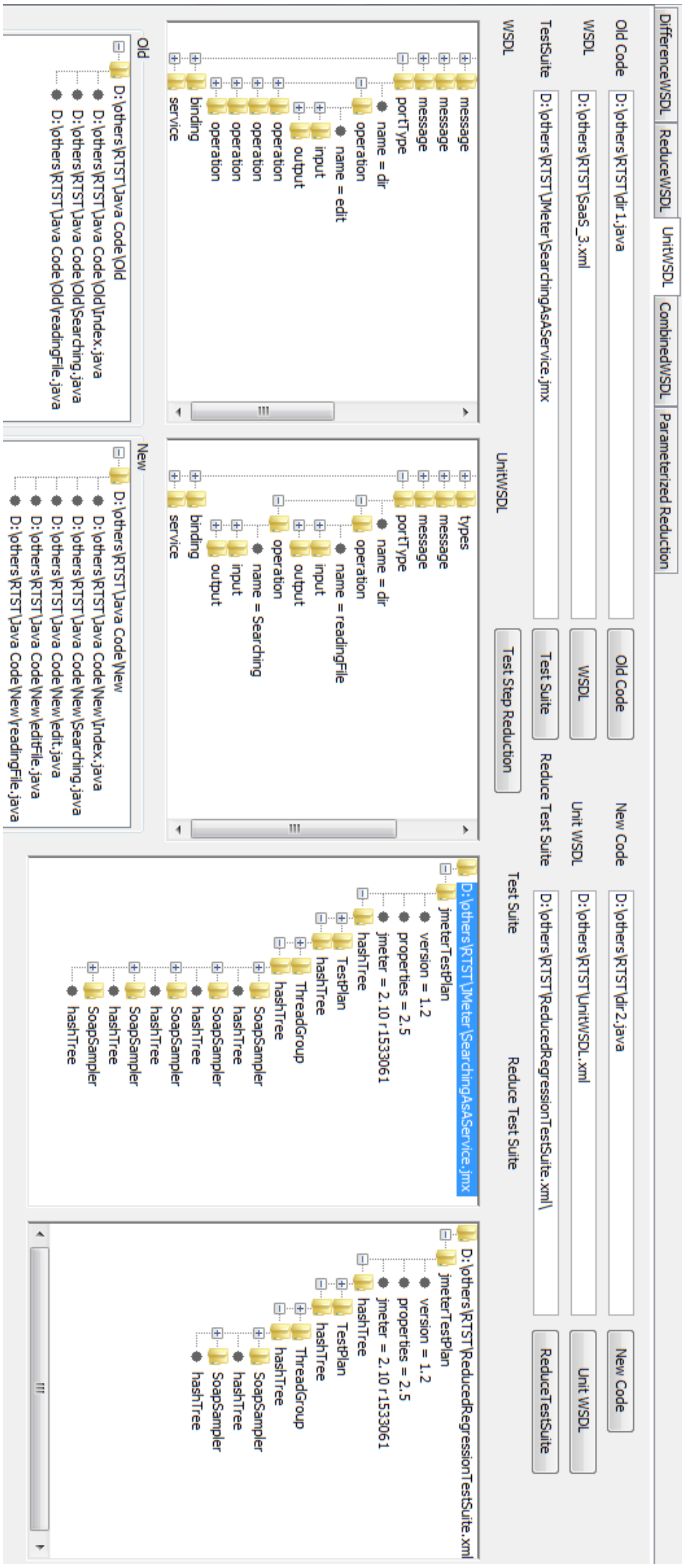

**Figure. 3-12 UnitWSDL of AWSCM for SaaS operations (editFile and Searching) modified at code to construct RRTS for JMeter old TS.**

## 3.5 Saving in Regression Testing Effort

According to the changing-link conversion of unit given in [54], “we multiply the original measurement by a conversion factor (a ratio of units that is equal to unity)”. Similarly, to calculate effort reduction estimation for regression testing of web service is technique we find out two types of empirical change information. In table 3-1 and 3-2, LW we stands for Lines of WSDL, Op stands for Operation and LoC stands for Lines of Code, WSDL_1 stands for WSDL version 1 and WSDL_2 stands for WSDL version 2.

First ratio between average lines of code in changed portion to the new version of web service. Table 3-1 given below shows effort reduction from ratio of number of line changed in WSDL and code with total lines of WSDL in the new version of the code. The size of code is very large, so we are incapable of finding out the number of lines undergone changes. To calculate average number of changed line (V’), multiple the average number of lines in operations to the number of operations that have undergone changes. To calculate average number of line in version 2 (V_2), multiple the average number of lines in operations to the number of operations in new version. To calculate effort required take the ratio of (V’) and (V_2).

Second, ratio between number of operations changed and total operations in new web service is described in Table 3-2. We can simply counted the number of operation in new WSDL as Y and the operations in Subset WSDL as Y-Z. Take the ratio of Y-Z/Y to calculate the average effort required.

In the Eucalyptus_CC code average code that its value is approx 8000, which comes to 8167, and we find 482.5 as the average changed line. In the Eucalyptus_CC Difference WSDL had 2 operations undergone through changes out of total 26 operations. The effort estimation with the technique 1 is almost equivalent to the technique 2.

**Table 3-1 Ratio of changes in number of lines.**

| *Variables* | *Quantity* | *Unit* | *SaaS* | *Eucalyptus_CC* |
|---|---|---|---|---|
| Number (No.) of lines in WSDL_1 | L1 | LW | 77 | 1712 |
| No. of lines in WSDL_2 | L2 | LW | 115 | 1492 |
| Change in line of WSDLs | C = L1 - L2 | LW | 38 | 220 |
| line of WSDL_1, WSDL_2 / Average number of operation | X1, X2, Xav = (X1+X2)/2 | LW / Op | 26, 23, 24.5 | 62, 61, 61.5 |
| Average Line of Code (LoC) per operation | Y1, Y2, Yav = (Y1+Y2)/2 | LoC / Op | 40,56, 48 | 216, 291, 253.5 |
| No. of LoC for every operations | C * Yav | Op | 38 * 48 | 220 * 253.5 |
| No. of Line of Code (to be tested) V' | C * Yav / Xav | LoC | 38*48 / 24.5 | 117*253.5 / 61.5 |
| No. of Line of Code in V_2 | L2 * X2 / Y2 | LoC | 115*56 / 23 | 1712*291 / 61 |
| Effort required | V' / V_2 | | 74.4 / 280 | 482.5 / 8167 |
| Percentage effort required | (V' / V_2) * 100 | % | 0.26 * 100 | 0.0600 *100 |
| Percentage effort reduction | 100 – (V' / V_2) * 100 | % | 100 – 26.57 | 100 – 6.00 |

**Table 3-2 Ratio of operations changed**

| *Variables* | *Quantity* | *Unit* | *SaaS* | *Eucalyptus_CC* |
|---|---|---|---|---|
| No. of operations in WSDL_1 | X | Op | 3 | 24 |
| No. of operations in WSDL_2 | Y | Op | 4 | 26 |
| No. of operations in DWSDL | Z | Op | 1 | 2 |
| No. of operations in DWSDL | Y-Z | Op | 1 | 2 |
| Percentage reduction in operation by DWSDL | ((Y-Z)/Y) *100 | Op | 75% | 92.3% |

## 3.6 Thread to Validity

Our focus is on the detection of diffs for web service revision of operations and source code whereas we missed some fine-grained diff. We can infer that this approach can be applicable to any big industrial project. However, in the case study, we are not able to demonstrate quantitative evaluation for industrial strength test case coverage reduction.

## 3.7 Summary

This chapter demonstrated a change impact analysis based Operationalized regression testing approach. Implementation of this approach is in chapter 5 as a part of AWSCM tool. AWSCM generates Subset WSDL interface descriptions to construct the Reduce Regression Test Suite (RRTS) for testing purposes. RRTS used for Operationalized regression testing of web service is found to be effective and efficient. In our case study on SaaS and Eucalyptus_CC we achieved 75% and 93% approx of changed based reduction in regression testing.

# CHAPTER 4

# PARAMETERIZED REGRESSION TESTING OF WEB SERVICE

A parameter in a web service demonstrates the flows of logic, control, and data of an operation in a web service. The Parameterized regression testing of web service uses interdependencies between the parameters of an operation. For dynamic analysis of web service, this chapter proposes a technique named as Parameterized Regression Testing of Web Service (PRTWS). In general, ORTWS can be applied to any types of operation in WSDL whereas PRTWS is applied to the operational whose parameters are interdependent on each other. The Parameterized regression testing can be used to streamline and reduce the regression testing efforts further. Parameterized analysis is an extension of the Unit WSDL in ORTWS. Parameterized analysis checks operation whose call goes to methods and operations of same or other web service. Parameterized analysis works in the context of inter-procedural call tracing, which is useful in the analysis of web service composition.

Although both inter-procedural and intra-procedural dependencies can be considered for parameterized analysis. In this chapter, we are considering only inter-procedural parameter call dependencies.

Simple example of an inter-procedural calling is shown in figure 4-1 (a)

```
main() {
    a = 6;
    R= F (a); // inter-procedural call
    a= R;
    c = F (a + 9); // inter-procedural call
}
```

```
F (int a) {
    if (a < 8)
    y = 0;
    else
    y = 1;
return R;
}
```

Another example, following inter-procedural call takes place between two Web services WS_main and WS_Called and one class ‘Class_Called’ as shown in figure 4-1 (b).

1. Operation ‘Operation_A’ of WS_Main calls method ‘Method_B’
2. Operation ‘Operation_A’ of WS_Main calls Operation ‘Operation_C’ of WS_Called.
3. Further ‘Operation_C’ calls method ‘Method_D’ of class ‘Class_Called’.

Suppose, after code review we deduce that inter-procedural call 1 and inter-procedural call 2 takes place due to parameter ‘Parameter_1’. Again, inter-procedural call 3 takes place due to parameter ‘Parameter_2’ and ‘Parameter_3’. Now we can easily deduce that ‘Parameter_1’ is a primary parameter and ‘Parameter_2’ and ‘Parameter_3’ are non-primary parameter.

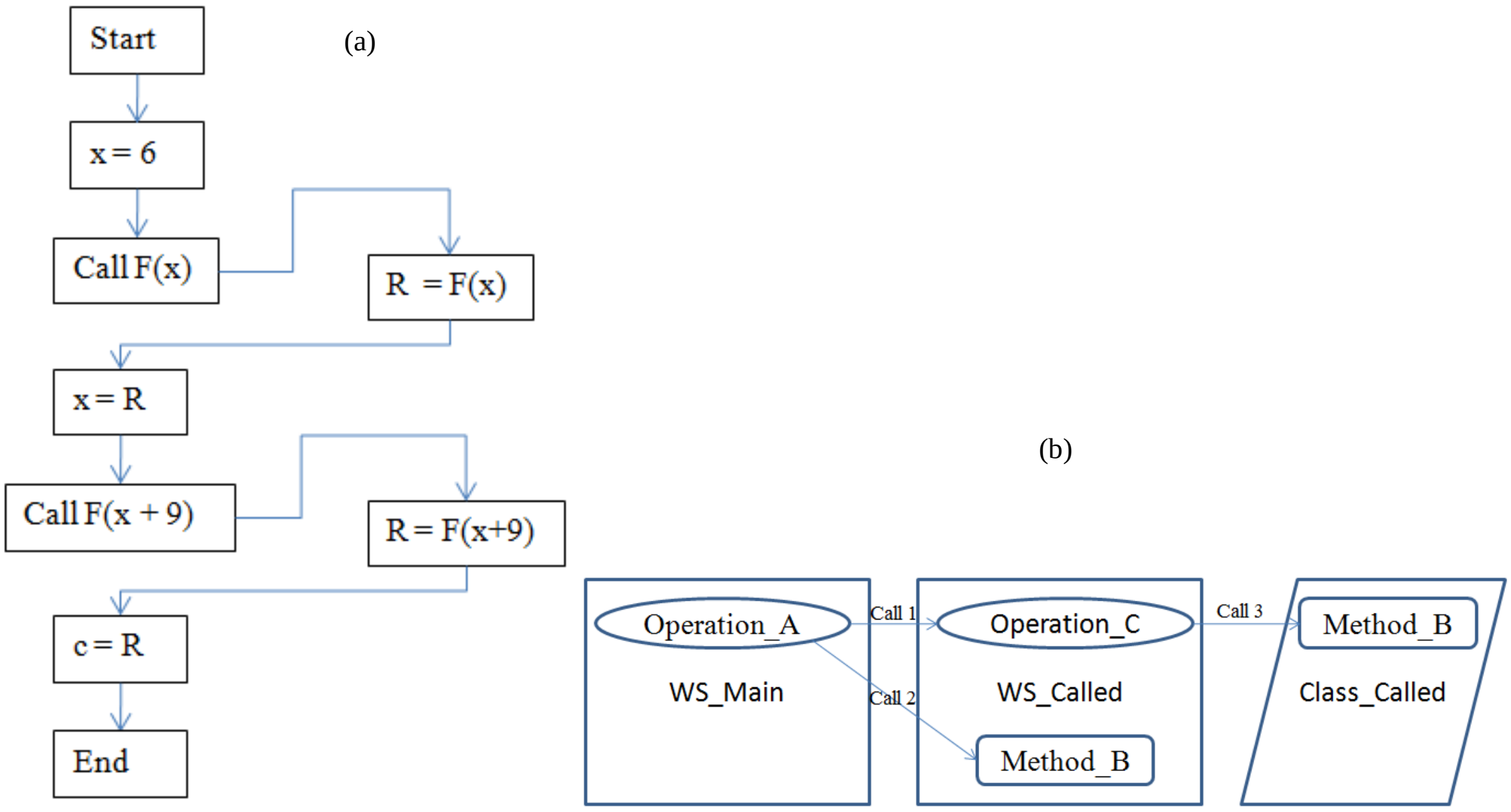

**Figure 4-1 (a) Simplest Inter-procedural call to the function F(x). (b) Inter-procedural call between two web services and one class.**

We considered code as dynamic database, which give output according to the input and code logic. Similar to the database design we had manually selected primary parameter for dependent parameter. It is based on testers need for the selection of primary parameters dependency. Dependency can be of many types, specifically depending of code like intra-procedural call or inter-procedural call.

We had selected small web services for our case study and for them we find out primary parameter. Although for all case study services we had manually selected primary parameter easily which helps in code analysis. Yes, there may be some false positive because of huge depth in method calling and if there is both fork/join in middle of the code. Depth of method call and fork/join combined may cause confusion in primary parameter selection because human may induce error in checking the dependency. We had not automated primary parameter selection step. It could be a research in itself to create an automated approach to find out the primary parameter in code.

## 4.1 The Approach

The Parameterized analysis comprehends web service by performing analysis for the operations whose parameters have dependencies on each other. In Parameterized analysis, primary parameter gives mapping between web service code and test suite as shown in figure 4-2. Effort reduction in web service testing can save huge cost in the maintenance and evolution of project. PRTWS is based on the mapping between web service code and the test cases.

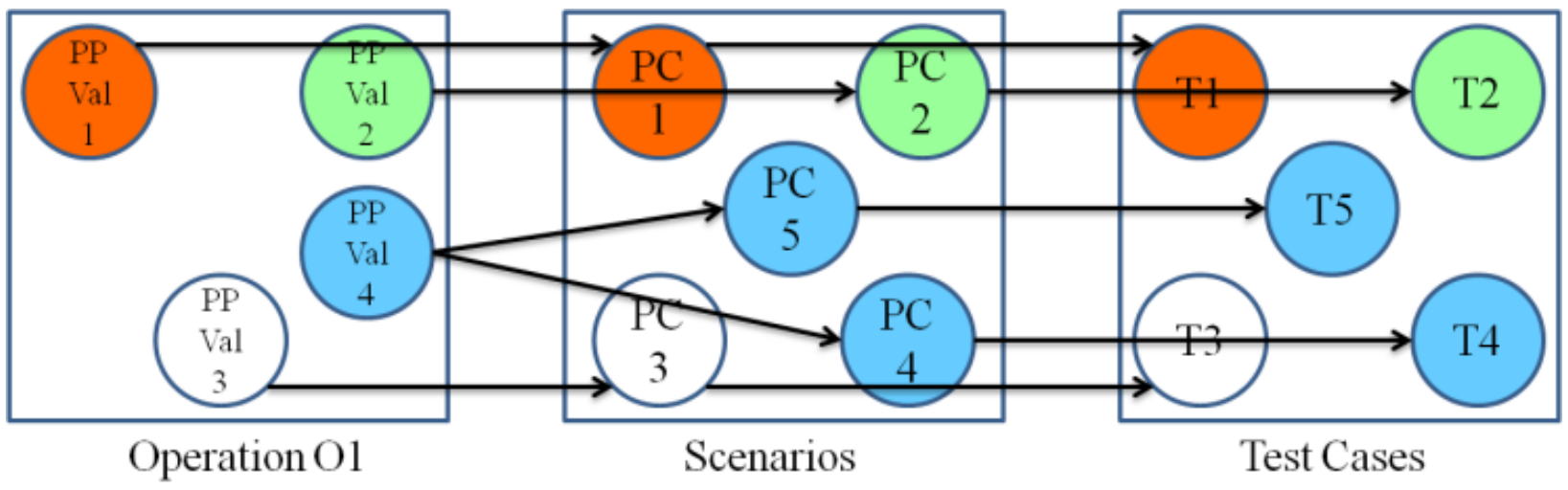

**Figure 4-2 Mapping of operations with test scenarios i.e. combination to capture dynamic flow, these scenarios are further mapped to the set of test case.**

In relational database tables, a database developer identifies 'primary' parameter in a given relation [5]. Certain kind of web services operation parameters may be interdependent on each other. Some of the parameters in a web service operation have capability for identifying other parameters for a given operation, and such parameter are called as the 'primary' parameter. Similar to the database whenever an operation has two or more non-redundant 'primary' parameter, one of them arbitrarily selected and called the primary parameter of that operation. We identify a 'primary' parameter inside an operation, which uniquely defines other parameters. 'Primary' parameter can combine with all the permutations of the other non-primary parameters for the re-testing of selected operations of the web service.

### 4.1.1 Primary Parameter

We used parameter dependency of operation to comprehend the change impact analysis in both white and black box code of web service. Maintenance of software is easier if the part of the real world is reflected in programming and testing [18]. Inspired with EF Codd rules in [3], with database concept of primary key author is constructing analogy. We find operations in web service are analogous to relation in database and message based web service calling is analogous to query.

A ‘primary’ parameter (or set of parameters) has values which uniquely identify other input/output parameters of the operation. Such a parameter (or set) called a ‘primary’ parameter. Whenever an operation has two or more non-redundant ‘primary’ parameter, one of them selected as ‘primary’ parameter of that operation.

A test case is built for the value of ‘primary’ parameter with its test sequences/steps built for combination of values of non-primary parameters. Test case is such that value of primary parameter is fixed for all its test sequences/steps whereas values of other parameters were varies as combination. These combinations were decided based on data flow and conditions inside operation code. If scenario is big then we can also build and assign whole test suite dedicatedly toward a primary parameter. In simple words, primary parameter can uniquely determine the rest of parameter and makes testing easier and efficient which further help in regression testing.

We can parse web service code and separate-out different operation and method. After separation process, we gathered the changed inter-procedural called operations/methods in the three ways.

1. Changed operations required to be re-tested.
2. Operations that are dependent on the changed method require to be re-tested because those methods may also affect the functionality of the operation.
3. Operations which are dependent on the operations gathered in the above two steps, are also required to be re-tested.

To perform web service analysis test cases selection is done based on automated change impact analysis and user selection. As described following.

*First*, Anjaneyulu Pasala et. al. in [37], is gathering change impact analysis using method interactions among components of software applications. They are identifying affected methods through a change impact analysis of programs for selecting a smaller test suite. Similarly, change impact analysis of web service can be gathered using differencing between operation and methods of two versions of source code. Change impact analysis done with the help of parsing the new source code, new WSDL, code changes, and test suite. We used java to parse using libraries of file and string manipulation. We used libraries like “java.io” for file manipulation; “java.util” for reading and writing and “java.util.collection” for storing and managing the data. We parse web service codes to separate out all the operations and methods to gather the inter-procedural flow dependencies paths using [55]. JDiff is used for automatic change impact analysis between operations and methods of two different versions. After separation and diff process, we find the affected flow dependencies between these methods and operations. These changed operations and methods will be used to find the mapping to the test sequences/steps in a test case. Mapping requires parsing of a test suite, which is build such that test sequences/steps will trace those flows. For example, gathering test step combination value in figure 4.4 (a) is map to the flow in figure 4.4 (c). This mapping is used to find out test sequences/steps relating to the changes in the flow of subsequent version of code. These gathered test sequences/steps combine to construct a test case of an operation, further such test cases combines to build RRTS.

*Second*, user selected test sequences/steps one by one to construct a test case, which combines to build RRTS. For example, it is selection of one of the combination path in figure 4.4 (c) according to the tester. PRTWS takes a test suite as input, to prompt user to select test case, then their test sequences/steps were selected to be stored. User can again repeat the selection procedure for other test sequences/steps inside test case.

Experiment of testing various web service projects conducted on the SoapUI and JMeter are of the best and standard web service testing software. SoapUI and JMeter used because it is easier to export a test suite and manipulation according to various testing efficient approaches. All this process conducted to construct the reduce test suite with the help of string manipulation using java.util and regex libraries.

### 4.1.2 Mapping Test Cases to the Web Service Code

Some of the web service has operation whose parameter dependent on each other, this interdependency can be used for mapping between test suite and code. This thesis introduced web service test suite, which help in regression testing of web service. When test suite is constructed for web service operations as shown in Figure 4.3 (b) and test cases are build for the primary parameter. Further test sequences/steps (test data for assertion) constructed by the inter-procedural or intra-procedural flow inside the code of an operation. Test sequences/steps are the combinational value of non-primary parameters for a fixed primary parameter. Combination of the test data (of non-primary parameter) in test sequences/steps denotes some flow of path within the operation code in figure 4.4 (c). Data and control flows exist within the code of operation can be used to fill test data inside test sequences/steps. These flows can be affected during changes and can be detect by change impact analysis based selective test sequences/steps for regression testing. For example SoapUI which performs WS testing in figure 4.4 (a). An example to this concept, figure 4.3 (b) shows all the non-prime parameters 2, 3, 4 are dependent on primary parameter 1. Physical meaning of this internal mapping is that, each flow path in figure 4.4 (c) is mapped to one of the test sequence/step (combination of parameter) in figure 4.4 (a).

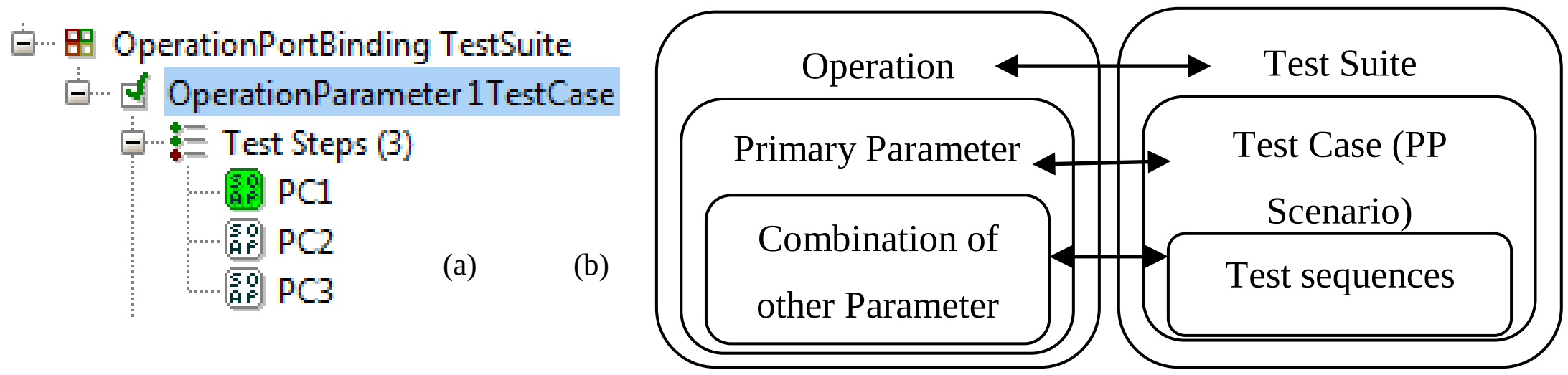

**Figure 4-3 Concept of PRTWS (a) Test suite snapshot of SoapUI (b) mapping between web service code and test suite.**

Regression testing of web service using PRTWS can done by generating a reduce regression test suite from dynamic analysis. Change impact analysis for control and data flow analysis can also be capture for dynamically executing web service operation.

This approach is prototyped with our proposed tool AWSCM, which works according to the algorithm proposed in chapter 5. RRTS as output of AWSCM construct by gathering user selected test sequences/steps using Regex [45] libraries. In PRTWS AWSCM, changes were identified in the code of the operation for inter-procedural data flow analysis.

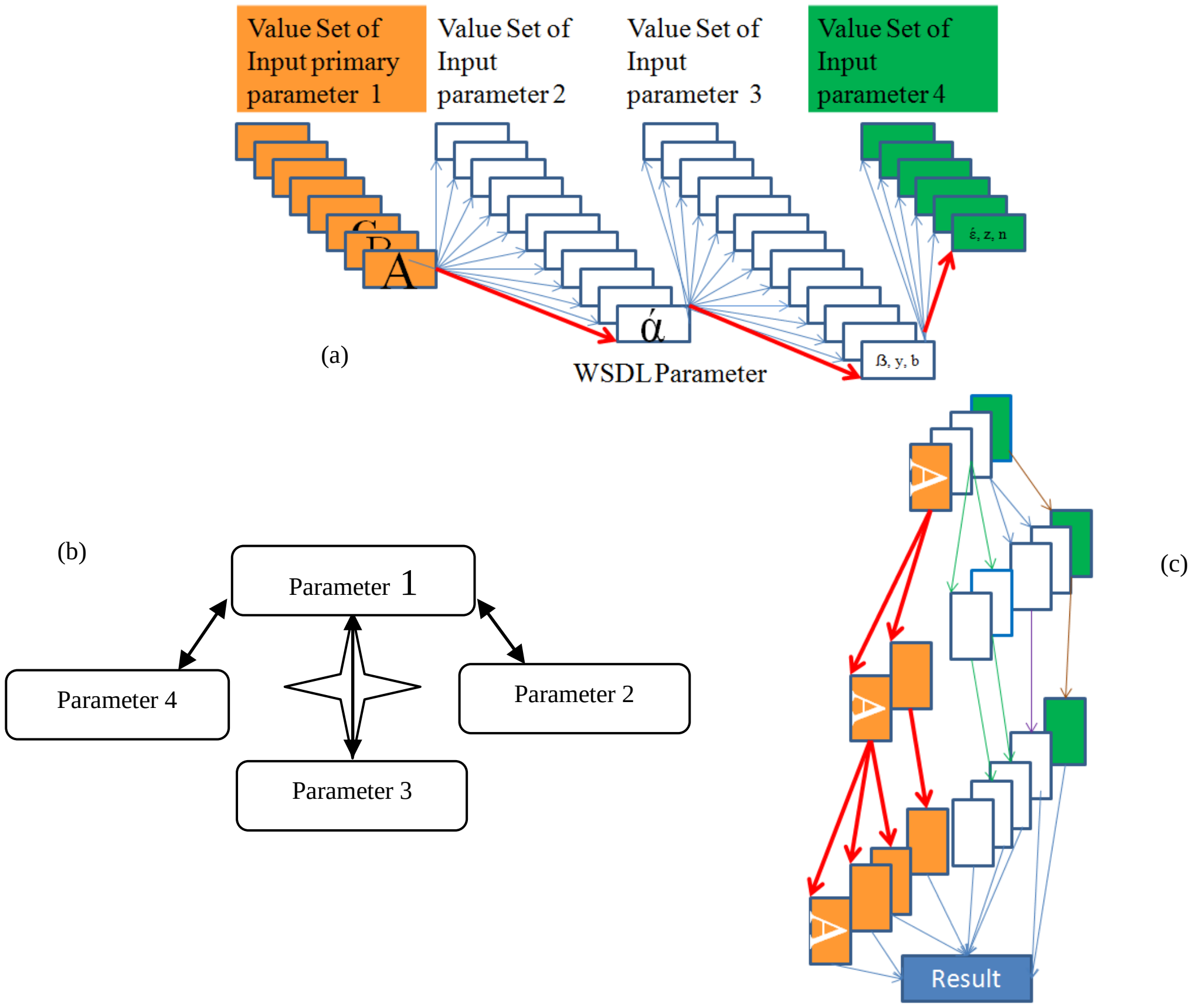

**Figure 4-4 (a) Mapping due to combination of parameters of operation (b) Parameter 1 as primary parameter and dependent Parameter 2, 3, 4 as non primary parameter (c) Flow of data inside the code of operation.**

### 4.1.3 Computing Parameter WSDL & Reduced Regression Test Suite

Test suite is given as input to AWSCM will automatically select test cases further test sequences/steps is selected. Test sequences/steps are selected according to the code flow combines to form a test case (based on primary parameter). These test cases combine to form a RRTS for PRTWS.

Regression testing will be more effective if changes in the code automatically identified. For WS we have done this using model described in Fig 4.5. Web service code is given as input to separated out all its operations and methods. Web service war is used as input to construct inter-procedural call graphs. We used, JDiff to generate patch between two versions of separated operations and methods. Two versions of operations and methods are given as input to JDiff to find the changes. Map the identified and extracted changed operations and methods to the inter-procedural flow graph against affected parameter. From this mapping, RRTS is constructed with selective test sequences/steps inside test cases for changed operations.

The input WSDL, code, change impact analysis and mapping are used for the construction of ParameterWSDL (PWSDL) and PRRTS. After extracting affected operations of following three types, search for the operation in WSDL.

1. Changed operations required to be re-tested.
2. Operations that are dependent on the changed method require to be re-tested because those methods may also affect the functionality of the operation.
3. Operations which are dependent on the operations gathered in the above two steps, are also required to be re-tested.

With the help of these operations, we constructed the PWSDL. This automatically brings operations of PWSDL in selected operation list. Further test sequence/step selected for the called inter-procedural operations and methods. PRRTS and PWSDL is constructed based on change impact analysis of the inter-procedural flow.

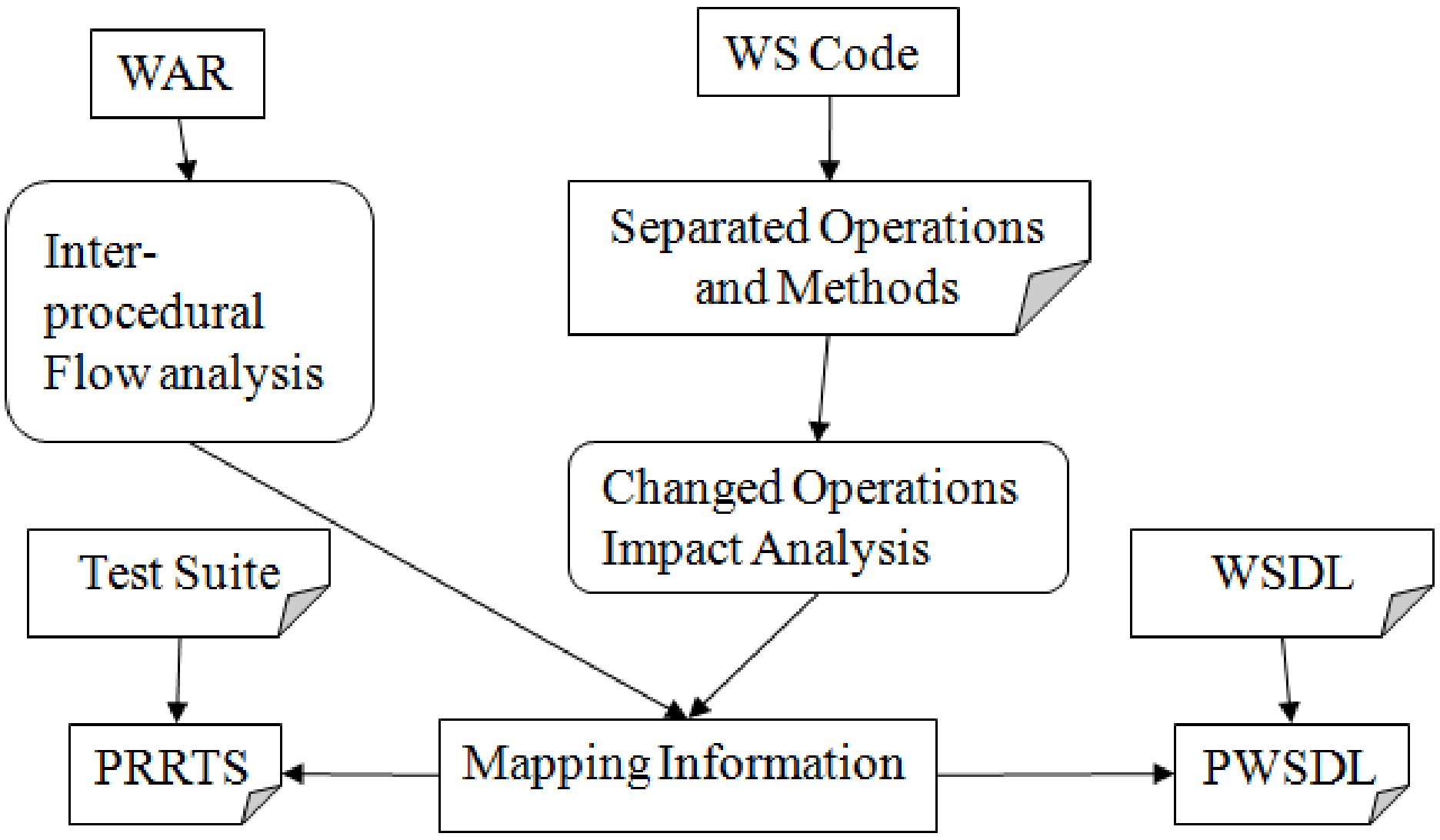

**Figure 4-5 Construction of RRTS and PWSDL while PRTWS.**

## 4.2 Illustrative Case Study

PRTWS case study, test sequence/step reduction for four dynamic web services, namely, currency conversion, Global weather, Sunset Sunrise service, Bible Web services. Using the tool, we constructed RWSDL, which further used to construct RRTS. The RRTS constructed according to the user selects test sequences/steps, which contain test sequences/steps inside test cases similar to the old test suite in Fig. 4.6 (a). AWSCM also does the code analysis of SaaS and Eucalyptus_CC (handler.c at Githhub) to perform automated inter-procedural flow analysis. AWSCM gathers operations affected indirectly due to inter-procedural flows to construct PWSDL and PRRTS with these operations. Change impact analysis using inter-procedural flow done to find out which operations and methods affected. Gathered impacts were map to the test sequences/steps of a test case for a particular operation in the test suite.

Using AWSCM in this thesis, we have conducted two types of case studies i.e. white box and black box. White box analysis on ‘SaaS’ and ‘BookService’, both of them are our own developed java web service and deployed on glassfish server. In black box, testing four case studies is conducted on publically exposed web services for Currency conversion, Global weather, Bible and Sunset-Sunrise. Experiment conducted based on user selective test sequences/steps combined to form a test case. These test cases further combines to form test suites of PRTWS. Selection of particular test steps according to primary parameter scenario helps in the execution of certain flow using certain combination of non-primary parameter values.

We found this test suite build according to PRTWS technique makes easy to conduct reduced regression testing similar to [3][4]. Because this makes test case structure more formal, systematic, and analytical by properly mapping test sequence/step to the code flow. The case study of PRTWS on real world web service is in Table 4-1 where Y stands for experiment conducted and X stands for experiment not conducted.

**Table 4-1 Case study of web service for PRTWS**

| Case Study Web Service Projects | PRTWS | |
|---|---|---|
| | ***Dynamic Black Box Testing*** | ***Static White Box Testing*** |
| Eucalyptus | X | Y* |
| SaaS | Y | Y |
| BookService | Y | Y |
| Amazon Web Service | Y* | X |
| Sunset Sunrise service | Y | X |
| Bible Web Service | Y | X |
| Currency Conversion Web Service | Y | X |
| Weather Web Service | Y | X |

* *Threat to validity* because we do not tested on proper test data for Eucalyptus and AWS.

The case study on SaaS was conduct by building test suites for changed operations. Test sequences/steps were built based on flow of data and control according to the PRTWS. Exported test suite was given as input to the prototype of PRTWS. The tool give us option to select test sequences/steps inside selected test cases to construct RRTS.

Dynamic flow analysis in PRTWS: Four black box case studies to build the reduced test suite based on proposed PRTWS. The Parameterized analysis case study was conducted over four black box web services, namely, Currency Conversion, Global Weather, Sunset Service, and Bible web services. It can easily infer that certain permutation and combination of values for parameter execute with the certain flow. For that certain flow, we build test suite according to primary parameter scenario.

Firstly, Currency Convertor web service has operation 'CurrencyConvertor', in experiment 'FromCurrency' is chosen as the primary parameter and 'ToCurrency' is chosen as non-primary parameter. Test case is made for a particular currency, for example, 'IndianRupee' as the primary parameter with combination for test sequences/steps of other currency like 'JapanYen', 'AustralianDollar' etc as the non primary parameter. Each test sequences/steps of 'IndianRupee' has value 'INR' of 'FromCurrency' whereas non-primary parameter 'ToCurrency' changes according to currency ('JPY', 'AUD' etc). In figure 4.6 (a) reduced test suite constructed for Currency Conversion web service.

Secondly, RRTS for Global weather service as shown in figure 4.6 (b), ‘CountryName’ is chosen as primary parameter whereas ‘CityName’ is chosen as non-primary parameter. City name vary with each test sequence/step inside test case of the particular ‘CountryName’.

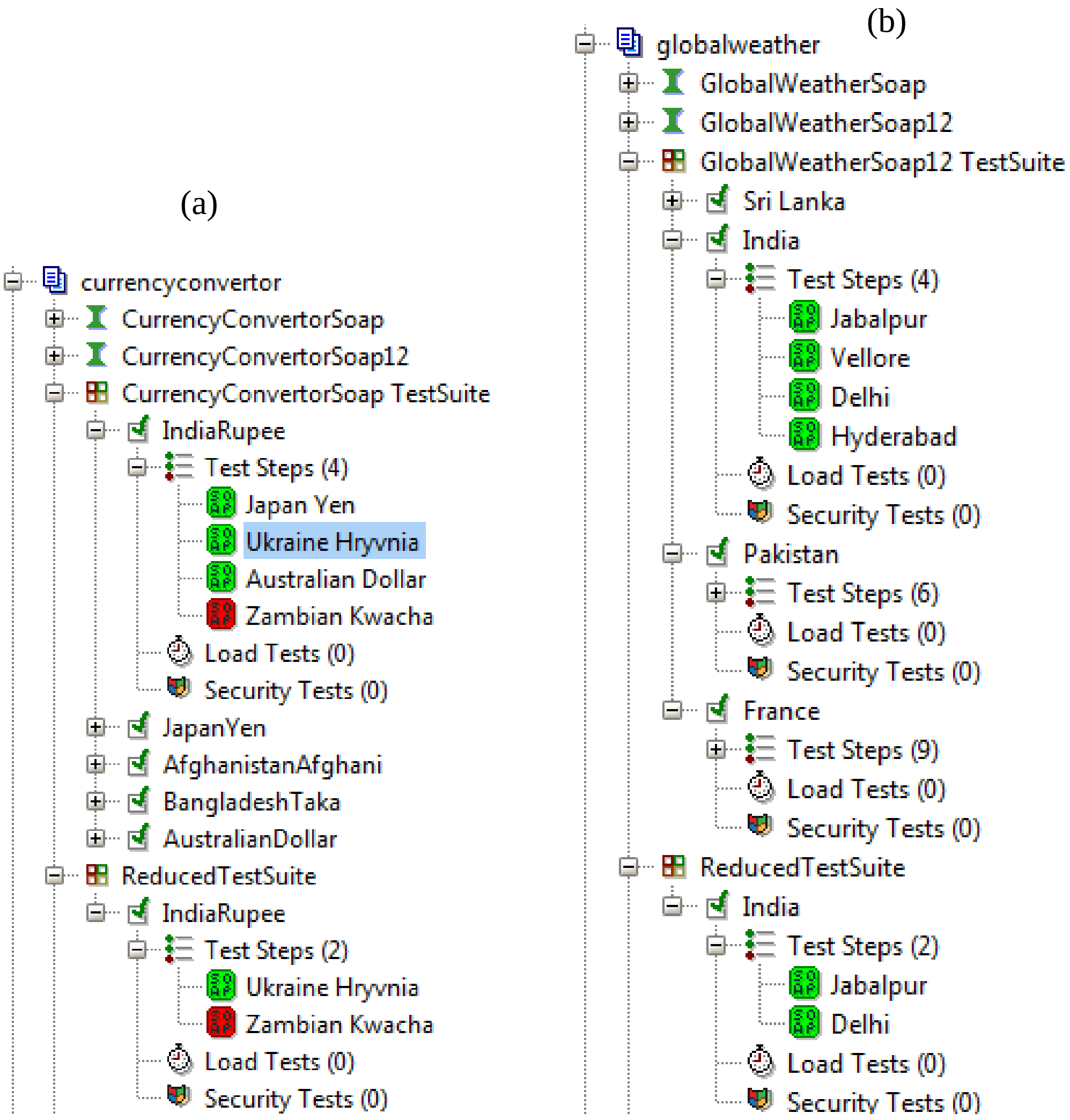

**Figure 4-6 Snapshots of SoapUI for the test suites of PRTWS (a) Reduce test suite of currency conversion web service. (b) Global weather web service test suite.**

Thirdly, in Bible web service ‘BookTitle’ as primary parameter with the combination of ‘ChapterName’ and ‘Verse’ as the non-primary parameter.

Fourthly, in special case of Sunset Sunrise web service, two primary parameters ‘Latitude’ and ‘Longitude’ are used to uniquely determine the other non-primary parameter (SunsetTime, SunRiseTime, TimeZone, Day, Month and Year in figure 4.7).

In the last row of table 4-2, the addition operation of 3 digit (a, b and c), we found all parameters as primary parameter same as we do in the database. Table 4-2 shows how the web service operations relate to 'primary' and non-primary parameter.

**Table 4-2 'Primary' parameter selection in operation of various web service projects**

| **Project** | **Primary parameter** | **Non – Primary Parameter** |
|---|---|---|
| Currency Convertor web service | FromCurrency | ToCurrency |
| Global weather service | CountryName | CityName |
| Bible web service | BookTitle | ChapterName and Verse |
| Sunset Sunrise web service | Latitude and Longitude | SunsetTime and SunRiseTime |
| Addition of 3 digit | a, b and c | ---------- |

In figure 4.6 both examples requires 50% less execution of test sequence/step. PRTWS saved the effort as compared to re-test all. Our study was very small but for the real world project like in industry, where thousands of test sequences/steps were required to be executed again and again, PRTWS will have its significant impact on effort saving.

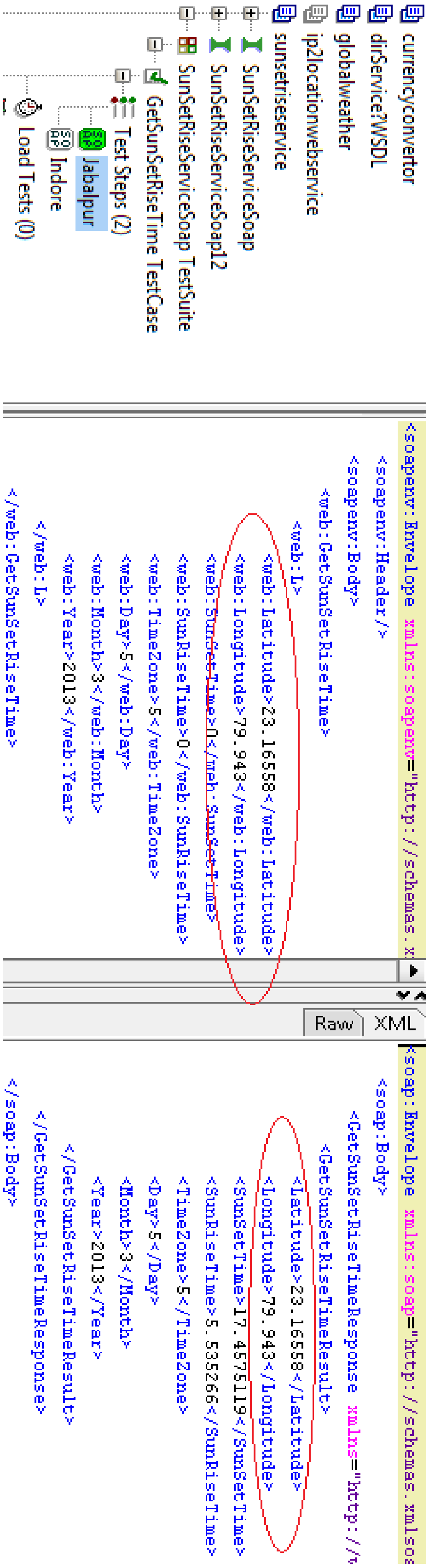

**Figure 4-7 Two parameters (latitude and longitude) combine to form the Primary parameter to uniquely identify other parameter for particular Soap response of Sunset and Sunrise in the service.**

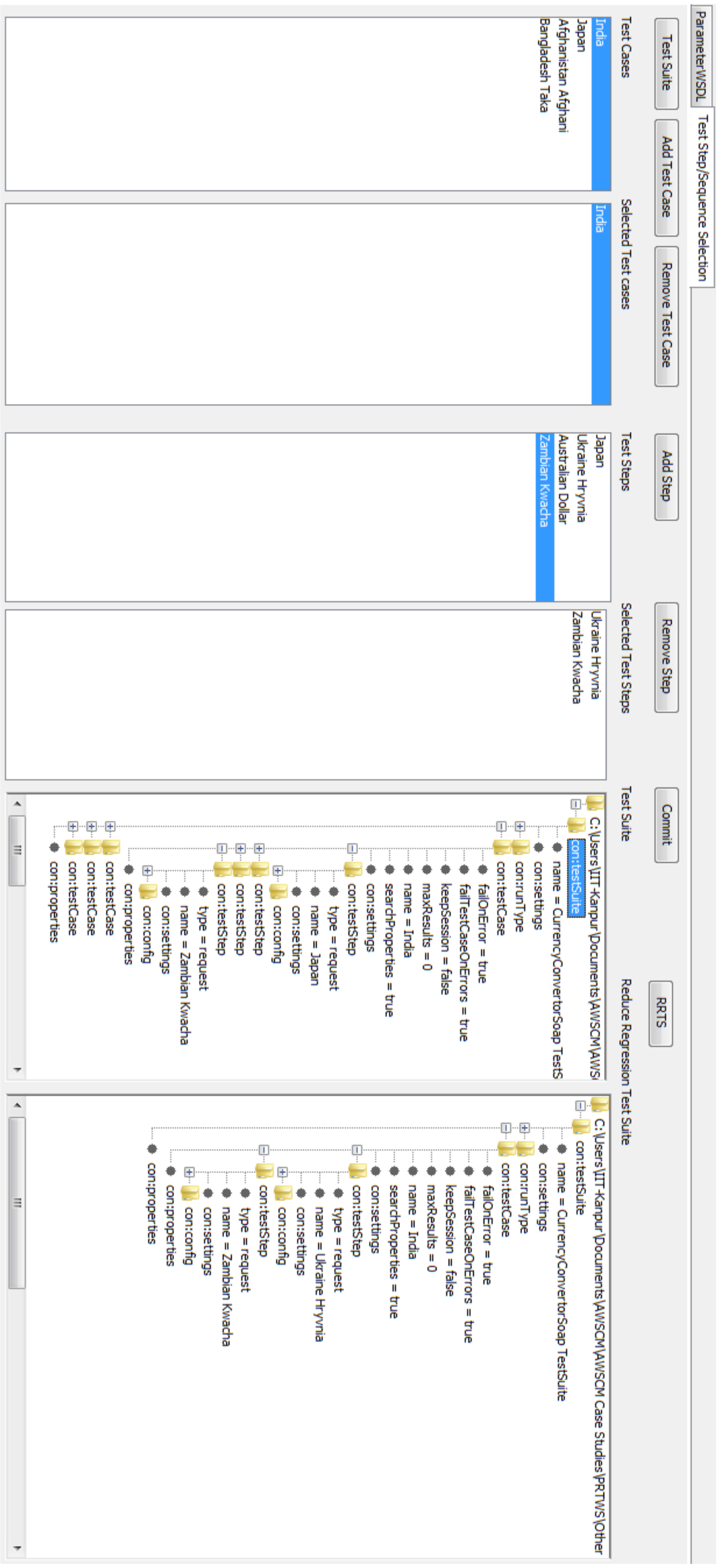

**Figure 4-8 Snapshot of prototype tool PRTWS for user selective test sequence/step inside test cases for a test suite of SoapUI in figur 4.5 snapshot of currency conversion example.**

## 4.3 Thread to Validity

Fine-grained changes at the code level for white box can be improved further. For black box, one can also check for trade of between accuracy and cost optimization. This paper works for the cost optimization whereas accuracy can be improved as much as tedious, dedicated, and refined changes can be measured.

Majorly we had worked on WSDL operation in ORTWS approach, so we had not captured the deeper semantics of relation for input/output. Limitation of our implementation is that we had not constructed deeper control flow and data flow graph at the code. In UWSDL of ORTWS implementation, required to explore more for change impact analysis based on fine-grained intra-procedural dependency analysis.

Nevertheless, in PRTWS approach we built Parameterized test suite such that test case are built for the value of primary parameter and test steps for non-primary parameter. Parameterized test suite helps in proper input/output mapping between WS test cases and WS code. We had also captured inter-procedural method and operation calls. For this, we had used call graph for web archive generated using open source code.

## 4.4 Combining ORTWS and PRTWS

Both ORTWS and PRTWS can be combined to reduce the regression testing efforts further in the following manner. For testing of a web service project, we have to construct test suite according to both the techniques. ORTWS selects the specific operations that have undergone any changes, by constructing the Reduce Regression Test Suite (RRTS) using the mapping between old test suite and one of the Subset WSDLs as described in chapter 3. PRTWS helps with the selection of specific test sequences/steps for the flow change in the web service as described in above in this chapter. We have done this for our own 'BookService' web service.

### 4.4.1 BookService

'BookService' is our own developed web service built on Java and deployed on glassfish server. 'BookService' is a web service composition with multiple calls of operations and methods. It is build to get information of two books 'Bhagavad Gita' (BG) and 'Bible' as shown in figure 4.9 and 4.10. Book Service contains of four operations, namely, 'findBookNumber', 'getAbstractOfChapter', 'getAllVerseByBookAndChapterNumber' and 'getVerseByBookAndChapterNumberAndVerseNumber'.

1. 'findBookNumber' gives the name and number of book in the database as output.

2. 'getAbstractOfChapter' give the abstract of the chapter as output for two input arguments 'bookNumber' and 'chapterNumber'. The operation get the abstract of the chapter by calling operations 'bgWSAbst', 'bibileWSAbst' in the BGWS and BibleWS.

3. 'getAllVerseByBookAndChapterNumber' give all verses as output for two input arguments 'bookNumber' and 'chapterNumber'. The operation retrieve all the verse of the chapter by calling operations 'bgAllVerse' and 'bibileAllVerse' in the BGWS and BibleWS respectively. Further calls go to the methods, repeatedly for the first to the last value of verse number, 'bgChapterN' and 'bibileChapterN' in BGVerse and BibileVerse classes respectively.

4. 'getVerseByBookAndChapterNumberAndVerseNumber' give the verse as output for three arguments 'bookNumber', 'chapterNumber and 'verseNumber'. The operation get the verse by calling operations 'bgWS', 'bibileWS' in the BGWS and BibleWS respectively. Further calls go to methods 'bgChapterN' and 'bibileChapterN' in BGVerse and BibileVerse classes respectively.

The Parameterized analysis case study, conducted over four operations of BookService. We found some parameters as primary and some as non-primary. Table 4-3 shows how the operations relate to 'primary' and non-primary parameter.

**Table 4-3 ‘Primary’ parameter in an operation in BookService**

| Operations | Primary parameter | Non–Primary Parameter |
|---|---|---|
| 1st findBookNumber | --- | --- |
| 2nd getAbstractOfChapter | ‘BookNumber’ | ‘ChapterNumber.’ |
| 3rd getAllVerseByBookAndChapterNumber | ‘BookNumber’ | ‘ChapterNumber’ |
| 4th getVerseByBookAndChapterAndVerseNumber | ‘BookNumber’ | ‘VerseNumber’ and ‘ChapterNumber’ |

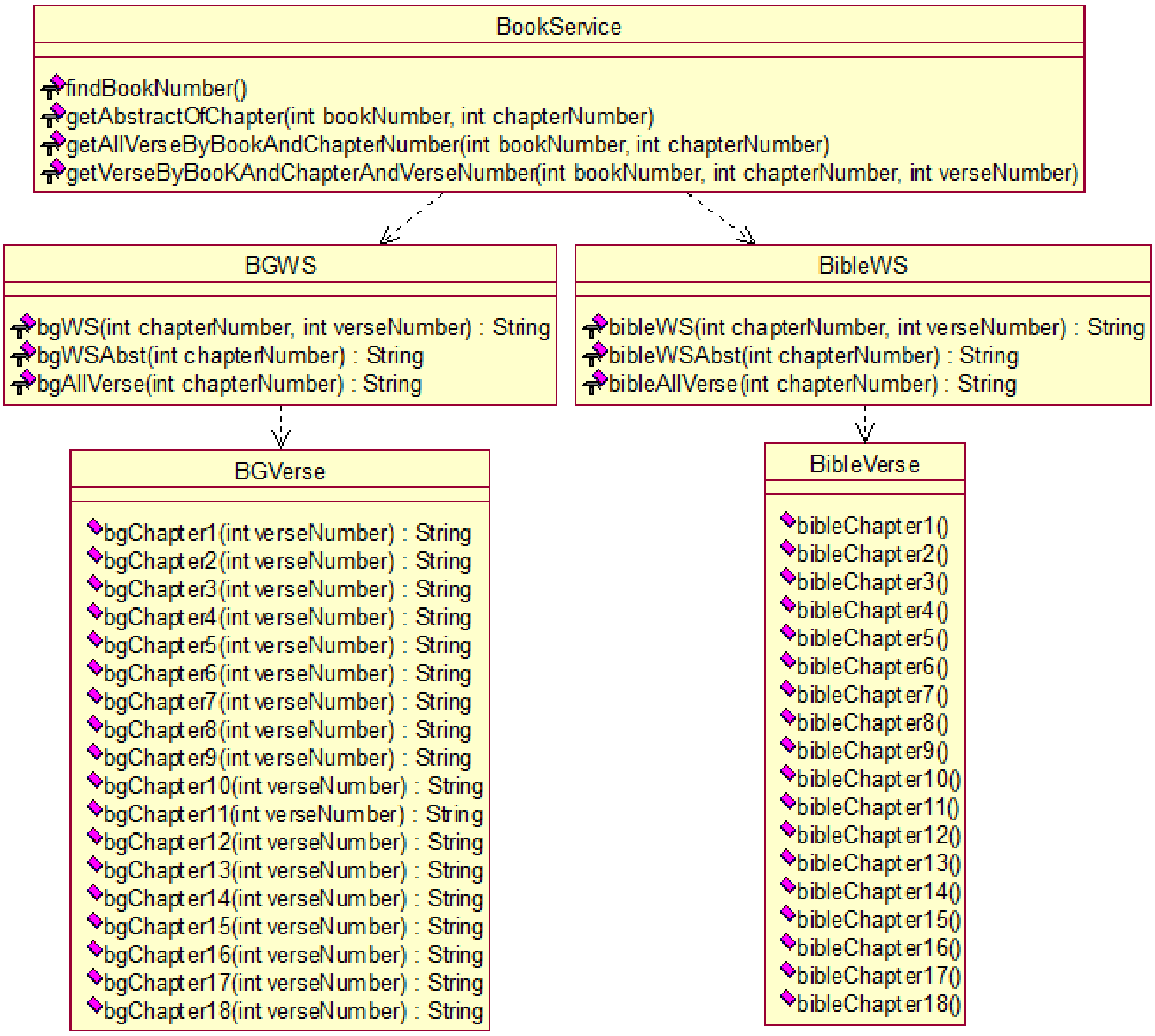

**Figure 4-9 ‘BookService’ service hierarchy and call to BGWS and BibleWS service with all the method and operation signatures.**

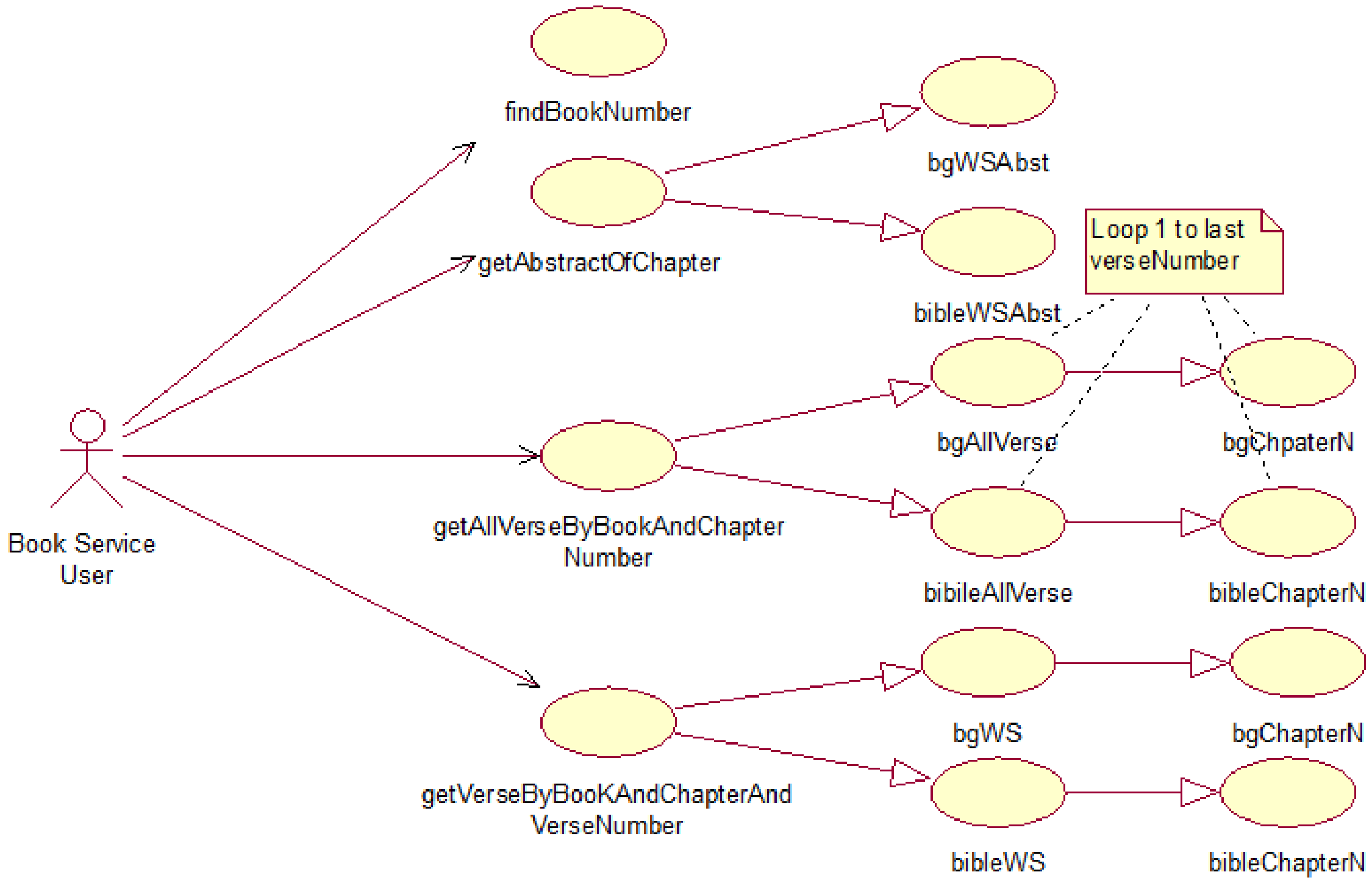

**Figure 4-10 BookService user functionality by using operations and method calls.**

An operation 'getAllVerseByBookAndChapterNumber' in Figure 4-11 and 'getAbstratofChapter' in Figure 4-12 has two parameters 'bookNumber' as primary parameter and 'chapterNumber' as non-primary parameter. We can easily see that to create test suite could have two test case named on books 'Bhagvad Gita' and 'Bibile'. For all the test step/sequences in a test case of 'BhagvadGita' and 'Bible' has the fixed value of primary parameter 'bookNumber'. There could be several test steps based on chapter numbers, which are the values of non-primary parameter.

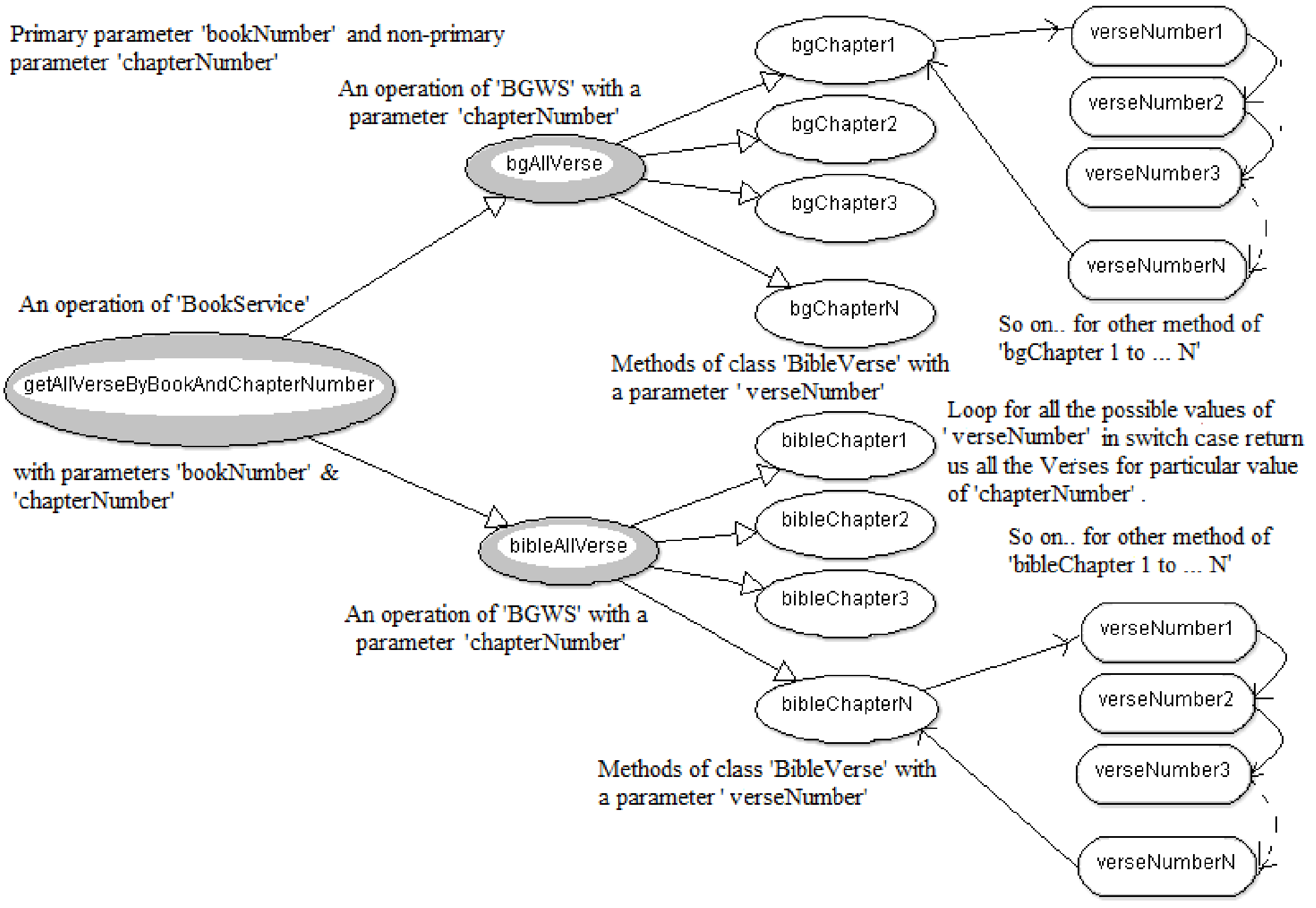

**Figure 4-11. An operation 'getAllVerseByBookAndChapterNumber'**

Similarly, an operation ‘getVerseByBookAndChapterAndVerseNumber’ in figure 4-13 has three parameters ‘bookNumber’ as primary parameter, ‘chapterNumber’, and ‘verseNumber’ as non-primary parameters. We can easily see that to create test suite could have two test case named on books ‘Bhagvad Gita’ and ‘Bibile’. For all the test step/sequences in a test case of ‘BhagvadGita’ and ‘Bible’ has the fixed value of primary parameter ‘bookNumber’. There could be several test steps based on ‘chapterNumber’ and ‘verseNumber’ as the values of non-primary parameter.

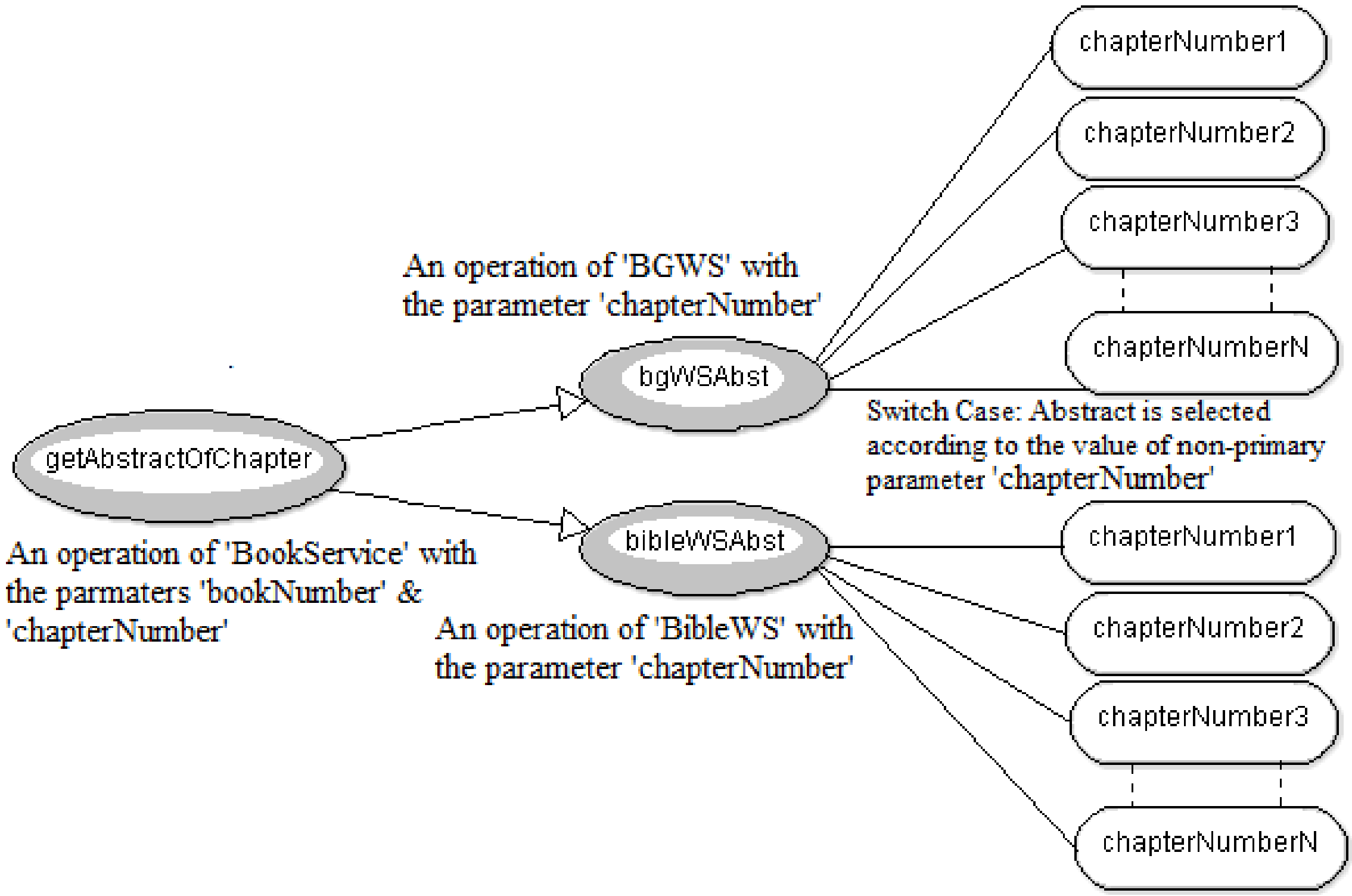

**Figure 4-12 An operation 'getAbstractOfChapter'.**

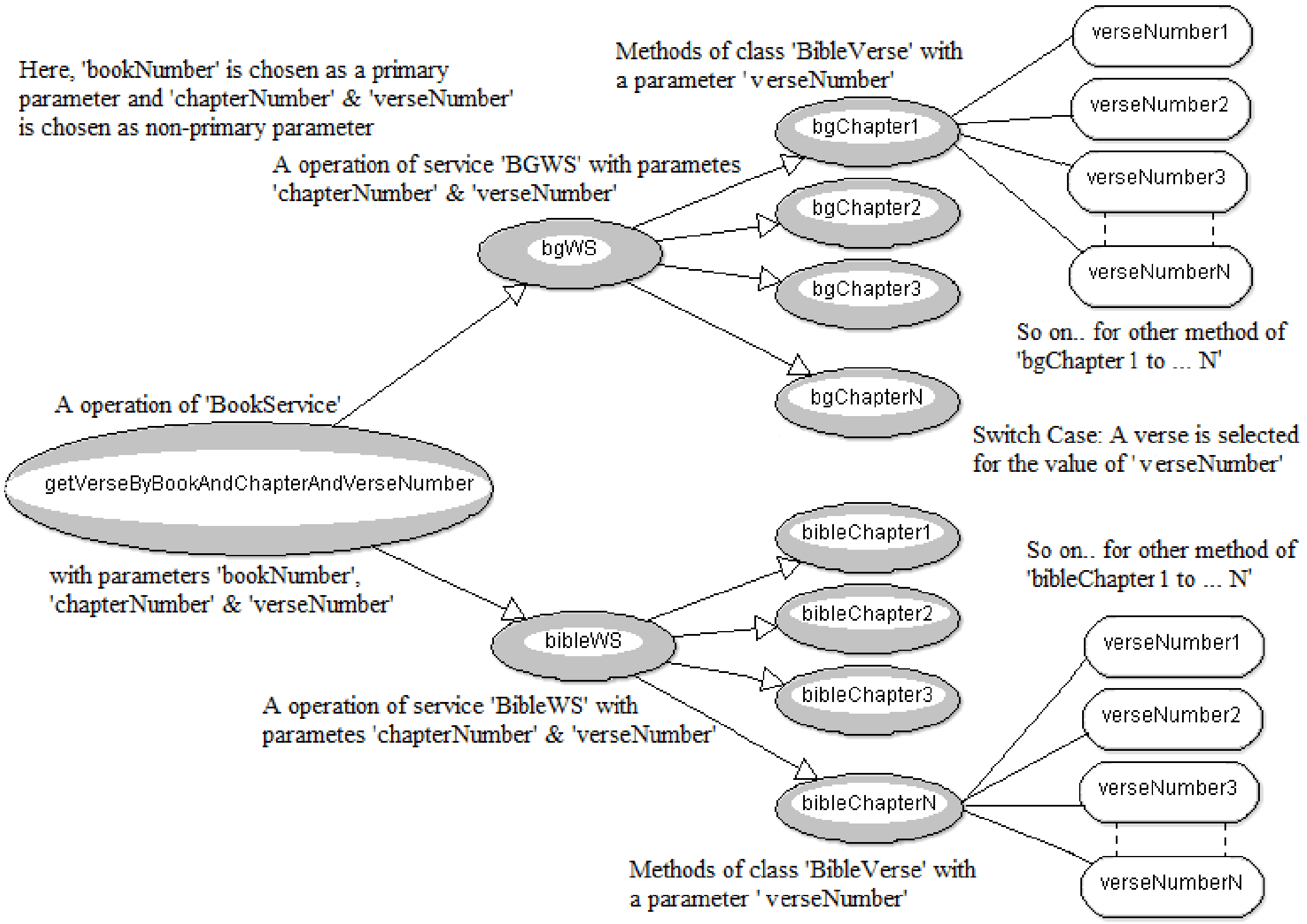

**Figure 4-13. An operation 'getVerseByBookAndChapterAndVerseNumber'.**

To perform the experiment an intentional change has been introduce in the ‘bgChapter2’ and ‘bgChapter3’. These changes are captured by AWSCM, additionally it has also captures the operation which is calling these changed methods. Further, the information of changes and operation call trace is used to construct the Parameter WSDL and PRRTS as shown in figure 4.14. We executed the RRTS build by AWSCM on the SoapUI shown in figure 4.15. The SopaUI test suite, show first test suite according to ORTWS, second test suite for operation ‘GetAbstractOfChapter’ with 8 test sequences/steps which was reduce to 2 in third test suite for PRRTS.

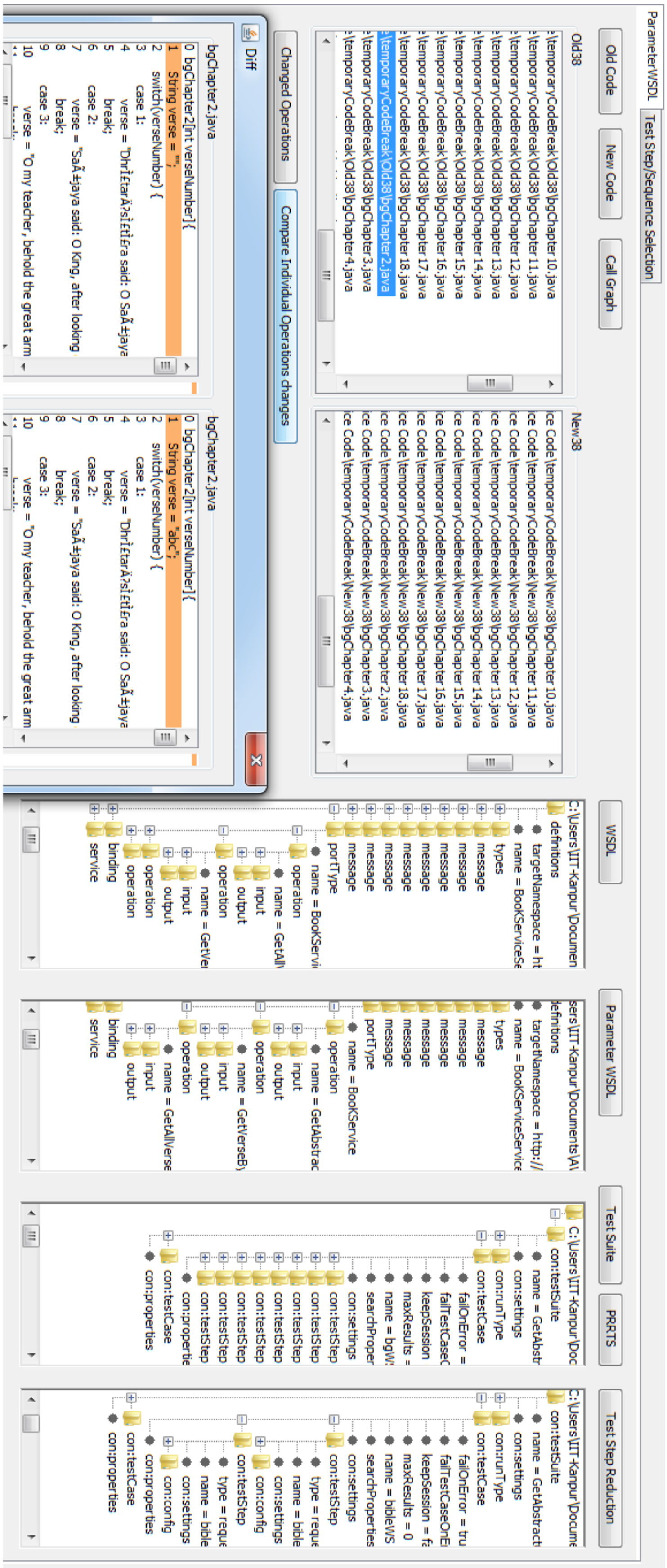

**Figure 4-14 AWSCM for construct RRTS of ‘BookService’ for operation ‘GetAbstractofChapter’**

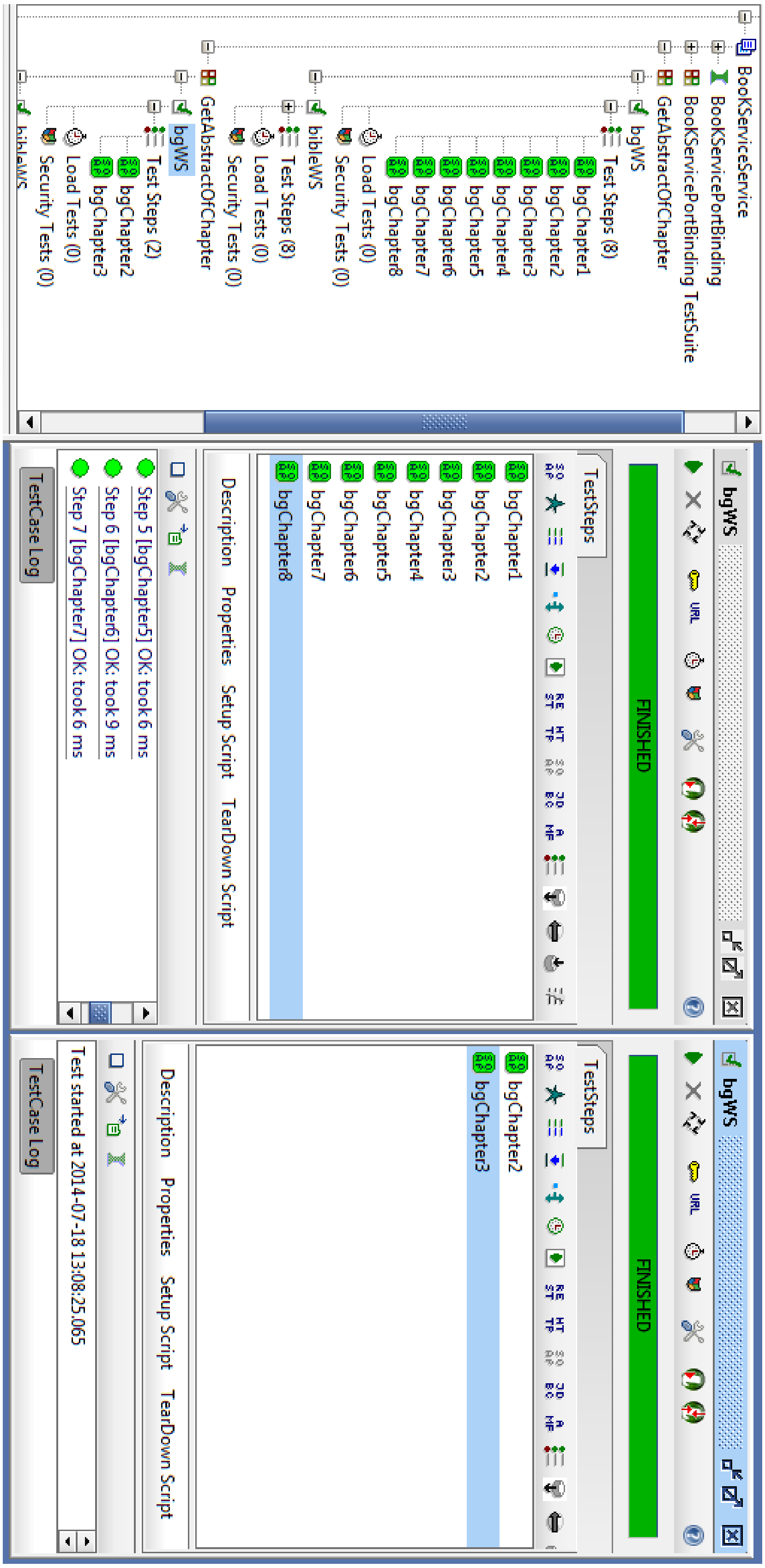

**Figure 4-15 Execution of Reduced Regression Test Suite of ‘BookService’ for operation ‘GetAbstractOfChapter’.**

## 4.5 Summary

In this chapter, we have described an approach PRTWS, which is prototyped as a tool for test sequence/step selection based on change impact analysis and user selective from old test suite. We had retrieved the modified primary parameter (i.e. test case) to construct the Reduced Regression Test Suite. RRTS required to be executed upon upgrade of versions for regression testing only on the modified code. For all the case studies, we achieved effort reduction.

# CHAPTER 5

# AUTOMATED WEB SERVICE CHANGE MANAGMENT TOOL

## 5.1 The Overview

The two regression testing approaches proposed in the previous chapters, i.e., ORTWS and PRTWS are prototyped as a tool, named as Automatic Web Service Change Management (AWSCM). AWSCM helps in selecting the relevant test cases along with test sequences/steps from the old test suite. The illustrative examples on six web service projects obtained from web, namely, Eucalyptus, Amazon Web Service, Currency Conversion, Global Weather, Sunset Service, and Bible Web Service demonstrate the applicability of the proposed tool for the real world projects. **Tool link** for more detail https://sites.google.com/site/animeshchaturvedi07/research/awscm

The tool saves manual efforts by detecting the changes in the web service automatically and selecting the relevant test cases to perform efficient regression testing of web service. AWSCM has two modules ORTWS and PRTWS. ORTWS helps to find the operations required to be regression testing. PRTWS helps in quickly figuring out the combinations of the inputs to be exercise for an effective re-testing of the web service.

## 5.2 Tool Architecture

AWSCM contains two modules for regression testing, namely, ORTWS and PRTWS. WS code automated analysis and manipulation is normally performed by using WSDL as interface. In AWSCM as shown in figure 5.1, we have prototyped two approaches i.e. ORTWS and PRTWS. ORTWS is divided into four features DWSDL, UWSDL, RWSDL, and CWSDL, as described in the chapter 3. PRTWS have features of the Parameterized sequence reduction, as described in the chapter 4. AWSCM contains the intermediate artifacts shown in figure 5.1 (a). AWSCM contains five package and components as shown in fig 5.1 (b) (c) respectively, further detail are given in section 5.3.1 "Implementation of AWSCM".

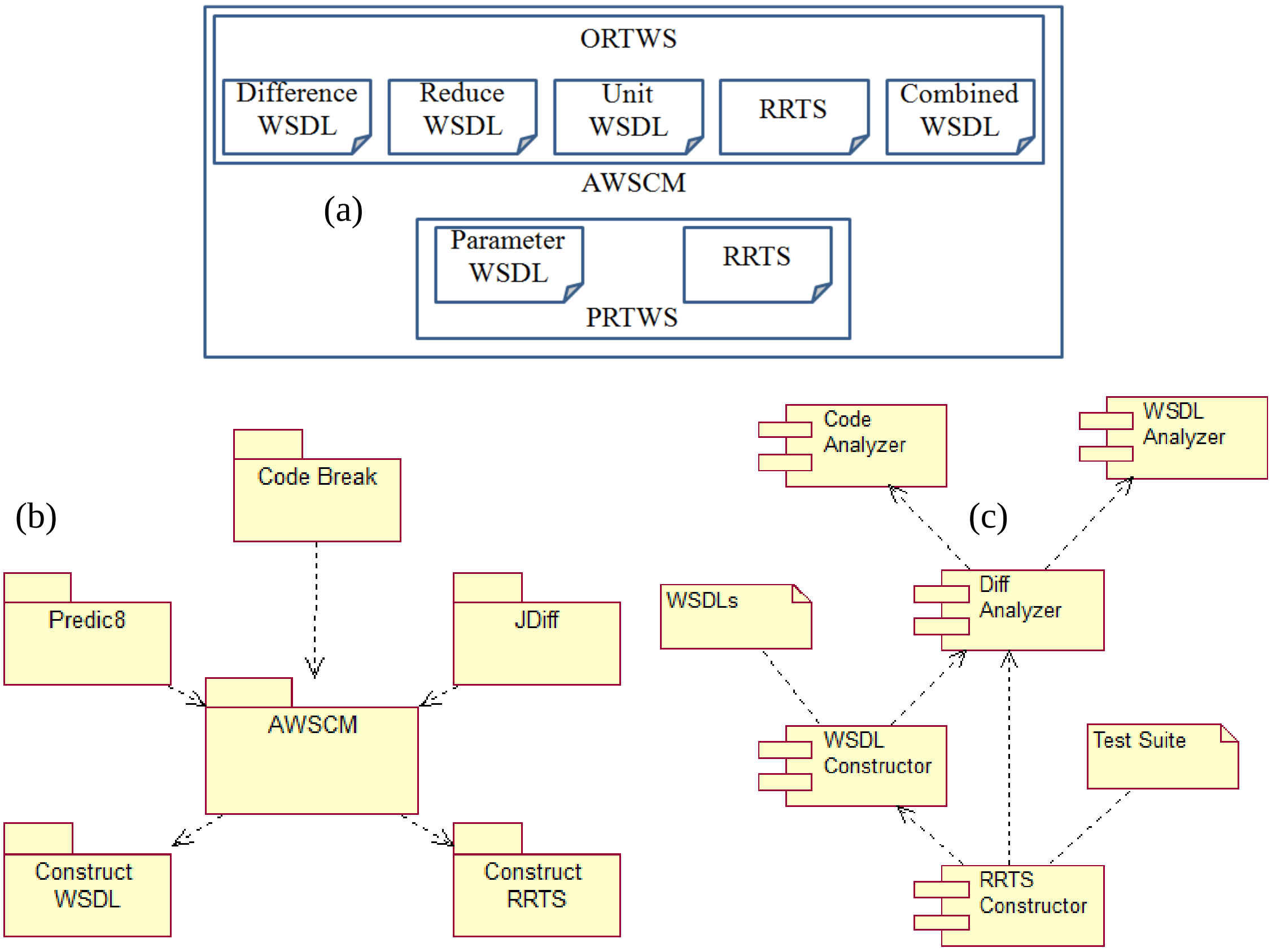

**Figure 5-1 Prototype AWSCM (a) Artifacts (b) Packages (c) Components**

## 5.2.1 ORTWS module

ORTWS module has three intermediate artifacts. Changes at WSDL are captured in the Difference WSDL (DWSDL) and changes at code are captured in the Unit WSDL (UWSDL). Additionally, we can also capture the need of selective re-testing using another WSDL, known as Reduced WSDL (RWSDL). These Subset WSDLs combined to form a Combined WSDL (CWSDL), which is used for the construction of Combined RRTS (CRRTS) as shown in figure 5.2.

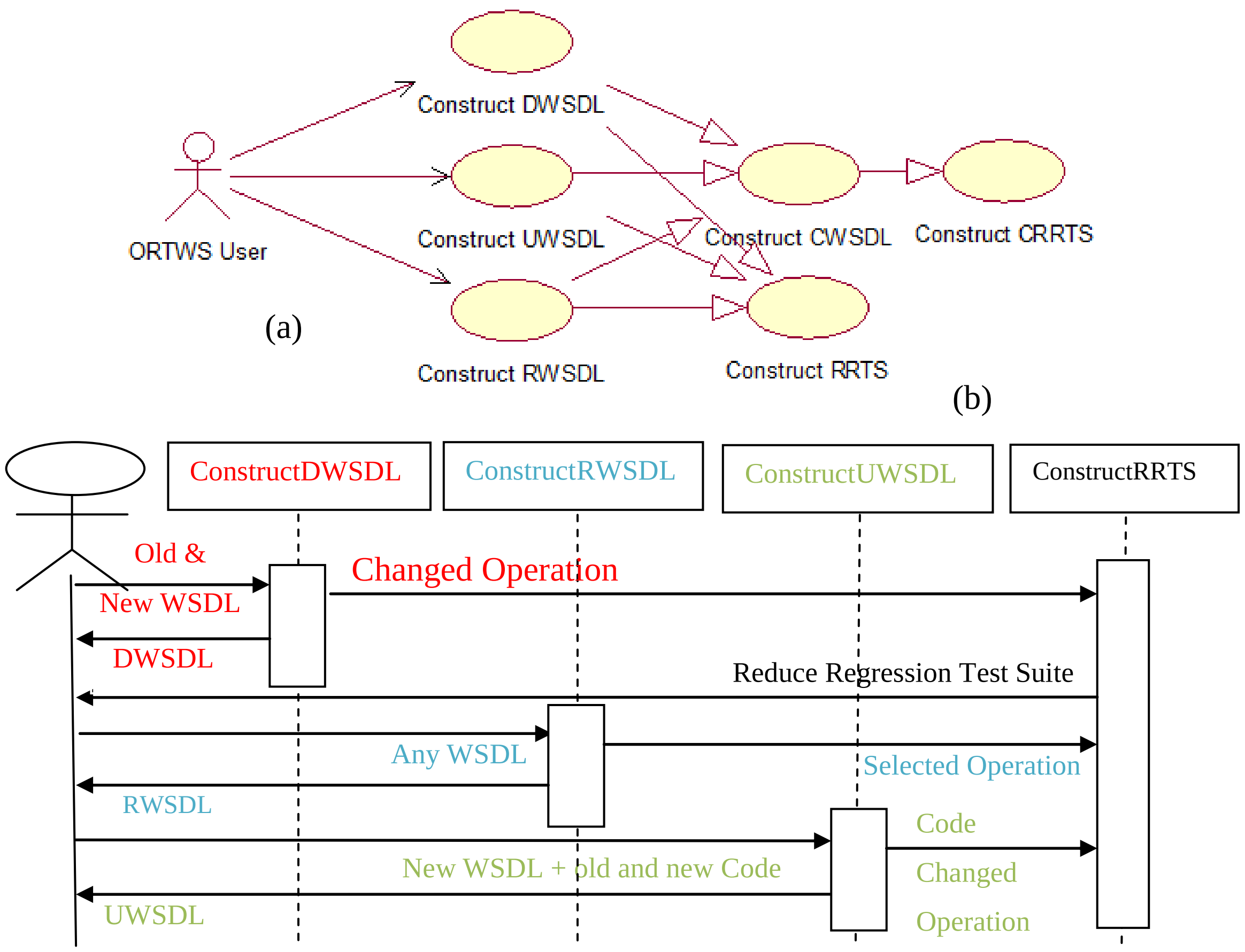

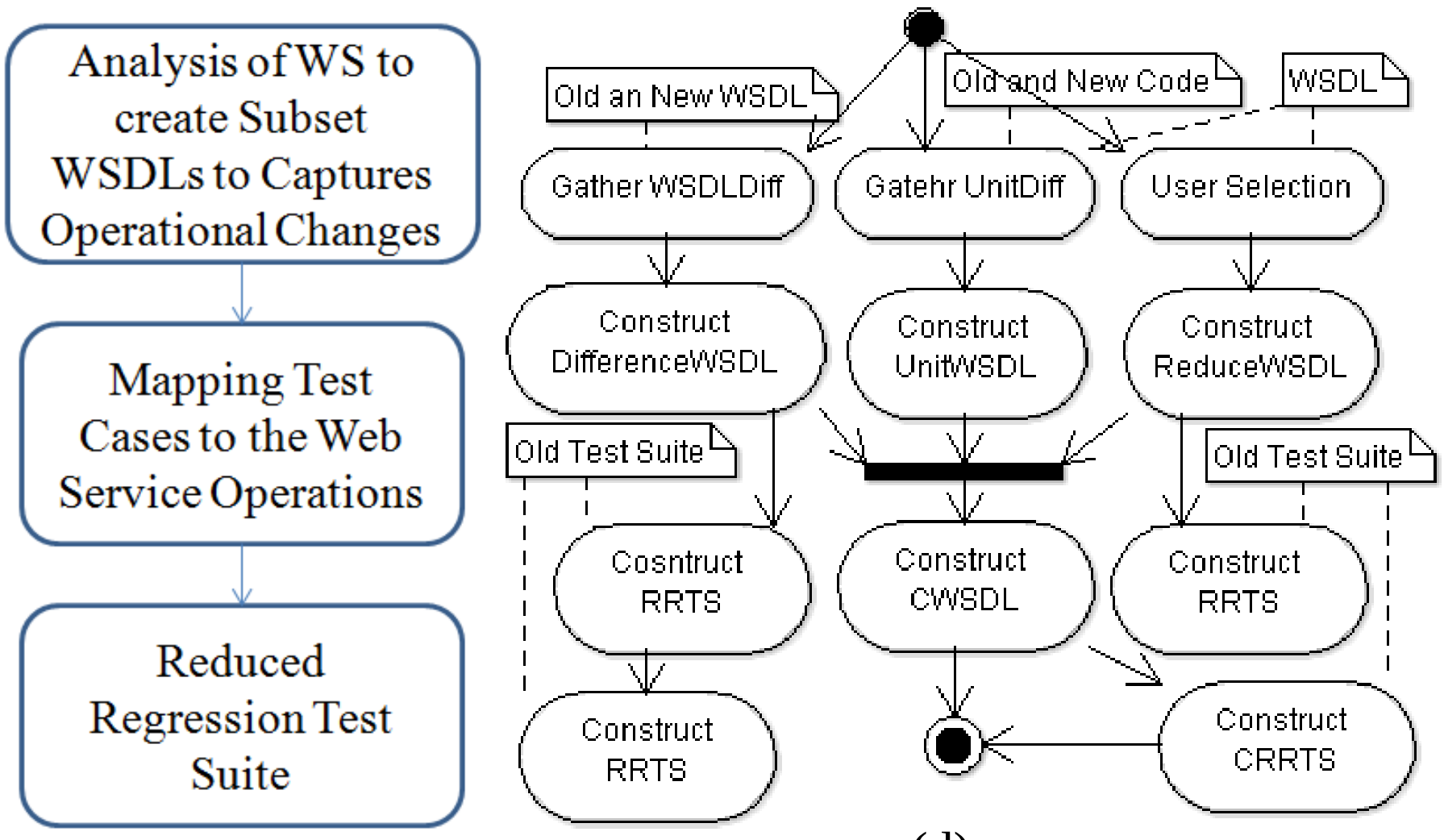

(c) (d)

**Figure 5-2 ORTWS module of AWSCM's (a) Use Case (b) Sequence Diagram (c) ORTWS process (d) Activity Diagram**

The Subset WSDLs must be in proper structure for maintaining communication with the web service logic. We have followed WSDL standard to construct these Subset WSDLs in AWSCM. Constructed WSDL must have correct XSD types, Message part, Operation, Port type, Binding, Port, and Service according to input WSDL. Further, we build functionality in the tool for the Reduce Regression Test Suite (RRTS) generation using old test suite and above constructed Subset WSDLs. RRTS returned as output, which contained test cases for only those operations, which were in the input WSDLs figure 5.2.

**Applicability of (D/U/R) WSDL:** If the changes to the web service are at the code then changes either reflected or not reflected on the WSDL in Table 5-1.

Firstly, if changes on WSDL, then they can be captured by (D/R) WSDL and they are helpful wherever we have access only to the WSDLs and do not have any details of web service code. Limitation with (D/R) WSDL is that they do not identify the changes at code. This WSDL change based regression-testing (black box) which cannot done by using (D/R) WSDL.

Secondly, if coding changes done to the web service, which is not on the WSDL, then they can be capture by UWSDL. This code change based regression testing (white box) can be done by using UWSDL.

**Construction of (D/U/R) WSDL:** AWSCM constructs (D/U/R) by gathering changes using semantic textual difference utilities and libraries, namely, JDIff [36], Predic8 [44] and Regex [45].

**Table 5-1 Changes captured by DWSDL, RWSDL, UWSDL**

| Scope of changes handling by utilities | |
|---|---|
| Utilities | Capture changes at WSDL or at code |
| DWSDL | Capture changes at WSDL for operation deletion, insertion or I/O modification |
| RWSDL | Uses WSDL to select operations to be test |
| UWSDL | Code of operation changes are captured |

## 5.2.2 PRTWS module

PRTWS is also useful to reduce test sequences/steps in the test suite according to the code flow or user selection for the combination of parameter scenario. According to the PRTWS, we built the web service test suite according to the primary parameter scenario as described in Section 4.1.1. A test suite given as input to the 'Test sequence/step selection' module of the AWSCM, we can select required test sequences/steps. We captured the flow within the operational code in figure 5.3.

PRTWS module has the Parameter WSDL (PWSDL) as an intermediate artifact. The Parameter WSDL is an extension of the Unit WSDL for WS compositions and inter-procedural call interaction. The Parameter WSDL is created with the operation whose inter-procedural parametric flow is affected i.e. if a called method that have undergone changes. In the approach, we had captured the flow within the operational code in figure 5.3 (d), (e).

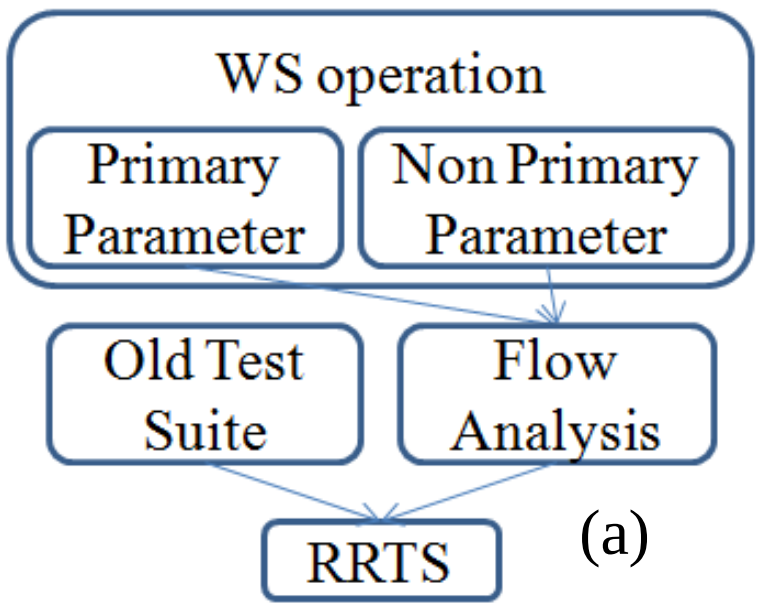

(a)

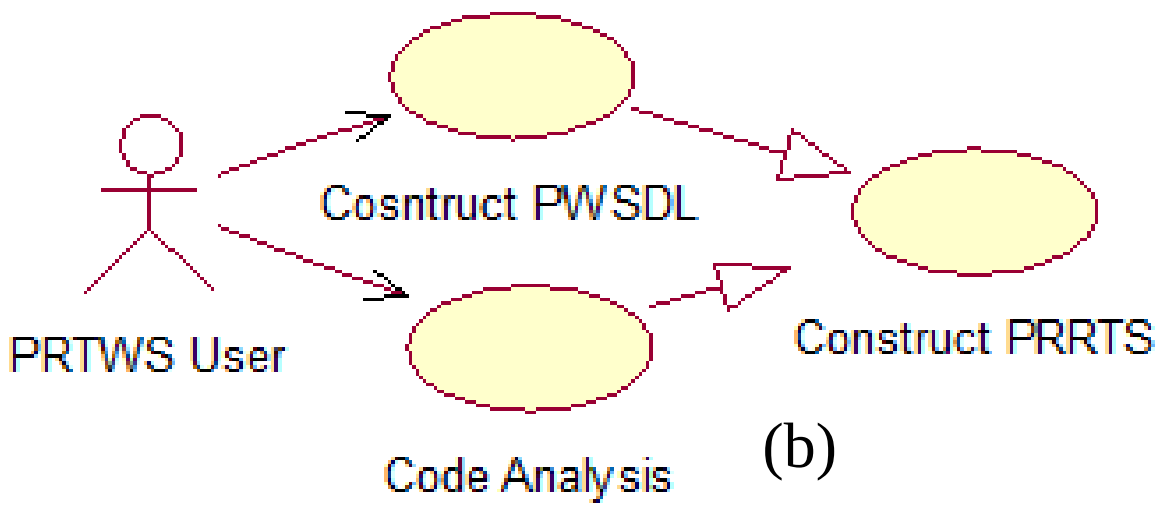

(b)

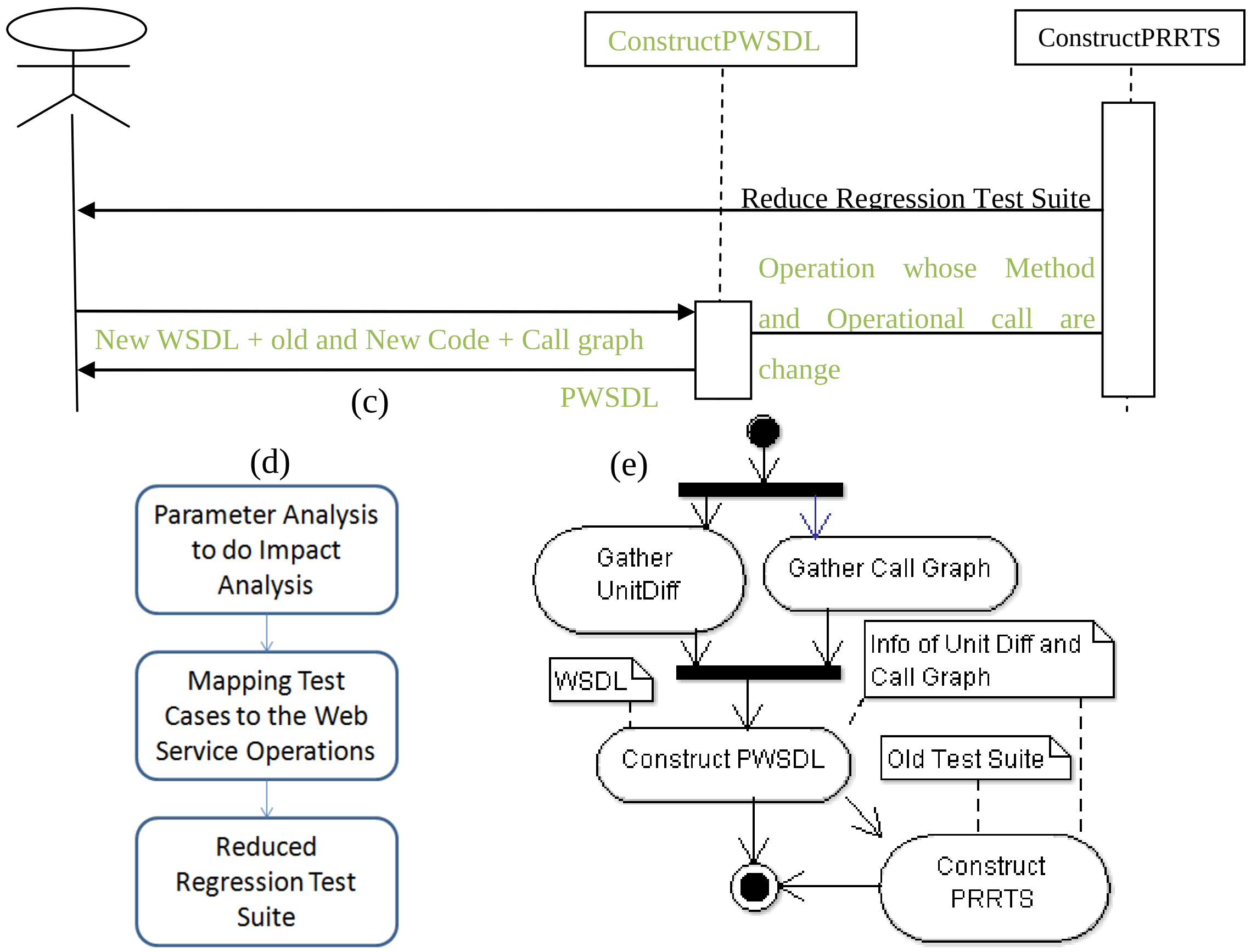

**Figure 5-3 PRTWS module of AWSCM's (a) Use Case (b) Sequence Diagram**

## 5.3 Implementation of AWSCM

AWSCM contains following five components shown in fig 5. 1 (c).

*A 'Code Analyzer'*: It is used for separating method and operation. All the separated operations and methods are kept in different file named on the operation and methods.

*B 'WSDL Analyzer'*: It is used for change impact analysis and operational analysis.

*C 'Diff Analyzer'*: It is uses both 'Code Analyzer' and 'WSDL Analyzer' to gather change impact analysis on separated operations.

*D 'WSDL Constructor'*: It uses the info of 'Diff Analyzer' and input WSDL to construct various Subset WSDLs.

*E 'RRTS Constructor'*: It is used for the construction of various RRTS according to the information of change impact analysis from 'Diff Analyzer' and the old test suite.

Various algorithms were described in details.

### 5.3.1 Algorithm inside 'WSDL Constructor'

We can gather difference between two WSDL by searching inserted, modified operation name (operation2') for example in figure 5.4. After gathering difference, DifferenceWSDL is constructed by manipulating version_1 and version_2 of WSDL for a web service. Algorithm gathers various parts of WSDL to construct a new Subset WSDL.

WSDL is the description of a web service. WSDL parsing is performed for automated web service testing. It can logically infer that to perform automated regression testing uses WSDL. Hence, it is required to find automated difference of web service based on WSDL changes. Difference of WSDL is same as difference in code as described for reduced regression testing. This difference helps in detecting the changes, which map for reduced regression testing.

All WSDL is constructed by AWSCM are in proper structure and compliance with the WSDL standards. Thus, web service client can maintain communication with the web service logic via WSDL. Special care is considered during the construction of (D/U/R/C) Subset WSDL, with the proper semantic, syntactical, XML tags, and data of WSDL. That is (D/U/R/C) WSDL must have the same data of port, service, binding and schema (input/outputs of operation) as in the inserted WSDL. (D/U/R/C) WSDL have the correct XSD types, message part, operation, port type, binding, port, and service according to input WSDL.

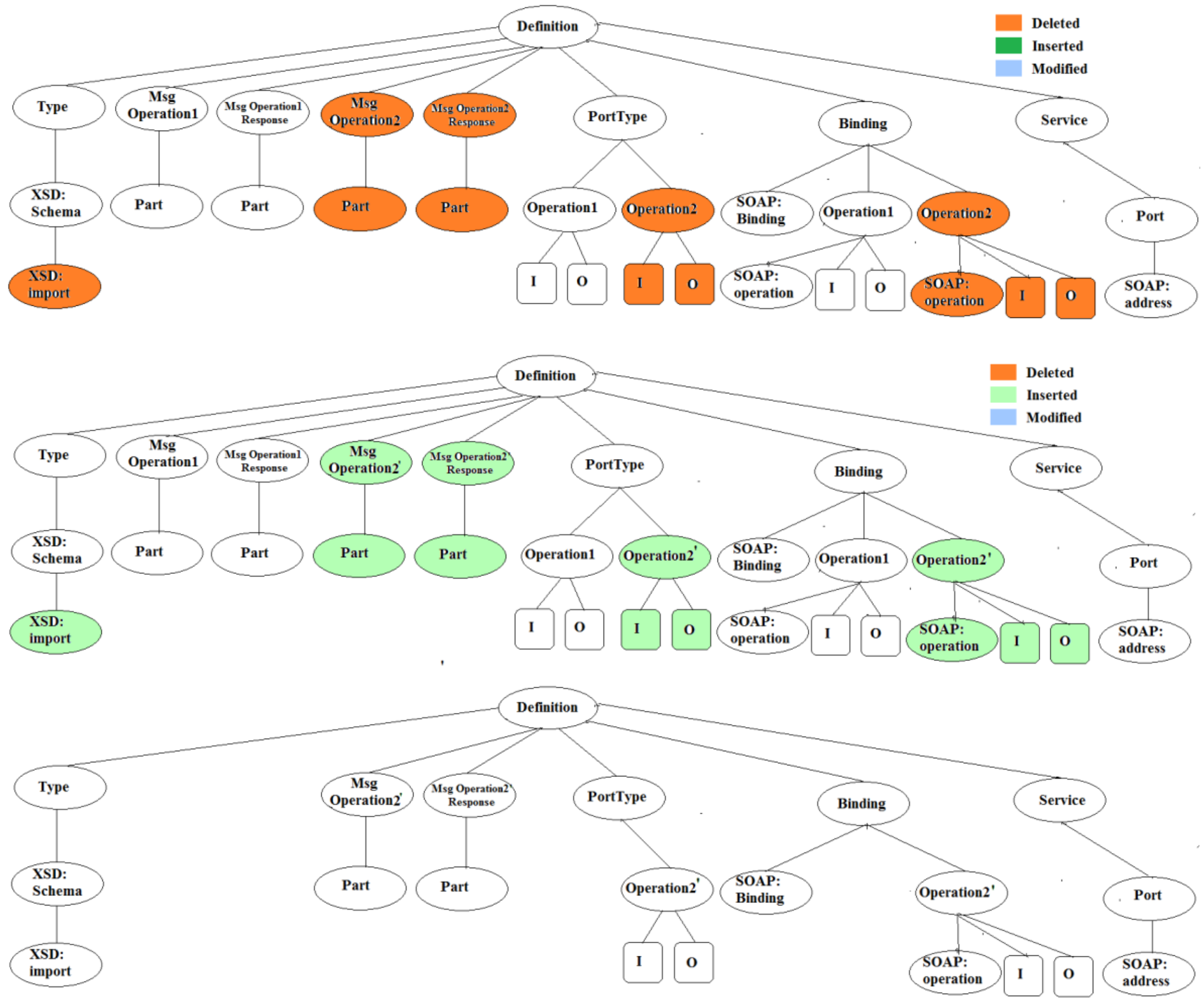

**Figure 5-4 Construction of WSDL for operation2' from new WSDL.**

To construct (D/U/R/C) WSDL with the help of algorithms are used after gathering the differences. In the section, a tool is constructed based on proposed algorithm. Many type of difference measuring criteria, technique, and libraries as described in [9]. All the sub functions calling algorithm for the construction of D/U/R/C WSDL to the graph are explained one by one. Algorithm for parts of WSDL is demonstrated in the subsections given below.

### *5.3.1.1 constructStartDefinition(newWSDL)*

Policies of web service defined at starting of definition parts of WSDL. DifferenceWSDL must have same policy as in the latest version of WSDL because policies are the condition for working of web service.

Extraction of xml starting tag with <?xml….?> which describe a xml file version and encoding technique. Thereafter extraction of Definition tag <definitions…> (can also be represented as <wsdl:definitions….> depending on WSDL standard). Definition describes standard of xml entities (xsd, soap, and wsdl), location of target namespace, and name of the application. Same thing in the algorithmic form is given below

String constructStartDefinition(File *newWSDL*) {

String *buffer* = Read(*newWSDL*)

*cStartDef* = Retrieve String from tag<?xml ………?>

//store String in *cStartDef*

*cStartDef* = *cStartDef* + <wsdl:definition……./>

// store String in *cStartDef*

return *cStartDef* }

#### *5.3.1.2 constructXSD(newWSDL, Schema)*

Schema for input/output of operation was present in XSD of WSDL. If schema input or output changed for the operation, then that operation must added in DifferenceWSDL.

Extraction of all the Types for the required operations, Type describes schema (XSD: data structure for Input / Output) inside <types> element tag of WSDL. The entire nested schema must consider for the construction of new schema in new WSDL. Same thing in the algorithmic form is given below

String **constructXSD**(File *newWSDL,* String[] *differenceOperation*) {

String *buffer* = Read(*newWSDL*)

For(scan till end of the file) {

*cXSD* = Retreive String from <xs:complextype name = "differenceOperation[i]Type">……..<xs:complexType>

*cXSD* = *cXSD* + Retreive String<xs:complextype name = "differenceOperation[i]ResponseType">

……..<xs:complexType >

} **return** *cXSD* } // instead of complex one could also have simple type according to requirement

### *5.3.1.3 constructMessage(newWSDL, requiredOperation)*

Extraction of the message part for the 'requiredOperation' in between the tags <message…>…. </message>. Each operation has its corresponding unique message, which describes the information necessary to perform the operation call. Every Message consist of one or more logical parts (i.e. part corresponds to a unique message type attribute in a Type). Same thing in the algorithmic form is given below

String **constructMessage**(File *newWSDL, differenceOperation*) {

String *buffer* = Read(*newWSDL*)

For(scan till end of the File) {

*cMsg* = Retrieve String from <wsdl:message name = " + s[i] +

"Response">……. </wsdl:message>

} **return** *cMsg* }

### *5.3.1.4 constructPort(newWSDL, requiredOperation)*

For Operation difference present in port type, binding of operation of web service is inserted, deleted, or modified. Operation inserted means operation added to new version of web service. Operation deletion means operation in old version of web service. Operation modification means operation modified either by changing its name or by its input output parameter. Extraction of all the Ports required for the operation in between <portType ……</portType>. PortType contains address or connection point, for communication of operations with the messages. Same thing in the algorithmic form is given below

String **constructPort**(File *newWSDL,* String[] *differenceOperation*) {

String *buffer* = Read(*newWSDL*)

*cPort* = Retrieve String from <wsdl:portType…>

For(scan till end of the File) {

```
        cPort = cPort + Retrieve String <wsdl:operation name = " + s[i] + "
                >……</wsdl:operation >
    }
    cPort = cPort + String </wsdl:portType >
    return cPort }
```

### *5.3.1.5 constructBinding(newWSDL, differenceOperation)*

Extraction of all the bindings required for the operation to name its input/output used in its type. Binding defines the operations and its SOAP binding form (RPC/Document) and transport (SOAP Protocol). Same thing in the algorithmic form is given below

```
String constructBinding(File newWSDL, String[] differenceOperation) {
    String buffer = Read(newWSDL)
    cBinding = Retrieve String from <wsdl:binding………..>
    For(scan till end of the File) {
        cBinding = cBinding + Retrieve String <wsdl:operation name = " +            s[i]
+ " >…………</wsdl:operation >
    }
    cBinding = cBinding + String </wsdl:binding >
    return cBinding }
```

### *5.3.1.6 construcService(newWSDL)*

For Service URL links difference present in Service in (D/U/R/C) WSDL service tag must be same as in new version of WSDL because service condition for collection of related ends of web service. Extraction of the service location parameter required for web service functions that are exposed to the Web based protocols (client communicates to its web service via URL and Port Binding). Same thing in the algorithmic form is given below

```
String constructService(File newWSDL) {
```

String *buffer* = Read(*newWSDL*)

*cService* = Retrieve String from tag <wsdl:service name = “ ”>……… </wsdl:service > //store in *out*

**return** *cService* }

#### *5.3.1.7 constructEndDef(newWSDL)*

WSDL end with </definition> tag, so extract the end of Definition </definitions>. Same thing in the algorithmic form is given below

String constructEndDef(File *newWSDL*) {

**return** *cEndDef* = String </wsdl:definitions > }

Note for all the algorithms given in sub section. Some terminology are different between WSDL1.1 and WSDL2.0 i.e. PortType change with Interface. In WSDL 2.0 message is not used.

Various algorithms inside the AWSCM tool for WSDL is construction are as follows

### 5.3.2 Algorithm for Computing Difference WSDL

Changes at WSDLs are mapped automatically by identifying using textual differencing in our tool. For example operation2’ in figure 5.3, is an inserted operation between two subsequent versions of web service. To gather changes between two versions of web service the Difference WSDL feature takes two inputs, namely, new and old WSDL. Using these changed operations, AWSCM semantically constructs DWSDL.

For the construction of ‘DifferenceWSDL’ as output, three inputs are required ‘*oldWSDL*’, ‘*newWSDL*’ and ‘*Difference*’ (can be gathered using utility, libraries etc). Construction of DWSDL is a three-step process.

First, gather the operations ‘*differenceOperation', ' differenceSchemaOperation'* which are present in ‘*Difference*’ as impact of the changes.

Second, construct the various parts of WSDL (Definition, XSD, Message, Port, Binding and Service), by using the operations gathered in first step. Definition and Service are constant (i.e. independent of our variable operations), whereas others are dependent on the operations.

Third, combine all the parts in a proper sequence to form a Difference WSDL.

File ConstructDifferenceWSDL(File *oldWSDL*, File *newWSDL*, *Difference*) {

String *cStartDef, cXSD, cMsg, cPort, cBinding, cService, cEndDef*

Array of String[] *differenceOperation, differenceSchemaOperation*

[*differenceOperation* ∈ *Difference* | *differenceOperation* = Operation for which input/*out*put is modified]

[*differenceSchemaOperation* ∈ *Difference* | *differenceSchemaOperation* = Operation which is inserted in *newWSDL*]

*cStartDef* = **constructStartDefinition**(*newWSDL*)

*cXSD* = **constructXSD**(*newWSDL, differenceSchema*)

*cMsg* = **constructMessage**(*newWSDL, differenceOperation*)

*cPort* = **constructPort**(*newWSDL, differenceOperation*)

*cBinding* = **constructBinding**(*newWSDL, differenceOperation*)

*cService* = **construcService**(*newWSDL*)

*cEndDef* = **constructEndDef**(*newWSDL*)

*DifferenceWSDL= cStartDef+cXSD+ cMsg+ cPort+ cBinding+ cService +cEndDef*

**return** *DifferenceWSDL* }

### 5.3.3 Algorithm for Computing Unit WSDL

UWSDL of AWSCM takes three inputs: new, old code of operations and new WSDL. Tool semantically constructs UWSDL according to the semantics of new WSDL with the operations gone through the change at the code.

Functionality of UWSDL gathers change impact and maps them to the test suite in terms of test sequences/steps inside test case. We used JDiff for getting textual changes (line by line). In AWSCM, changes are identifying in the code of the operations in following steps.

1. The operations are separated out from source code, using 'Code Analyzer' component in AWSCM. Separated code of operation is stored inside the folders ‘New’ and ‘Old’. Each file has same name of the operations and contains code of that operation.
2. Existing code of JDiff is applied to above separated operation to gather the changed operations. JDiff indentifying line-by-line changes between the entire operation file in New and Old folder. This find file (operations) which are changed.
3. We constructed UWSDL using 'WSDL Constructor' component of AWSCM.
4. These operation are mapped to old test suite to construct reduce regression test suite. Gather mapping for test sequences/steps in the test cases according to the changes.

We had implemented 'Code Analyzer', 'WSDL constructor' and mapping function between UWSDL operations to the test cases. We can also use intra-procedural analysis techniques for change impact analysis based on commonly known control and data flow analysis.

For the construction of ‘UnitWSDL’ as output, three inputs are required ‘*oldWSDL*’, ‘*newWSDL*’ and ‘*codeDifference*’ (can be gathered using utility, libraries etc). Construction of UWSDL is a three-step process.

First, gather the operations ‘*codeDifferenceOperation'*, which are present in ‘*codeDifference*’ as impact of the changes.

Second, construct the various parts of WSDL (Definition, XSD, Message, Port, Binding and Service), by using the operations gathered in first step. Definition and Service are constant (i.e. independent of our variable operations), whereas others are dependent on the operations.

Third, combine all the parts in a proper sequence to form a Unit WSDL.

File ConstructUnitWSDL(File *oldWSDL*, File *newWSDL*, *codeDifference*) {

String *cStartDef, cXSD, cMsg, cPort, cBinding, cService, cEndDef*

Array of String[] *codeDifferenceOperation*

[*codeDifferenceOperation* ∈ *codeDifference* | *codeDifferenceOperation* = Operation for which source code is modified]

*cStartDef* = **constructStartDefinition**(*newWSDL*)

*cXSD* = **constructXSD**(*newWSDL, oldSchema*)

*cMsg* = **constructMessage**(*newWSDL, codeDifferenceOperation*)

*cPort* = **constructPort**(*newWSDL, codeDifferenceOperation*)

*cBinding* = **constructBinding**(*newWSDL, codeDifferenceOperation*)

*cService* = **construcService**(*newWSDL*)

*cEndDef* = **constructEndDef**(*newWSDL*)

*UnitWSDL = cStartDef + cXSD + cMsg + cPort + cBinding + cService + cEndDef*

return *UnitWSDL* }

### 5.3.4 Algorithm for Computing Reduce WSDL

The Reduce WSDL of AWSCM takes any WSDL, gather its operations and then prompt user to select operations that are required for performing regression testing of web service. Using these selected operations, AWSCM semantically constructs RWSDL.

Web service operations are the most abstract component, so we have used them to make selection. Further internal components depend on the availability of code. We had not selected the internal component.

For the construction of 'Reduce WSDL' as output, two inputs are required '*newWSDL*' and '*Selection*' (done by using utility, libraries etc). Construction of RWSDL is a three-step process.

First, gather the operations '*selectedOperation'*, which are present in '*Selection*' as selection done by tester.

Second, construct the various parts of WSDL (Definition, XSD, Message, Port, Binding and Service), by using the operations gathered in first step. Definition and Service are constant (i.e. independent of our variable operations), whereas others are dependent on the operations.

Third, combine all the parts in a proper sequence to form a Reduce WSDL.

File ConstructRedueWSDL(File *newWSDL, Selection*) {

String *cStartDef, cXSD, cMsg, cPort, cBinding, cService, cEndDef*

Array of String[] *selectedOperation*

[*selectedOperation* ∈ *Selection* | *selectedOperation* = Operation for which is selected by user]

*cStartDef* = **constructStartDefinition**(*newWSDL*)

*cXSD* = **constructXSD**(*newWSDL, oldSchema*)

*cMsg* = **constructMessage**(*newWSDL, selectedOperation*)

*cPort* = **constructPort**(*newWSDL, selectedOperation*)

*cBinding* = **constructBinding**(*newWSDL, selectedOperation*)

*cService* = **construcService**(*newWSDL*)

*cEndDef* = **constructEndDef**(*newWSDL*)

*ReduceWSDL = cStartDef +cXSD +cMsg + cPort + cBinding + cService +cEndDef*

**return** *ReduceWSDL* }

### 5.3.5 Algorithm for Computing Combined WSDL

Combined WSDL is constructed by taking the union of the operations in the D/U/R WSDL such that it contains only unique and non-redundant operations. Basis for taking union is to neglect redundant operations present in different subset WSDLs. Thus, it is better to take union of the three sets to reduce the processing time by removing redundant and useless operation while constructing Combined WSDL. For example DifferenceWSDL_1 has {A, B, C} operations, ReduceWSDL has {C, D} operations and UnitWSDL has {D, E, F} operations. Here, {C, D} operations appeared twice. Combined WSDL must have {A, B, C, D, E, F} operations instead of {A, B, C, C, D, D, E, F} operations.

For the construction of 'Combined WSDL' as output, two inputs are required '*newWSDL*' and '*uniqueOperation*' (union of the operations present in D/U/R WSDL). Construction of CWSDL is a three-step process.

First, gather the operation '*uniqueOperation'* that are present in '*newWSDL*Operation' as operations required for regression testing.

Second, construct the various parts of WSDL (Definition, XSD, Message, Port, Binding and Service), by using the operations gathered in first step. Definition and Service are constant (i.e. independent of our variable operations), whereas others are dependent on the operations.

Third, combine all the parts in a proper sequence to form a Combined WSDL.

File ConstructCombinedWSDL(File *newWSDL, uniqueOperation*) {

String *cStartDef, cXSD, cMsg, cPort, cBinding, cService, cEndDef*

Array of String[] *uniqueOperation*

[*uniqueOperation* ∈ *newWSDLOperation* | *uniqueOperation* = Operation collected once if it occurred in Difference, Unit , Reduce WSDL]

*cStartDef* = **constructStartDefinition**(*newWSDL*)

*cXSD* = **constructXSD**(*newWSDL, oldSchema*)

*cMsg* = **constructMessage**(*newWSDL, uniqueOperation*)

*cPort* = **constructPort**(*newWSDL, uniqueOperation*)

*cBinding* = **constructBinding**(*newWSDL, uniqueOperation*)

*cService* = **construcService**(*newWSDL*)

*cEndDef* = **constructEndDef**(*newWSDL*)

*CombinedWSDL = cStartDef +cXSD +cMsg +cPort +cBinding+cService +cEndDef*

return *CombinedWSDL* }

### 5.3.6 Algorithm for Computing Parameter WSDL

The Parameter WSDL is constructed by taking the change affected operations, such that its method and operations undergone changes which is called by the operation. Called operation can also be part of other web service i.e. this is also applicable in web service composition

For the construction of 'Parameter WSDL' as output, two inputs are required '*newWSDL*' and '*affectedOperation*' (operations who has inter-procedural called changed operation or method). Construction of PWSDL is a three-step process.

First, gather the operations '*affectedOperation'* that are present in '*newWSDLOperation*' as their method and operation calling required to regression testing.

Second, construct the various parts of WSDL (Definition, XSD, Message, Port, Binding and Service), by using the operations gathered in first step. Definition and Service are constant (i.e. independent of our variable operations), whereas others are dependent on the operations.

Third, combine all the parts in a proper sequence to form a Parameter WSDL.

File ConstructParameterWSDL(File *newWSDL, affectedOperation*) {

String *cStartDef, cXSD, cMsg, cPort, cBinding, cService, cEndDef*

Array of String[] *affectedOperation*

[*affectedOperation* ∈ *newWSDLOperation* | *affectedOperation* = Operation calling the changed method and operations (may be of other service)]

*cStartDef* = **constructStartDefinition**(*newWSDL*)

*cXSD* = **constructXSD**(*newWSDL, oldSchema*)

*cMsg* = **constructMessage**(*newWSDL, affectedOperation*)

*cPort* = **constructPort**(*newWSDL, affectedOperation*)

*cBinding* = **constructBinding**(*newWSDL, affectedOperation*)

*cService* = **construcService**(*newWSDL*)

*cEndDef* = **constructEndDef**(*newWSDL*)

*ParameterWSDL = cStartDef +cXSD +cMsg +cPort+cBinding+cService +cEndDef*

return *ParameterWSDL* }

### 5.3.7 Algorithm inside 'RRTS Constructor' for Computing Reduced Regression Test Suite

For generating the Reduce Regression Test Suite (RRTS), first gather the affected operation in the web service and store the '*requiredOperations*' from the diff applied to web service to construct a Subset WSDL, this is demonstrated in detail in chapter 3.

Secondly, pass the '*requiredOperations*' to the ReducedRegressionTestCase function to gather the required regression test cases. Description of algorithm to construct RRTS is as follows

1) T* = T-old: Reduced test case initially has old test cases.

2) The declarations of inserted, deleted, modified, and unmodified operations with their test cases.

3) If operations are deleted then delete its corresponding test cases.

4) If operations are inserted then add test cases templates for the new regions.

5) If operations are modified then add test cases for the selective test sequence/step for the modified regions.

6) T* = T* - tu were delete all the remaining unused test cases which are already executed in the testing of previous version of web service. These test cases are not required because component of those test cases in web service are already tested.

7) Return the T*(reduced test case) is calculate by algorithm is T* = ti + tm -td – tu.

WS – Web service version 1
WS* – Modified Web service version 2
T-old= Test Cases for the code of web service
T-new = Test Cases for the code of WS*
T* = Reduced test Cases

*requiredOperations* = Operations that are undergone changes.

**Function** ReducedRegressionTestSuite (*requiredOperations, T-old, T-new*)

{ **Assignment**

**1.** *T* = T-old* //Reduced test case i.e. T* must have T-old initially, then we can reduce test cases from T-old.

**2.** *Del*= **deleted operation** of web service
*Ins*= **inserted operation** of WS*
*Mod*= **modified operation** of WS*

*td:* {*td* ∈ *T-old* | *td* = Test cases of a **deleted** operation of WS}
*ti:* {*ti* ∈ *T-new* | *ti* = Test cases of a **inserted** operation of WS*}
*tm:* {*tm* ∈ *T-new* U *T-old*| *tm* = Selective test cases of a **modified** part of operation of web service to make WS*}
*tu:* {*tu* ∈ *T-old* | *tu* = Test cases of a **unmodified** operation of WS}
*requiredOperation:* {*requiredOperation* ∈ *operationInNewWSDL* | *operation* = Operation which is required to construct the reduced regression test suite]

**3.** **if(***requiredOperations==Del***)** //search for the **deleted** operations of WS
{
*T** = *T** - *td* //reducing test cases of **deleted** operations
}

**4.** **if(***requiredOperations==Ins***)** // **inserted** operations of WS*
**{**
*T** = *T** + *ti* //inserting test cases of **inserted** operations of WS*
}

**5.** **if(***requiredOperations==Mod***)** // **modified** operations of web service in WS*
**{**
*T** = *T** + *tm* //inserting test sequences/steps of **modified** operations
}

**6.** *T** = *T** - *tu* //reducing test cases of **unmodified** operations

**7.** *TestSuite* = **constructTestSuiteSemantics**(T*) //{This can't be standardized!! because every testing tool have its own semantics to construct its Test Suites}

**8.** **Return** *TestSuite* //Return reduced regression test suite with test cases T*

} ***ReduceRegressionTestSuite Construction algo ends***

Three RRTS are constructed by automated selection of test cases of old test suite for the operation in the RWSDL, DWSDL and UWSDL respectively. Form the operations of CWSDL and PWSDL we can construct Combined RRTS (CRRTS) or Parameter RRTS (PRRTS) respectively. CRRTS and PRRTS consist all the unique test cases i.e. redundant test case were eliminated. We have reduced number of worthless test cases to perform efficient regression testing of web service.

According to paper [65], "an obsolete test case specifies an invalid input to new code that is an invalid input-output relation. These test cases required to be removed during regression test selection. If we cannot effectively determine test case obsolescence, we cannot effectively judge test case correctness." Therefore, in both our approaches ORTWS and PRTWS we had not selected test cases for the operations that do not exist in new version.

New test case templates were created for inserted operation, which requires test data to be filled. We had not created any automated test data or assertions. Traditionally in regression testing, claim of reduce test case is true in normal cases. But for exceptional case if there are more number of insertions then more number of test data required to be filled in every test template either automatically or manually. Test suite grows because changed code portion grows due to many changes in requirement i.e. software code is incremented significantly.

We had reused the old test suite to generate new Reduce Regression Test Suite (RRTS). The previously generated RRTS_1 can be also be reused as input to generate another RRTS_2. All SoapUI shows the reuse of old test suite to generate RRTS.

We cannot ensure complete perfectness that we had not missed anything. However, we verified manually we gathered almost all the change after using our code integrated with other diff api. Assuring the perfection of gathering changes dependence on the fine-grained analysis diff utility.

To prove perfectness of our approach we had demonstrated the case studies in which we made intentional changes in our own web service (BookService and SaaS). AWSCM automatically gathered all the changes and mapped these change to the test suites. For other case studies (Eucalyptus), we check manually that whether AWSCM gathered the changes perfectly or not.

## 5.4 Integration to SoapUI and JMeter

Tool GUI, having too many buttons look complex interface but this can be considered because the tool is a prototype to demonstrate our proposed approach. In all the frames of GUI, ‘new and old WSDL’, ‘Difference, Reduce, Unit and Combined WSDL’ and ‘test suite’ is shown as XML tree viewer, to show the working of this approach. These descriptions of XML files are not require when these approaches implemented in web service testing tools like JMeter and SoapUI.

If ORTWS is implemented as the part of testing tool JMeter and SoapUI then following button are not required, ‘Test suite’, ‘new WSDL’, ‘Difference, Reduce, Unit and Combined WSDL’ and ‘RRTS’. The JMeter and SoapUI can internally compensate buttons ‘Test suite’ and ‘New WSDL’ as test suite, new WSDL is already present in web service testing environment. Buttons ‘Difference, Reduce, Unit and Combined WSDL’ and ‘RRTS’ can also be reduced to one button named on ‘RRTS’ for concept of ‘change impact analysis based regression testing of web service’.

PRTWS is implemented as the part of web service testing tools like JMeter and SoapUI then, the button ‘Test suite’ is not required, because a test suite is already their inside testing environment.

As ‘Difference’, ‘Reduce’, and ‘Unit’ WSDL is an intermediate process. Thus, the buttons ‘Difference, Reduce, Unit, and Combined WSDL’ are replacing by only one button ‘Combined WSDL’ when our approach integrated to the JMeter and SoapUI. An actual input require is ’New Code’, ‘Old Code’ and ‘Old WSDL’ and actual output is ‘CRRTS’ which can be further reduced using and sequence reducer approach in figure 5.5.

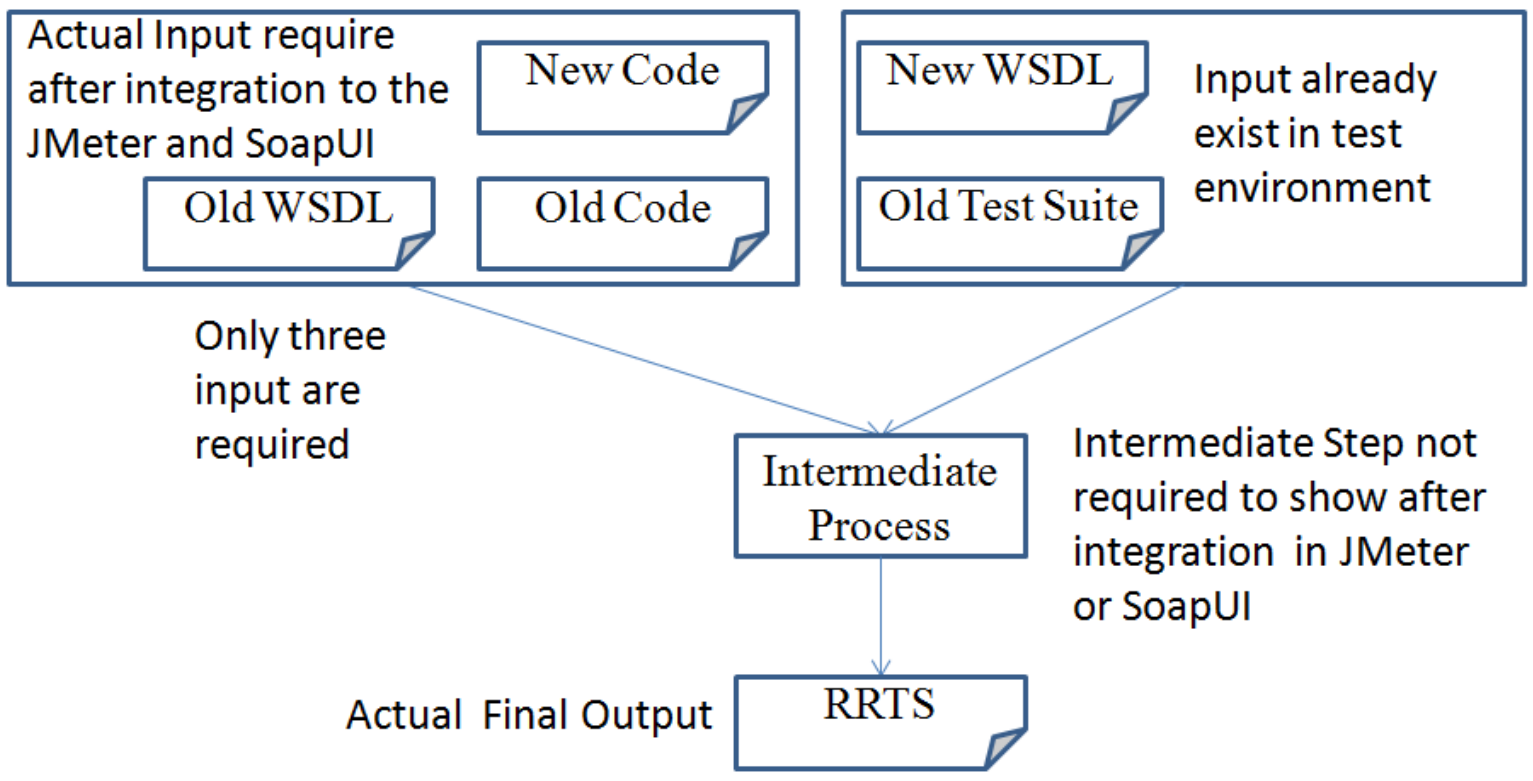

**Figure 5-5 AWSCM intergration with SoapUI and JMeter**

## 5.5 Applications of AWSCM

We find that AWSCM can be useful in regressions testing, selective retesting, and top-down development of web service in figure 5.6. Regression testing can be performed with the two approaches ORTWS and PRTWS in Chapter 3 and Chapter 4 respectively. The user can do selective retesting by selecting the required test cases as well as their test sequences/steps.

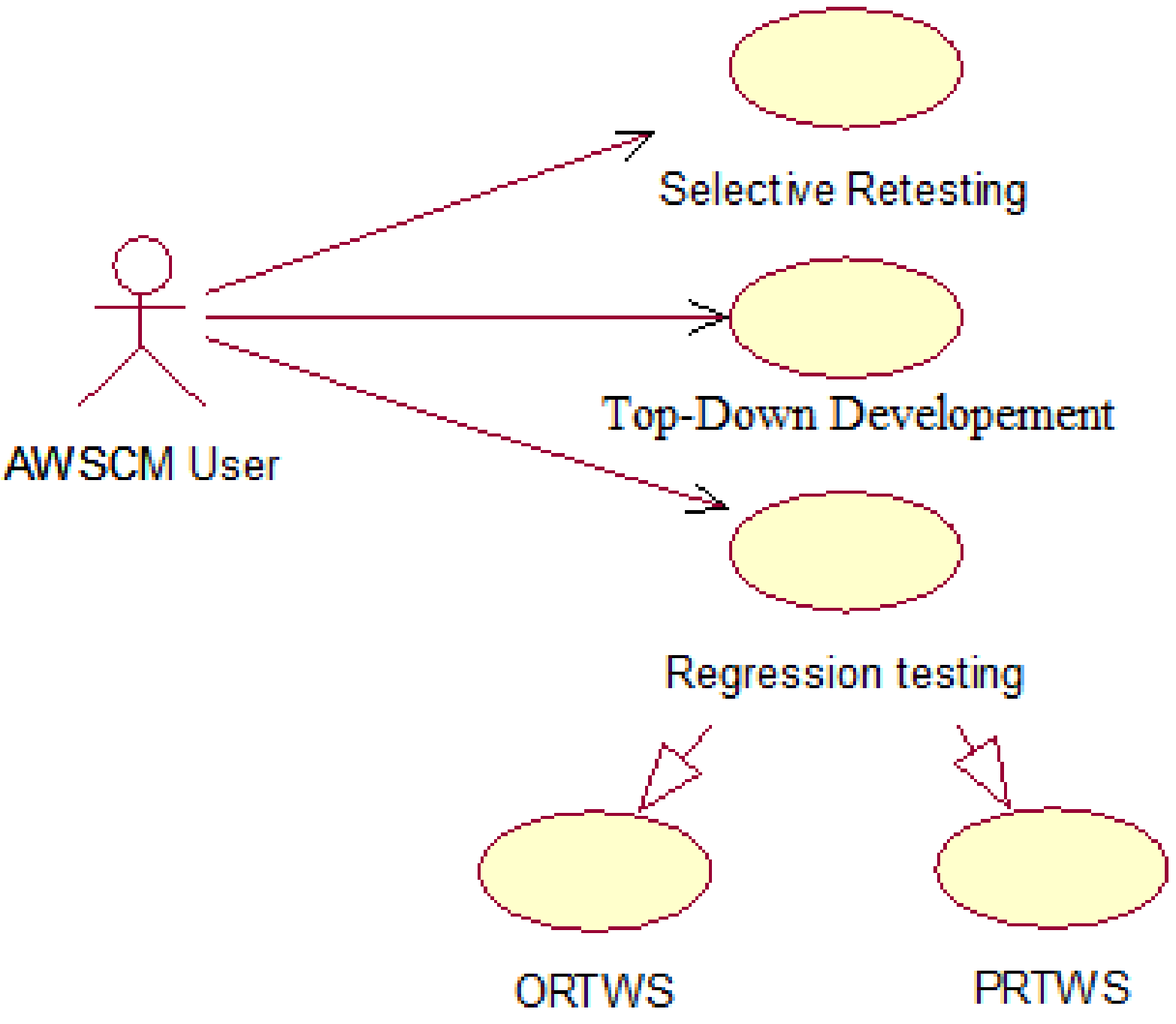

**Figure 5-6 AWSCM use case**

### 5.5.1 AWSCM in Regression Testing of Web Service

From the study of this thesis in web service testing, if test case were saved with certain well-defined setting then we can easily map them to changes in those scenarios for change impact analysis. The case study of ORTWS and PRTWS on real world web service is in Table 5-2.

**Table 5-2 Web service based case studies of ORTWS & PRTWS**

<table>
<tr><th rowspan="2">Case Study web service Projects</th><th colspan="3">CWSDL for ORTWS</th><th colspan="2">PRTWS</th><th rowspan="2">Type of Changes</th></tr>
<tr><th>DWSDL</th><th>RWSDL</th><th>UWSDL</th><th>Dynamic Black Box Testing</th><th>Static White Box Testing</th></tr>
<tr><td>Eucalyptus</td><td>Y</td><td>Y</td><td>Y</td><td>X</td><td>Y*</td><td>Available on Github</td></tr>
<tr><td>SaaS</td><td>Y</td><td>Y</td><td>Y</td><td>Y</td><td>Y</td><td>We did</td></tr>
<tr><td>BookService</td><td>Y</td><td>Y</td><td>Y</td><td>Y</td><td>Y</td><td>We did</td></tr>
<tr><td>Amazon web service</td><td rowspan="5">Different versions of WSDL is not available</td><td>Y</td><td rowspan="5">Code is not available</td><td>Y*</td><td rowspan="5">Code is not available</td><td rowspan="5">No changes</td></tr>
<tr><td>Sunset Sunrise service</td><td>Y</td><td>Y</td></tr>
<tr><td>Bible web service</td><td>Y</td><td>Y</td></tr>
<tr><td>Currency Conversion web service</td><td>Y</td><td>Y</td></tr>
<tr><td>Weather web service</td><td>Y</td><td>Y</td></tr>
</table>

* According to the inference from our work in thesis, we can say that it is easy to perform these actions but as we are not able to show this in the thesis because we do not have any test data for Eucalyptus and AWS.

### 5.5.2 AWSCM in Top-Down development of Web Service

Top-Down development can done for the operation we want by both selectively and automatically for the operations undergone changes in subsequent version. Our analysis for the application of AWSCM gives us interesting result in top-down development of web service with the help of Subset WSDL i.e. D/U/R/C WSDL. We used WSDLs of web services Eucalyptus, Amazon and various others. WSDLs of different version given as input to our tool AWSCM for the construction of Subset WSDLs, for changed and selected operation of input WSDLs. These WSDLs were given as input to the Eclipse and Netbeans, to generate proper development template. Development template required to be re-written with its code. Again, acceptance of these WSDL by Eclipse and Netbeans justifies that these WSDLs are semantically and structurally correct according to the WSDL standard. Hence, top down development could do to black box web services with selective operations.

The top down development template example shown in figure 5-7, for the operation captured in the Combined WSDL of figure 5-9. Similarly, top down development template example in figure 5-8 for the three operation captured in the Parameter WSDL out of four operation of input WSDL shown in figure 5-10.

```java
@WebService(serviceName = "BooKServiceService", portName = "BooKServicePort",
endpointInterface = "bookservice.BooKService",
targetNamespace = "http://BooKService/",
wsdlLocation = "WEB-INF/wsdl/CombineWSDLReEngineering/CombinedWSDL.wsdl")
public class CombineWSDLReEngineering {

    public String getVerseByBooKAndChapterAndVerseNumber(int bookNumber,
                                            int chapterNumber, int verseNumber) {
        //TODO implement this method
        throw new UnsupportedOperationException("Not implemented yet.");
    }
}
```

**Figure 5-7. Top-Down development for CWSDL of fig 5-9.**

```java
@WebService(serviceName = "BooKServiceService", portName = "BooKServicePort",
endpointInterface = "bookservice.BooKService", targetNamespace = "http://BooKService/",
wsdlLocation = "WEB-INF/wsdl/NewWebServiceFromWSDL/ParameterWSDL.wsdl")
public class NewWebServiceFromWSDL {

    public java.lang.String getAbstractOfChapter(int bookNumber, int chapterNumber) {
        //TODO implement this method
        throw new UnsupportedOperationException("Not implemented yet.");
    }

    public java.lang.String getVerseByBooKAndChapterAndVerseNumber(int bookNumber,
                                            int chapterNumber, int verseNumber) {
        //TODO implement this method
        throw new UnsupportedOperationException("Not implemented yet.");
    }

    public java.lang.String getAllVerseByBookAndChapterNumber(int bookNumber,
                                                      int chapterNumber) {
        //TODO implement this method
        throw new UnsupportedOperationException("Not implemented yet.");
    }
```

**Figure 5-8 Template of PWSDL in fig 5-10 with 3 operations out of 4 operations are undergone changes at calling interaction level.**

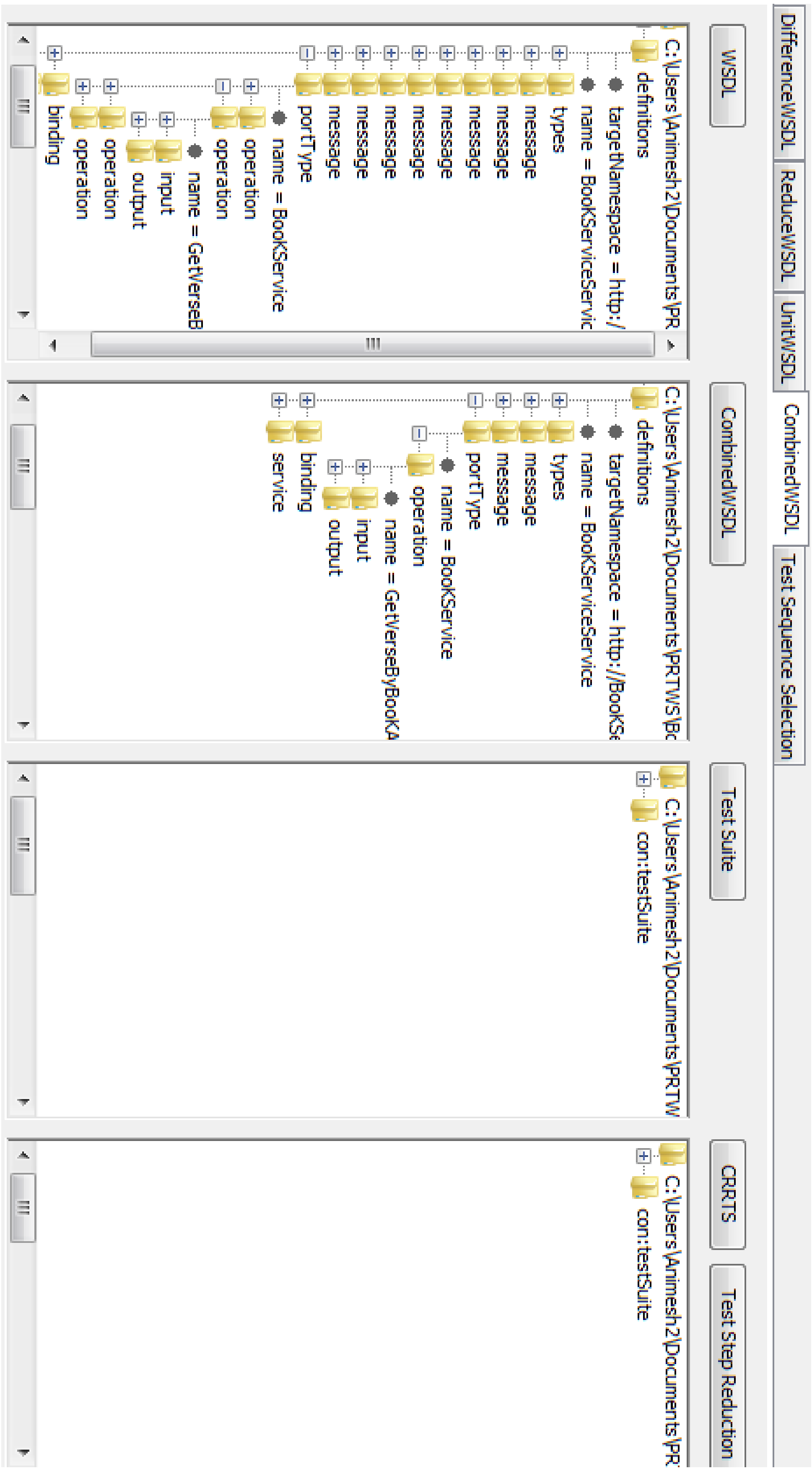

**Figure 5-9 Combined WSDL and CRRTS construction using the intermediate information.**

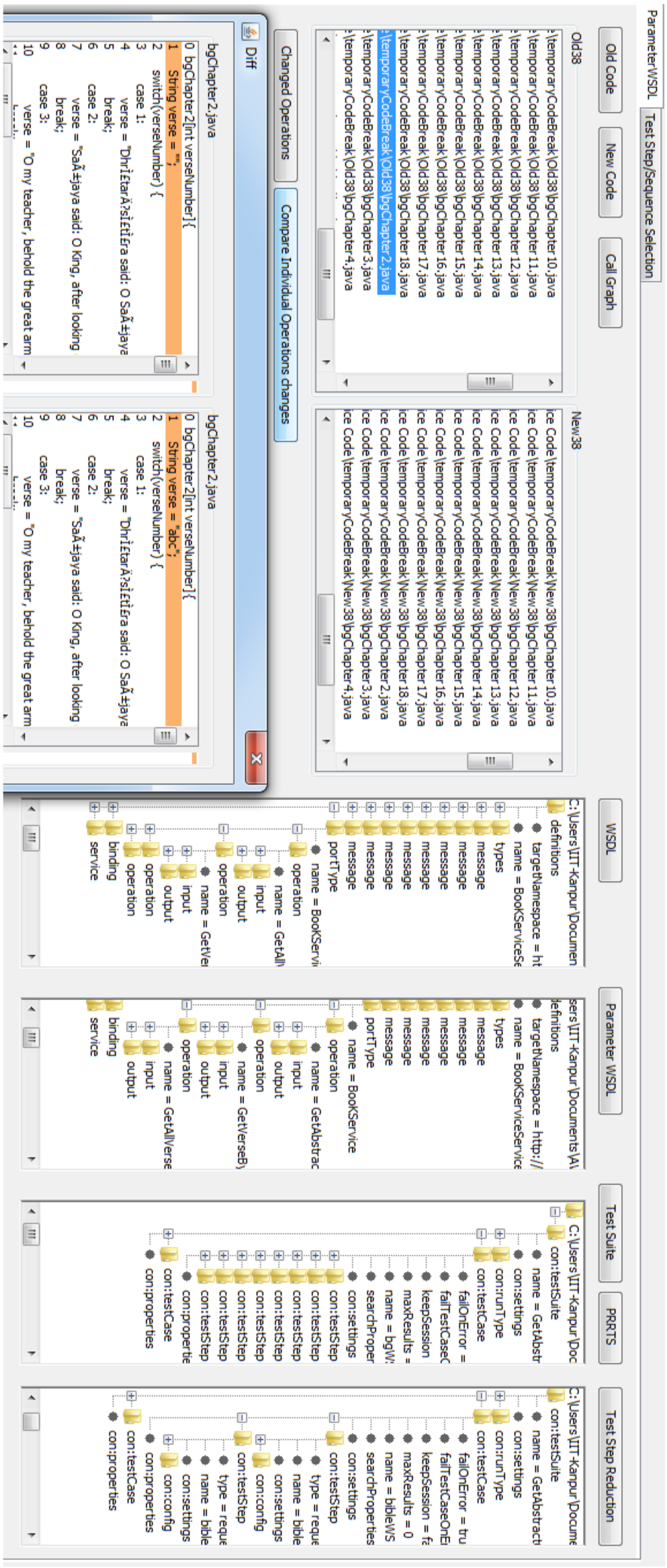

**Figure 5-10 PRTWS module of AWSCM with intermediate information of capturing the inter-procedural operations and method change for the 'BookService'.**

### 5.5.3 AWSCM in Selective Re-testing

Additionally for test sequence/step reduction for four dynamic web services, namely, Currency conversion, Global weather, Sunset Sunrise service, Bible web services and Amazon web service. Using the tool in figure 5.11, our constructed RWSDL further used to construct RRTS. We also used AWSCM for the code analysis of SaaS and Eucalyptus_CC (handler.c at Githhub) to perform data flow analysis. In this feature, AWSCM gathers affected parameter and based on these parameter, tool construct UWSDL with these operation parameters. Gathered impacts mapped to the test sequences/steps, which were inside test cases for a particular operation of the test suite. Selective regression testing of web service is performed by selecting the targeted operations to construct RWSDL from WSDL of Amazon web service as shown in Figure 5.12 (a). The RRTS constructed for selective scenarios for Currency Conversion, which contain test sequences/steps in test cases similar to the old test suite in Figure 5.12 (b). The web service projects along with developed Subset WSDLs in our case studies shown in Table 5-3. The case study on Bible web service for operation name ‘GetBibleWordsByBookTitleAndChapter’ shown in figure 5-13 and operation ‘GetBibleWordsByChapterAndVerse’ shown in figure 5-14. The case study on Bible web service for operation name ‘GetSunSetRiseService’ shown in figure 5-15 (a) and figure 5.15 (b) for Jabalpur and Indore city respectively.

**Table 5-3 Web service projects and construction of proposed WSDL**

| **Case Study web service Projects** | **CWSDL** | | |
|---|---|---|---|
| | ***DWSDL*** | ***RWSDL*** | ***UWSDL*** |
| Eucalyptus | Y | Y | Y |
| SaaS | Y | Y | Y |
| BookService | Y | Y | Y |
| Amazon web service | Different versions of WSDL was not available | Y | Code was not available |
| Sunset Sunrise service | | Y | |
| Bible web service | | Y | |
| Currency Conversion web service | | Y | |
| Global Weather web service | | Y | |

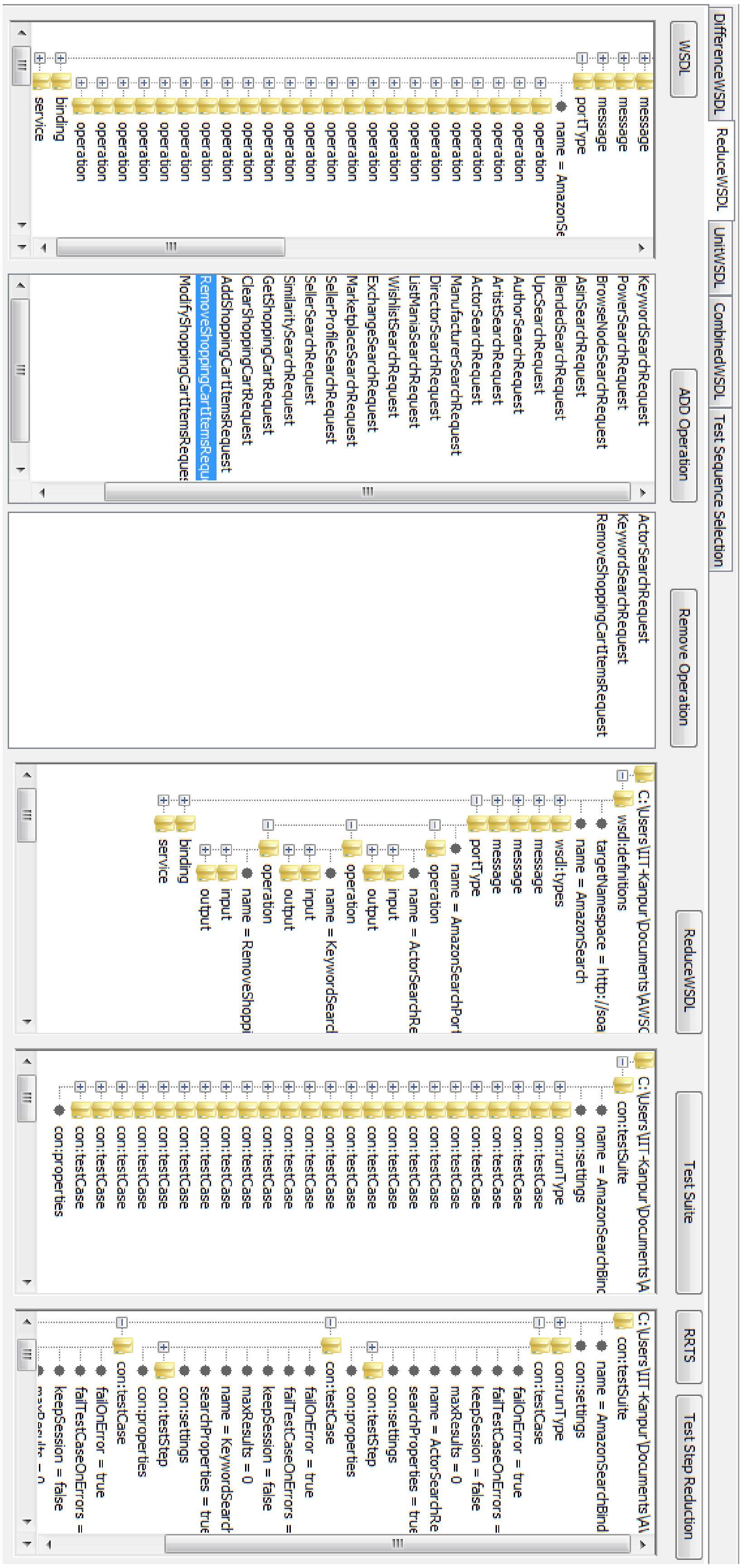

**Figure 5-11 ReduceWSDL of AWSCM for Amazon WSDL.**

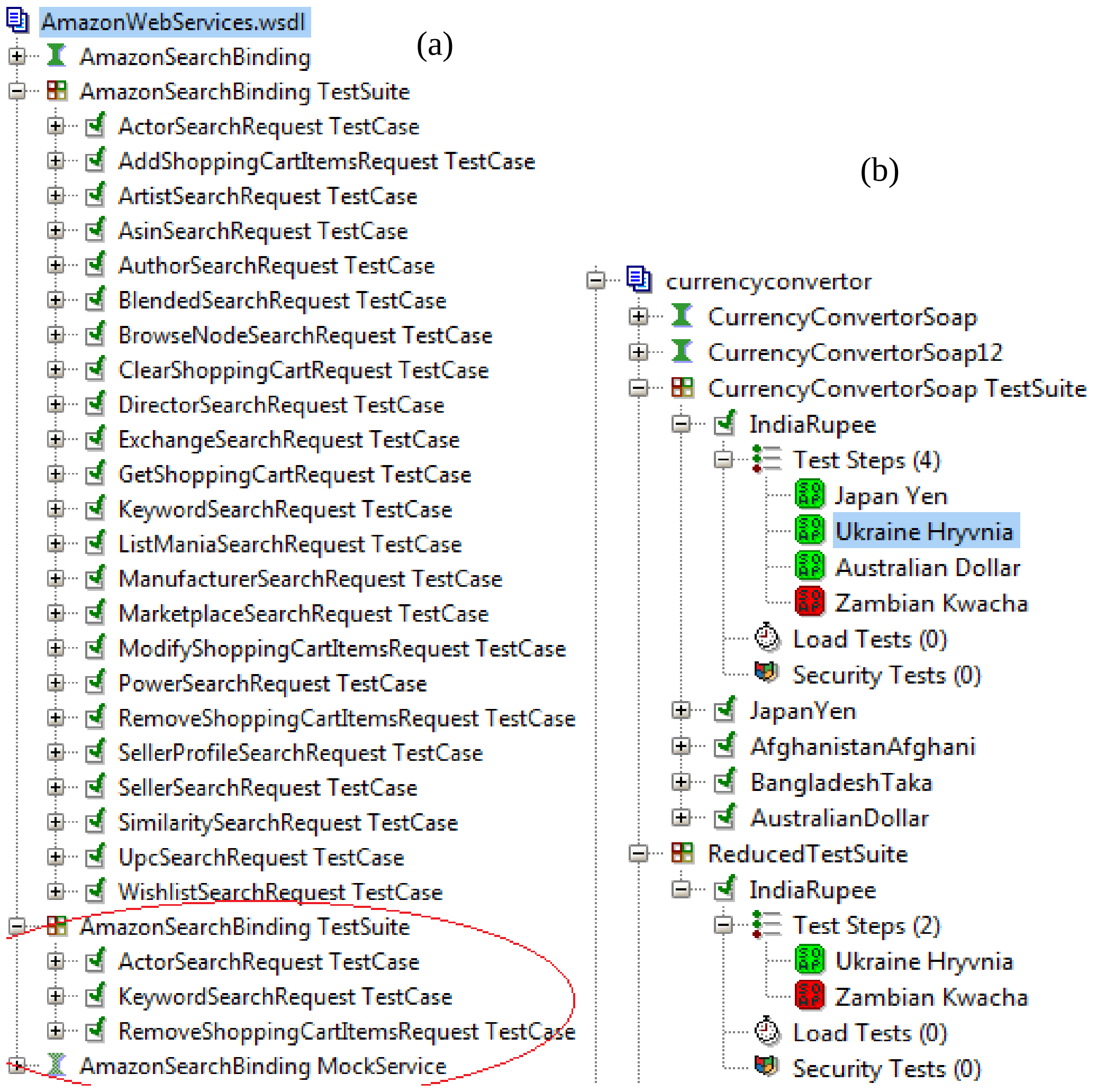

**Figure 5-12 Snapshot of SoapUI showing the old test suite and RRTS (a) Amazon consumer WSDL (b) test sequences/steps of 'India'.**

We performed an analysis on SaaS was our own developed java web service, deployed on glassfish and used for manipulation about 200 lines of code. Firstly using ORTWS we gathered changed operation (using DWSDL and UWSDL) while SaaS_3 was changed to SaaS_4. Tool gathered three operations are changed. AWSCM selected the three test cases to construct RRTS as shown in figure 5.16. Secondly, we further reduced test sequences/steps using PRTWS technique. RRTS generated by ORTWS is given as input to the PRTWS feature for further test sequence/step reduction as shown in figure 5.17. Effort reduction is estimation with respect to the re-test all. ORTWS got 3 out of 4 as finally changed operations as shown in figure 5.16. PRTWS further reduce by selecting 2 out of 6 test sequences/steps for operation 'editFile' as shown in figure 5.17.

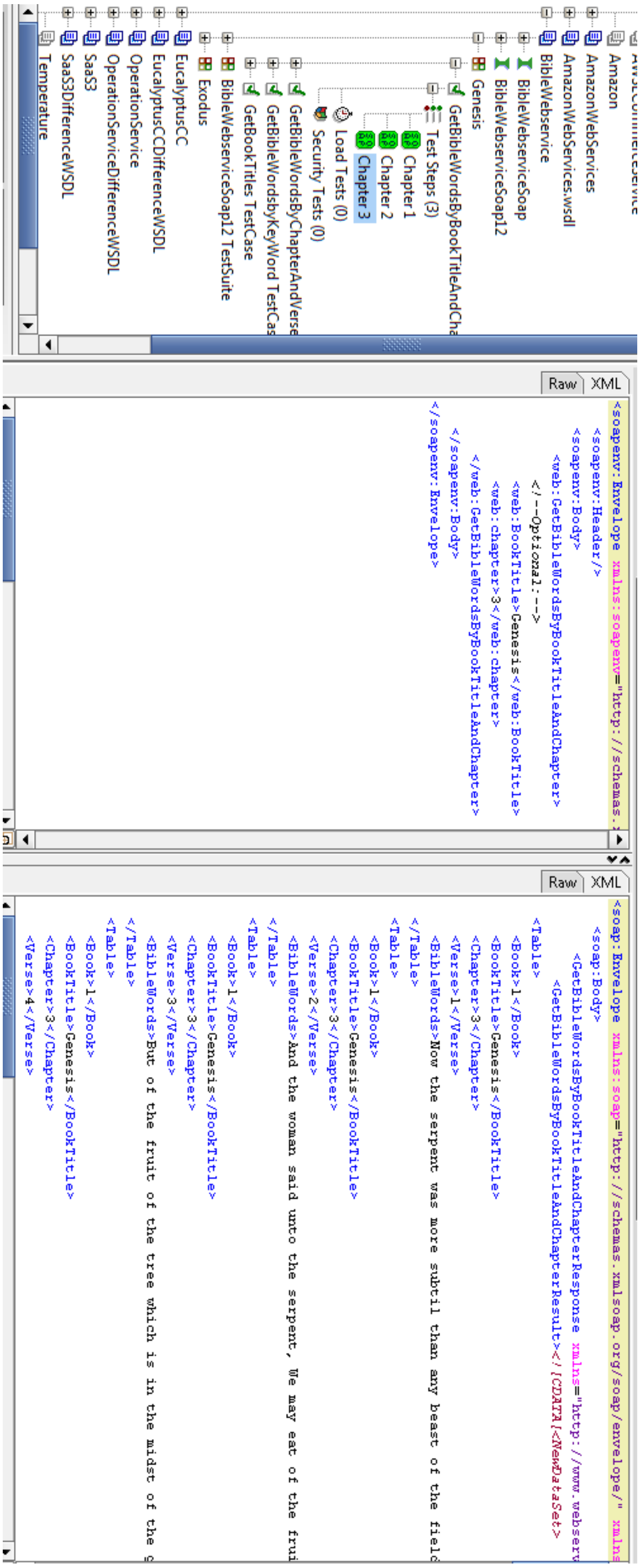

**Figure 5-13 Bible web service for geting verse by book title and chapter**

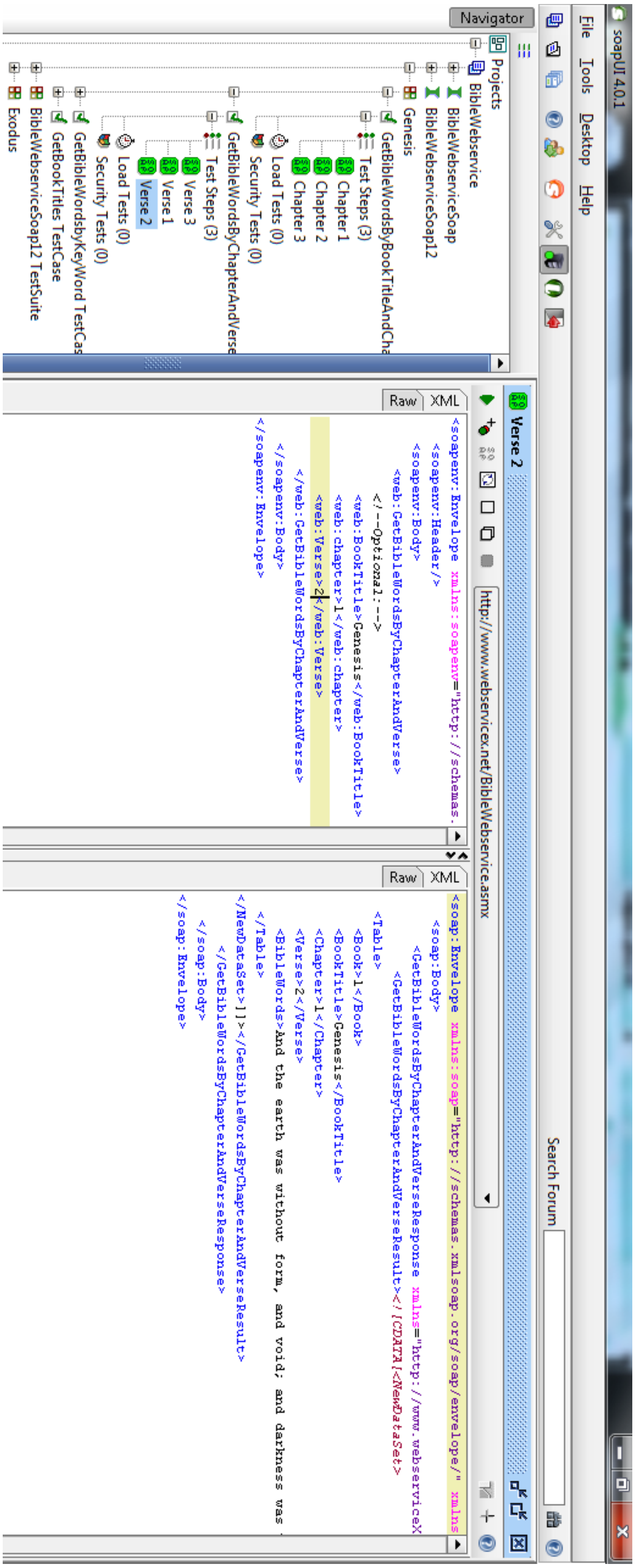

**Figure 5-14 Bible web service soap respone of verse number, book title, and chapter.**

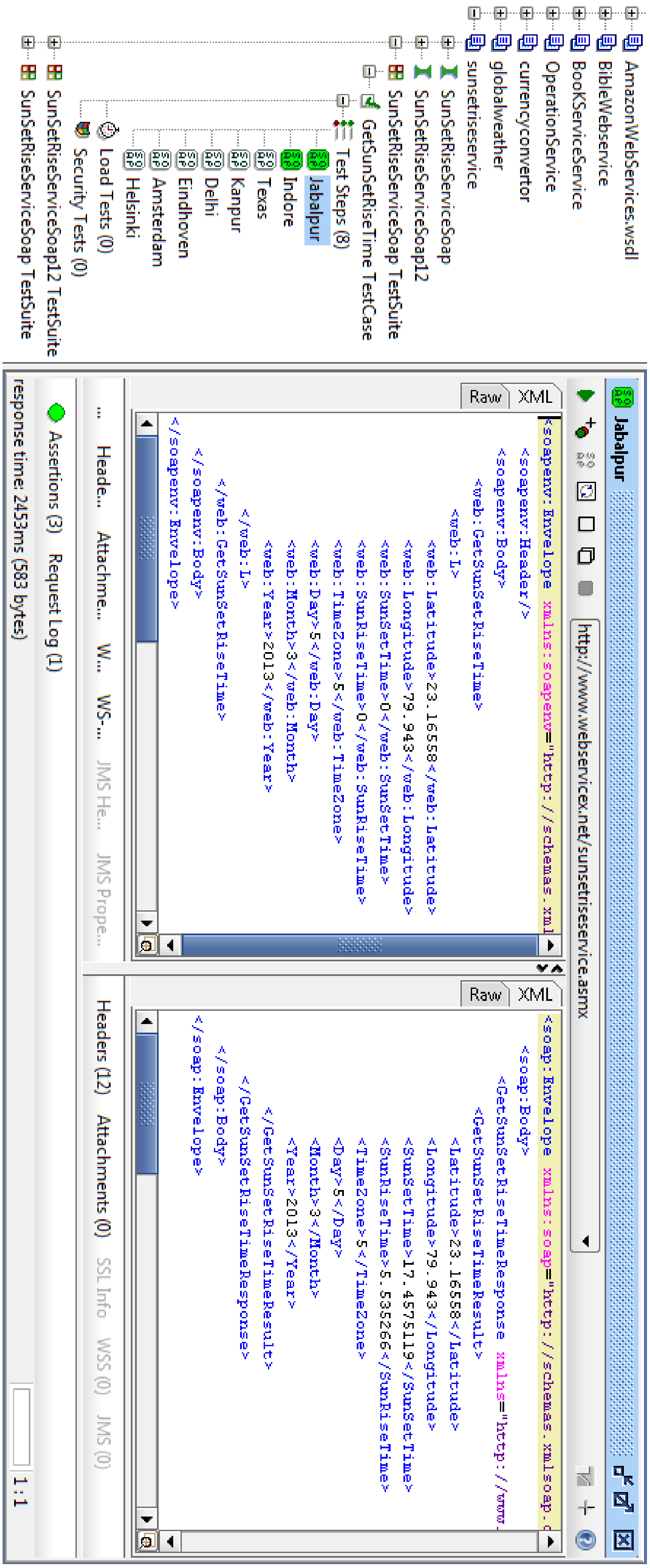
AmazonWebServices.wsdl
BibleWebservice
BookServiceService
OperationService
currencyconvertor
globalweather
sunsetriseservice
SunSetRiseServiceSoap
SunSetRiseServiceSoap12
SunSetRiseServiceSoap TestSuite
GetSunSetRiseTime TestCase
Test Steps (8)
Jabalpur
Indore
Texas
Kanpur
Delhi
Eindhoven
Amsterdam
Helsinki
Load Tests (0)
Security Tests (0)
SunSetRiseServiceSoap12 TestSuite
SunSetRiseServiceSoap TestSuite
Jabalpur
http://www.webservicex.net/sunsetriseservice.asmx
Raw
XML
<soapenv:Envelope xmlns:soapenv="http://schemas.xml
<soapenv:Header/>
<soapenv:Body>
<web:GetSunSetRiseTime>
<web:L>
<web:Latitude>23.16558</web:Latitude>
<web:Longitude>79.943</web:Longitude>
<web:SunSetTime>0</web:SunSetTime>
<web:SunRiseTime>0</web:SunRiseTime>
<web:TimeZone>5</web:TimeZone>
<web:Day>5</web:Day>
<web:Month>3</web:Month>
<web:Year>2013</web:Year>
</web:L>
</web:GetSunSetRiseTime>
</soapenv:Body>
</soapenv:Envelope>
Heade... Attachme... W... WS-... JMS He... JMS Prope...
Raw
XML
<soap:Envelope xmlns:soap="http://schemas.xmlsoap.o
<soap:Body>
<GetSunSetRiseTimeResponse xmlns="http://www.
<GetSunSetRiseTimeResult>
<Latitude>23.16558</Latitude>
<Longitude>79.943</Longitude>
<SunSetTime>17.457119</SunSetTime>
<SunRiseTime>5.535266</SunRiseTime>
<TimeZone>5</TimeZone>
<Day>5</Day>
<Month>3</Month>
<Year>2013</Year>
</GetSunSetRiseTimeResult>
</GetSunSetRiseTimeResponse>
</soap:Body>
</soap:Envelope>
Headers (12) Attachments (0) SSL Info WSS (0) JMS (0)
Assertions (3) Request Log (1)
response time: 2453ms (583 bytes)
1 : 1

(a)

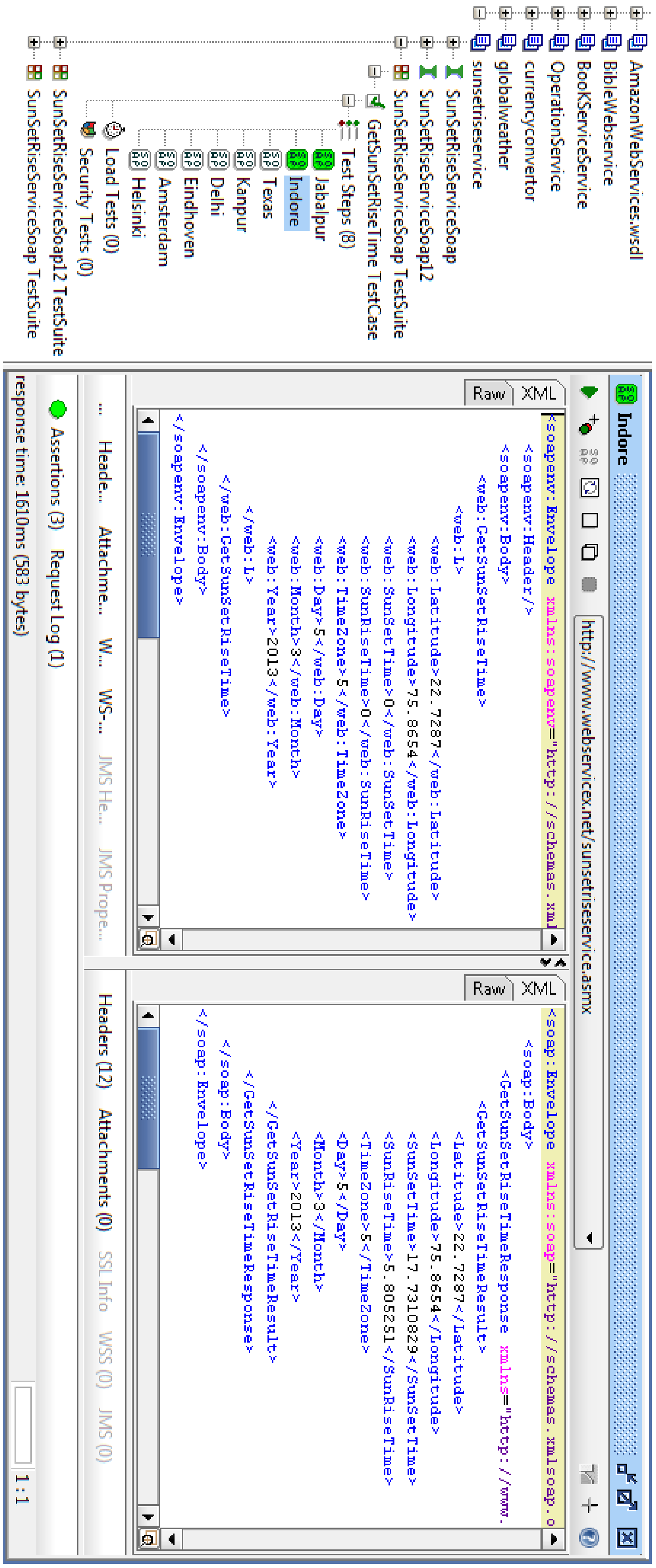

(b)

**Figure 5-15 Sunset and Sunrise web service for latitude & longitude of (a) Jabalpur (b) Indore city.**

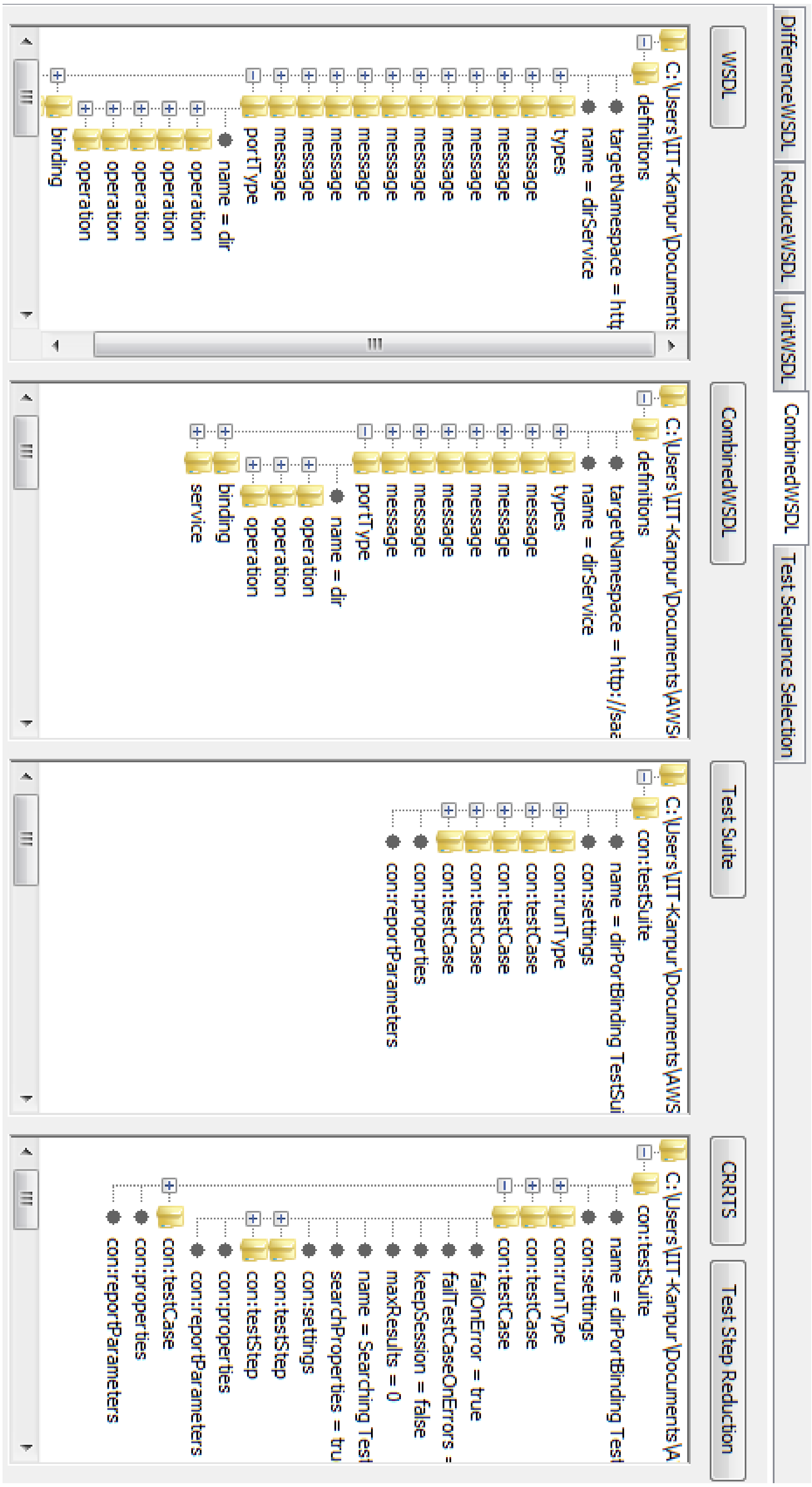

**Figure 5-16 Combined WSDL to construct RRTS for SaaS_3**

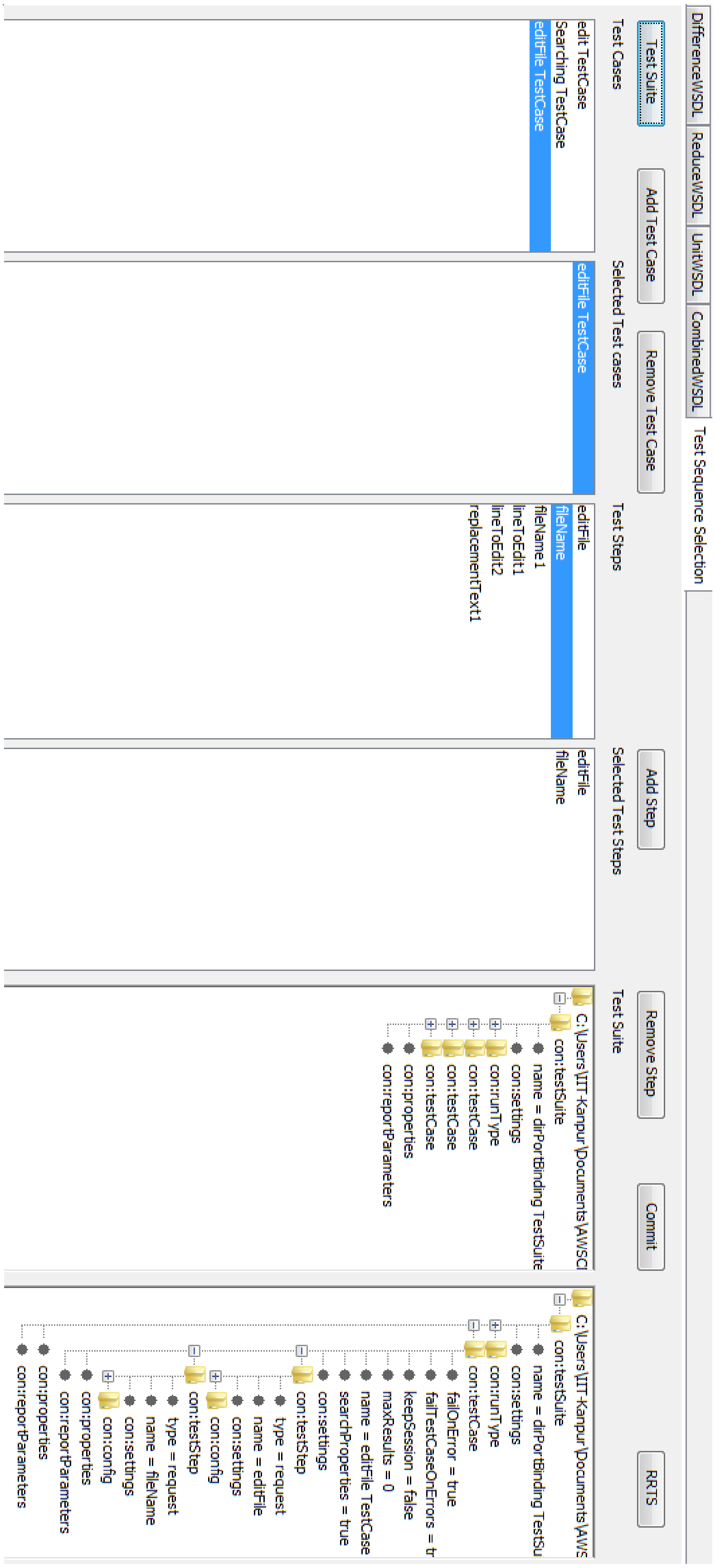

**Figure 5-17 PRTWS for the operation 'editFile' of SaaS_3**

## 5.6 Summary

AWSCM constructed functionality of generated four different type of SWSDL. To validate semantics and structure of SWSDL (i.e. D/U/R WSDL), SWSDLs were applied to the different software like SoapUI, JMeter, Eclipse and NetBeans. SWSDL perfectly performed regression testing and top down development for various WSDL based projects (Eucalyptus, AWS etc). In the future, many Subsets WSDLs can be generated for various domains/fields.

AWSCM is a prototype not an actual implementation of the proposed approach. Implementation can done when this prototype integrate with the standard web service testing tools like SoapUI and JMeter. After integration, GUI will be completely different. A description of the integration is in figure 5.5. In all the frames of GUI, 'new and old WSDL', 'Difference, Reduce, Unit and Combined WSDL' and 'test suite' is shown as XML tree viewer, to show the working of the approach. These descriptions of XML files are not require when our approaches is integrated with web service testing tools like JMeter and SoapUI.

# CHAPTER 6

# CONCLUSION & FUTURE WORK

Regression testing of web service can be optimized by doing selective testing of inserted and modified portions of the web service code, with enough assurance that new and old code would work according to changed requirements. In this thesis, we have described two approaches of efficient regression testing, namely, Operationalized Regression Testing of Web Service (ORTWS) and Parameterized Regression Testing of Web Service (PRTWS) along with our developed tool-support named AWSCM, that prototypes these approaches.

The ORTWS approach finds the changed operations and use this information for reducing the efforts in regression testing. The tool AWSCM constructs three types of Subset WSDLs, namely, DWSDL, RWSDL, and UWSDL that can be used to generate Reduce Regression Test Suites (RRTS). CWSDL is constructed by combining the unique operations in (D/U/R) WSDL. Efficient regression testing of web service is performed using Combined RRTS (CRRTS) generated from CWSDL. We have estimated the reduction in regression testing efforts based on the change ratio between the sizes of code that undergone changes and the new web service.

The proposed approach of PRTWS helps in formalizing, analyzing, and generalizing of dynamic analysis of web service testing. The two approaches are prototyped as an automated tool, in which selective test sequences/steps were used to generate reduced test suite for SoapUI. We require selecting the modified primary parameter scenario (i.e. test case), to construct has reduced test suite for the modified code. This observation is exploited in the context of web service where we used a primary parameter based web service regression testing technique for efficient regression testing of web services.

In general, ORTWS can be applied to any type of operations in WSDL where as PRTWS can be applied only to the special cases of operational behavior where some of the parameters are interdependent on each other. The main contribution of this thesis is that it captures the changes at both WSDL and code, and maps these changes to smaller Subset WSDL. By facilitating the standard WSDL parsing techniques, the operations in Subset WSDL are further mapped to their respective test cases to perform efficient regression testing of web service. In general, PRTWS approach works in tracing inter-procedural calls to method and operation based change impact analysis to map these changes into parameter based test suites. This thesis explores the mapping between the code and WSDL and facilitates WSDL based web service testing even with or without having any detailed knowledge of code.

We have used a 'Diff' utility to capture the changes between two WSDLs and code. A more structured form of captured changes will allow to perform a more fine grained mapping of the changes with the test cases. This may help to improve the regression testing efforts further.

We had developed the tool AWSCM only for two approaches based on three components, namely, operation, inter-procedural calls and XSD parameters. Thus, development is required for other layers of Web service analysis, namely, message posting, and client calling.

We performed small empirical study but detailed empirical study on AWSCM is still a future work.

# PUBLICATIONS

1. Animesh Chaturvedi and Atul Gupta. "A Tool Supported approach to Perform Regression Testing of Web Service". In the Proceeding of the 7th IEEE International Symposium on the Maintenance and Evolution of Service-Oriented and Cloud-Based Systems (IEEE MESOCA), pp 50-55. IEEE Xplore, TCSE, IEEE Cloud Computing, and IEEE Computer Society, Eindhoven, Netherlands, Sept. 2013.

2. Animesh Chaturvedi. "Reducing Cost in Regression Testing of Web service". In the Proceeding of CSI 6th International Conference on Sixth Software Engineering (CONSEG), IEEE Xplore, Indore, India, Sept 2012.